\pdfoutput=1

\documentclass[11pt,twoside,a4paper,cmspaper,final,collab]{cms-tdr}
\def\svnVersion{873eb5f-D}\def\svnDate{2020/02/24}\def\cmsCernNoTag{CERN-EP-2019-154}\def\cmsCernDate{\today}\def\cmsMessage{"Published in the European Physical Journal C as \href{http://dx.doi.org/10.1140/epjc/s10052-020-7739-7}{\doi{10.1140/epjc/s10052-020-7739-7}.}"}

\begin{document}\cmsNoteHeader{SUS-18-006}

\hyphenation{had-ron-i-za-tion}
\hyphenation{cal-or-i-me-ter}
\hyphenation{de-vices}

\newlength\cmsFigWidth
\ifthenelse{\boolean{cms@external}}{\setlength\cmsFigWidth{0.49\textwidth}}{\setlength\cmsFigWidth{0.65\textwidth}}
\ifthenelse{\boolean{cms@external}}{\providecommand{\cmsLeft}{upper\xspace}}{\providecommand{\cmsLeft}{left\xspace}}
\ifthenelse{\boolean{cms@external}}{\providecommand{\cmsRight}{lower\xspace}}{\providecommand{\cmsRight}{right\xspace}}
\ifthenelse{\boolean{cms@external}}{\providecommand{\cmsTableNoResize}[1]{#1}}{\providecommand{\cmsTableNoResize}[1]{\resizebox{\textwidth}{!}{#1}}}
\providecommand{\cmsTable}[1]{\resizebox{\textwidth}{!}{#1}}
\providecommand{\NA}{\ensuremath{\text{---}}}
\newlength\cmsTabSkip\setlength{\cmsTabSkip}{1ex}
\newcommand{\staul}{\ensuremath{\PSGt_{\mathrm{L}}}\xspace}
\newcommand{\staur}{\ensuremath{\PSGt_{\mathrm{R}}}\xspace}
\newcommand{\etau}{\ensuremath{\Pe\tauh}\xspace}
\newcommand{\mutau}{\ensuremath{\PGm\tauh}\xspace}
\newcommand{\tautau}{\ensuremath{\tauh\tauh}\xspace}
\newcommand{\leptau}{\ensuremath{\ell\tauh}\xspace}
\newcommand{\dxy}{\ensuremath{d_{xy}}\xspace}
\newcommand{\dz}{\ensuremath{d_{z}}\xspace}
\newcommand{\irel}{\ensuremath{I_{\text{rel}}}\xspace}
\newcommand{\dR}{\ensuremath{\Delta R}\xspace}
\newcommand{\wjets}{\ensuremath{\PW}+jets\xspace}
\newcommand{\ztautau}{\ensuremath{\PZ\to\PGt^{+}\PGt^{-}}\xspace}
\newcommand{\zmumu}{\ensuremath{\PZ\to\PGm^{+}\PGm^{-}}\xspace}
\newcommand{\DZ}{\ensuremath{D_{\zeta}}\xspace}
\newcommand{\sumMT}{\ensuremath{\Sigma\mT}\xspace}
\newcommand{\nj}{\ensuremath{N_{\text{j}}}\xspace}
\newcommand{\npv}{\ensuremath{N_{\mathrm{PV}}}\xspace}

\cmsNoteHeader{SUS-18-006}
\title{Search for direct pair production of supersymmetric partners to the \PGt lepton in proton-proton collisions at $\sqrt{s}=13\TeV$}
\titlerunning{Search for direct pair production of supersymmetric partners to the \PGt lepton at 13\TeV}

\date{\today}

\abstract{
A search is presented for \PGt slepton pairs produced in proton-proton collisions at a center-of-mass energy of 13\TeV. The search is carried out in events containing two \PGt leptons in the final state, on the assumption that each \PGt slepton decays primarily to a \PGt lepton and a neutralino. Events are considered in which each \PGt lepton decays to one or more hadrons and a neutrino, or in which one of the \PGt leptons decays instead to an electron or a muon and two neutrinos. The data, collected with the CMS detector in 2016 and 2017, correspond to an integrated luminosity of 77.2\fbinv. The observed data are consistent with the standard model background expectation. The results are used to set 95\% confidence level upper limits on the cross section for \PGt slepton pair production in various models for \PGt slepton masses between 90 and 200\GeV and neutralino masses of 1, 10, and 20\GeV. In the case of purely left-handed \PGt slepton production and decay to a \PGt lepton and a neutralino with a mass of 1\GeV, the strongest limit is obtained for a \PGt slepton mass of 125\GeV at a factor of 1.14 larger than the theoretical cross section.
}

\hypersetup{
pdfauthor={CMS Collaboration},
pdftitle={Search for direct pair production of supersymmetric partners to the tau lepton in proton-proton collisions at sqrt(s)=13 TeV},
pdfsubject={CMS},
pdfkeywords={CMS, physics, SUSY}}

\maketitle

\section{Introduction}
\label{sec:intro}
Supersymmetry (SUSY)~\cite{Ramond:1971gb,Golfand:1971iw,Neveu:1971rx,Volkov:1972jx,Wess:1973kz,Wess:1974tw,Fayet:1974pd,Nilles:1983ge} is a possible extension of the standard model (SM) of particle physics, characterized by the presence of superpartners for SM particles. The superpartners have the same quantum numbers as their SM counterparts, except for the spin, which differs by half a unit. One appealing feature of SUSY is that the cancellation of quadratic divergences in quantum corrections to the Higgs boson mass from SM particles and their superpartners could resolve the fine tuning problem~\cite{Gildener:1976ai,Veltman:1976rt,'tHooft:1979bh,Witten:1981nf}. Another feature is that the lightest supersymmetric particle (LSP) is stable in SUSY models with $R$-parity conservation~\cite{Farrar:1978xj}, and could be a dark matter (DM) candidate~\cite{Goldberg:1983nd,Ellis:1983ew,darkmatter}.

The hypothetical superpartner of the \PGt lepton, the \PGt slepton (\PSGt), is the focus of the search reported in this paper. Supersymmetric models where a light \PSGt is the next-to-lightest supersymmetric particle are well motivated in early universe \PSGt-neutralino coannihilation models that can accommodate the observed DM relic density~\cite{WMAP,Griest:1990kh,Vasquez,King,Battaglia:2001zp,Arnowitt:2008bz}. The existence of a light \PSGt would enhance the rate of production of final states with \PGt leptons in collider experiments~\cite{Belanger:2012jn,Arganda:2018hdn}.

\begin{figure}[htb]
\centering
\includegraphics[width=0.5\textwidth]{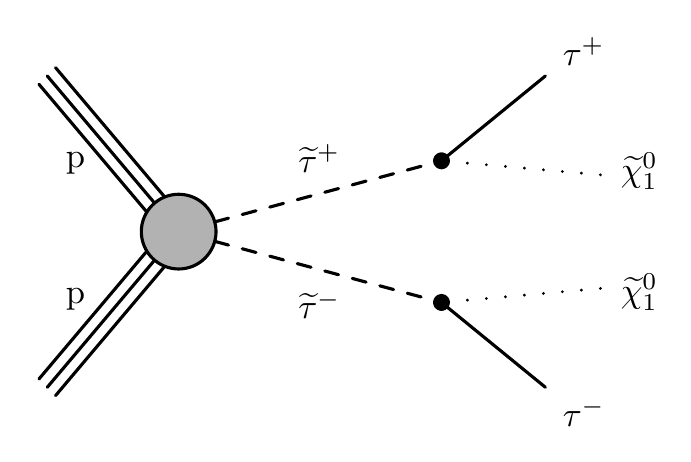}
\caption{\label{fig:diagram} Diagram for direct \PSGt pair production, followed by decay of each \PSGt to a \PGt lepton and a \PSGczDo.}
\end{figure}

In this analysis, we study the simplified model~\cite{Alwall:2008ag,Alwall:2008va,Alves:2011wf} of direct \PSGt pair production shown in Fig.~\ref{fig:diagram}. We assume that the \PSGt decays to a \PGt lepton and \PSGczDo, the lightest neutralino, which is the LSP in this model. The search is challenging because of the extremely small production cross section expected for this signal, as well as the large backgrounds. The most sensitive previous searches for direct \PSGt pair production were performed at the CERN LEP collider~\cite{Heister:2001nk,Abdallah:2003xe,Achard:2003ge,Abbiendi:2003ji}, excluding \PSGt masses at 95\% confidence level (\CL) up to $\approx$90\GeV for neutralino masses up to 80\GeV in some models. At the LHC, the ATLAS~\cite{Aad:2014yka,Aad:2015eda} and CMS~\cite{SUS14022} Collaborations have also performed searches for direct \PSGt pair production using 8\TeV data, and the CMS Collaboration has reported a search for direct \PSGt pair production in an initial sample of 35.9\fbinv at 13\TeV collected in 2016~\cite{Sirunyan:2018vig}. This paper presents a significant improvement in search sensitivity, which was limited by the small signal production rates, through the incorporation of improved analysis techniques and the inclusion of the data collected in 2017. The data used correspond to a total integrated luminosity of 77.2\fbinv.

Events with two \PGt leptons are used. We consider both hadronic and leptonic decay modes of the \PGt lepton, in which it decays to one or more hadrons and a neutrino, or to an electron or muon and two neutrinos, respectively. Independent analyses are carried out in the final states with two hadronically decaying \PGt leptons (\tautau) and with one \tauh and an electron or a muon (\leptau, where $\ell = \Pe$ or \PGm). The presence of missing transverse momentum, which can originate from stable neutralinos as well as neutrinos from \PGt lepton decays, provides an important source of discriminating power between signal and background.

We have introduced several improvements with respect to the analysis presented in Ref.~\cite{Sirunyan:2018vig} that are applied to both 2016 and 2017 data. We make use of dedicated machine learning techniques to enhance the search sensitivity. These include the incorporation of an improved \tauh selection method that makes use of a deep neural network (DNN) for the \tautau analysis, and of a boosted decision tree (BDT) for event selection in the \leptau analyses. Improvements have also been made to the background-estimation techniques and to the search region (SR) definitions. The incorporation of these enhancements is expected to improve the search sensitivity by up to 50\%, where the figure of merit considered is the 95\% \CL upper limit on the cross section for \PSGt pair production obtained with the data collected in 2016. The improvement is less significant than expected, since it is found that the estimated signal acceptance is reduced when the fast detector simulation that was previously used to model signal events is replaced in this search with the more realistic, full \GEANTfour-based detector simulation~\cite{geant4}. Differences in the signal acceptance for the fast and more accurate full detector simulations are mainly caused by differences in the reconstructed \tauh visible transverse momentum (\pt), which is found to have larger values in the case of the fast simulation.

We consider the superpartners of both left- and right-handed \PGt leptons, \staul and \staur. The cross section for \staul pair production is expected to be about a factor of three larger than for \staur pairs~\cite{Fuks:2013lya}. The experimental acceptance is also expected to be different for left- and right-handed assignments because of the differences in the polarization of the \PGt leptons produced in \staul and \staur decays. The decay products of hadronically and leptonically decaying \PGt leptons originating from \staur decays are predicted to have larger and smaller \pt, respectively, than those originating from \staul decays. Two simplified models are studied for direct \PSGt pair production. One model involves production of only \staul pairs and the other is for the degenerate case in which both \staul and \staur pairs are produced. No mixing is introduced between left- and right-handed states. We study models with \PSGt masses ranging from 90 to 200\GeV. The LEP limits~\cite{Heister:2001nk,Abdallah:2003xe,Achard:2003ge,Abbiendi:2003ji} place strong constraints on the allowed values of the \PSGt mass below this range, while the search sensitivity for \PSGt masses above this range is low as a result of the decrease in production cross section with increased mass. We also consider different assumptions for the \PSGczDo mass, namely 1, 10, and 20\GeV. The search sensitivity decreases when the mass difference between  the \PSGt and \PSGczDo becomes small, since the visible decay products in such cases have lower momentum, resulting in a loss of experimental acceptance for such signals.

\section{The CMS detector}
\label{sec:detector}
{\tolerance=800
The central feature of the CMS apparatus is a superconducting solenoid of 6\unit{m} internal diameter, providing a magnetic field of 3.8\unit{T}. A silicon pixel and strip tracker, a lead tungstate crystal electromagnetic calorimeter (ECAL), and a brass and scintillator hadron calorimeter, each composed of a barrel and two endcap sections, reside within the solenoid volume. Forward calorimeters extend the pseudorapidity ($\eta$) coverage provided by the barrel and endcap detectors. Muons are detected in gas-ionization chambers embedded in the steel flux-return yoke outside the solenoid. Events of interest are selected using a two-tiered trigger system~\cite{Khachatryan:2016bia}. The first level, composed of custom hardware processors, uses information from the calorimeters and muon detectors to select events at a rate of around 100\unit{kHz} within a time interval of less than 4\mus. The second level, known as the high-level trigger, consists of a farm of processors running a version of the full event reconstruction software optimized for fast processing, which reduces the event rate to about 1\unit{kHz} before data storage. A more detailed description of the CMS detector, together with definitions of the coordinate system and kinematic variables, can be found in Ref.~\cite{Chatrchyan:2008zzk}.
\par}

\section{Event reconstruction and simulation}
\label{sec:evtreco}
The event reconstruction uses a particle-flow (PF) algorithm~\cite{Sirunyan:2017ulk} that combines information from the tracker, calorimeter, and muon systems to identify charged and neutral hadrons, photons, electrons, and muons in an event. The missing transverse momentum vector, \ptvecmiss, is computed as the negative of the vector sum of the \pt of all PF candidates reconstructed in an event, and its magnitude \ptmiss is used in the search as a discriminator between signal and SM background. Events selected for the search are required to pass filters~\cite{Sirunyan:2019kia} designed to remove detector- and beam-related backgrounds, and must have at least one reconstructed vertex. Usually, more than one such vertex is reconstructed because of pileup, \ie, multiple proton-proton ($\Pp\Pp$) collisions within the same or neighboring bunch crossings. The mean number of interactions per bunch crossing was 27 in 2016, and increased to 37 in 2017, assuming a total inelastic $\Pp\Pp$ cross section of 80\unit{mb}. The reconstructed vertex with the largest value in summed object $\pt^2$ is selected to be the primary $\Pp\Pp$ interaction vertex (PV). These objects are defined by tracks associated with a given vertex that are clustered using a jet finding algorithm~\cite{Cacciari:2008gp,Cacciari:2011ma}, and a more restricted form of the vector missing transverse momentum that is calculated from these track-based jets.

Charged particles that originate from the PV, photons, and neutral hadrons are clustered into jets using the anti-\kt algorithm~\cite{Cacciari:2008gp} with a distance parameter of 0.4, as implemented in the \FASTJET package~\cite{Cacciari:2011ma}. The jet energies are corrected to account for the contribution from pileup interactions and to compensate for variations in the detector response~\cite{Cacciari:2011ma,pileup}. To mitigate issues related to noise in the ECAL endcaps that led to significantly worse modeling of the \ptmiss distribution, particularly for events with large values of \ptmiss in 2017 data, PF candidates that are clustered in jets in $2.65 < \abs{\eta} < 3.14$ with uncorrected $\pt<50\GeV$ are not used in the calculation of \ptvecmiss in 2017 data and simulation. Disagreements between the \ptmiss distributions in data and simulation ranging up to $>$100\% for $50 < \ptmiss < 170\GeV$ in DY+jets events, in which large values of \ptmiss arise mainly from mismeasurements, are reduced by this modification of the \ptvecmiss calculation. The modified \ptmiss distributions in simulated events and data agree within uncertainties.

Jets in the search are required to have their axes within the tracker volume of $\abs{\eta} < 2.4$. For the \tautau analysis, we use jets with $\pt>30\GeV$, while for the \leptau analyses, we veto events containing jets with $\pt>20$\GeV to provide efficient background rejection. Jets are required to be separated in $\eta$ and azimuthal angle ($\phi$) by $\dR \equiv \sqrt{\smash[b]{(\Delta\eta)^2 + (\Delta\phi)^2} } > 0.4$ from electron, muon, or \tauh candidates in order to minimize double counting of objects. Jets originating from the hadronization of {\cPqb} quarks are ``tagged" in the \tautau analysis through the DNN-based combined secondary vertex algorithm (DeepCSV)~\cite{Sirunyan:2017ezt} to reject events with {\cPqb} quark jets that are likely to originate from backgrounds with top quarks. The efficiency for tagging {\cPqb} quarks originating from top quark decays is about 84\%, while the misidentification rates for jets from charm quarks, and from light quarks or gluons, are about 41 and 11\%, respectively. In the \leptau analyses, the CSVv2 tagger~\cite{Sirunyan:2017ezt} is used to identify {\cPqb} quark jets for the selection of background-enriched control regions (CRs). The working point that is used corresponds to an efficiency of 63\% and misidentification rates of 12 and 0.9\% for jets from charm quarks and light quarks or gluons, respectively.

Electron candidates are reconstructed by first matching reconstructed tracks to clusters of energy deposited in the ECAL. Selections based on the spatial distribution of the shower, track--cluster matching criteria, and consistency between the cluster energy and the track momentum are then used in the identification of electron candidates~\cite{Khachatryan:2015hwa}. Muon candidates are reconstructed by requiring reconstructed tracks in the muon detector to be matched to the tracks found in the inner tracker~\cite{Sirunyan:2018}. We require the origin of electron and muon candidates to be consistent with the PV. Restrictions are imposed on the magnitude of the impact parameters of their tracks relative to the PV in the transverse plane (\dxy), and on the longitudinal displacement (\dz) of the point of closest approach. To ensure that electron or muon candidates are isolated from jet activity, we define a relative isolation quantity (\irel) as the ratio of the scalar \pt sum of hadron and photon PF candidates, in an $\eta$-$\phi$ cone of radius 0.3 or 0.4 around the candidate electron or muon, to the candidate \pt, requiring it to be below an upper bound appropriate for the selection. The quantity \irel is adjusted to account for the contributions of particles originating from pileup interactions. The electron and muon selection criteria applied in the analysis are the same as those described in Ref.~\cite{Sirunyan:2018vig}.

The \tauh candidates are reconstructed using the CMS hadrons-plus-strips algorithm~\cite{Sirunyan:2018pgf}. The constituents of the reconstructed jets are used to identify individual \PGt lepton decay modes with one charged hadron and up to two neutral pions, or three charged hadrons. The \tauh candidate momentum is determined from the reconstructed visible \PGt lepton decay products. The presence of extra particles within the jet that are  incompatible with the reconstructed decay mode is used as a criterion to discriminate jets from \tauh decays.  A multivariate-analysis (MVA) based discriminant~\cite{Sirunyan:2018pgf}, which contains isolation as well as lifetime information, is used to suppress the rate for quark and gluon jets to be misidentified as \tauh candidates. We employ a relaxed (``very loose") working point of this discriminant as a preselection requirement for the \tauh candidates selected in the \tautau analysis, as well as in the extrapolation used to estimate the contributions of events to the background in which quark or gluon jets are misidentified as \tauh candidates. This working point corresponds to an efficiency of $\approx$70\% for a genuine \tauh, and a misidentification rate of $\approx$1\% for quark or gluon jets. A DNN is used to improve the discrimination of signal \tauh candidates from background, as discussed in more detail below.  Two working points are used in the \leptau analysis: a ``very tight" working point for selecting signal \tauh candidates that provides stringent background rejection, and a ``loose" working point for the extrapolation procedure to estimate the misidentified \tauh background that provides higher efficiency and less background rejection. These working points, respectively, typically have efficiencies close to 45 and 67\% for a genuine \tauh, with misidentification rates of $\approx$0.2 and 1\% for quark or gluon jets. Electrons and muons misidentified as a \tauh are suppressed via criteria specifically developed for this purpose that are based on the consistency of information from the tracker, calorimeters, and muon detectors~\cite{Sirunyan:2018pgf}.

{\tolerance=800
The dominant background in the \tautau final state originates from misidentification of jets as \tauh candidates, mainly in SM events exclusively comprising jets produced through the strong interaction of quantum chromodynamics (QCD). These are referred to as QCD multijet events in what follows. To further improve the suppression of this background while retaining high signal efficiency, we have pursued a new approach for \tauh isolation in the \tautau analysis that is based upon the application of a DNN that is fed information about the properties of PF candidates within an isolation cone with $\dR < 0.5$ around the \tauh candidate. We refer to this as ``Deep Particle Flow" (DeepPF) isolation. Charged PF candidates consistent with having originated from the PV, photon candidates, and neutral hadron candidates with $\pt > 0.5$, 1, and 1.25\GeV, respectively, provide the inputs to the DeepPF algorithm. The list of observables incorporated for each PF candidate includes its \pt relative to the \tauh jet, $\dR$ between the candidate and \tauh, particle type, track quality information, and \dxy, \dz and their uncertainties, $\sigma(\dxy)$ and $\sigma(\dz)$. A convolutional DNN~\cite{Lecun98gradient-basedlearning} is trained with simulated signal and background events. Signal \tauh candidates are those that are matched to generator-level \PGt leptons from a mixture of processes that give rise to genuine \PGt leptons. Background candidates that fail the matching are taken from simulated \wjets and QCD multijet events. The DeepPF discriminator value is obtained by averaging the DNN output with the nominal MVA-based discriminant described above. The working point for DeepPF isolation is chosen to maintain a constant efficiency of $\approx$50\%, 56\%, and 56\% as a function of \pt for the three respective \tauh decay modes: one charged hadron, one charged hadron with neutral pions, and three charged hadrons. Since the \tauh candidate \pt distribution in signal events depends on the \PSGt and \PSGczDo masses, this choice of discriminator and working points allows us to maintain high efficiency for \PSGt pair production signals under a large range of mass hypotheses. The overall misidentification rate for jets not originating from \PGt leptons ranges from 0.15\% to 0.4\% depending on \pt and decay mode.
\par}

Significant contributions to the SM background originate from Drell--Yan+jets (DY+jets), \wjets, \ttbar, and diboson processes, as well as from QCD multijet events, where DY corresponds to processes such as $\qqbar\to\ell^{+}\ell^{-}$. Smaller contributions arise from single top quark production and rare SM processes, such as triboson and Higgs boson production, and top quark pair production in association with vector bosons. We rely on a combination of measurements in data CRs and Monte Carlo (MC) simulation to estimate contributions of each source of background. The MC simulation is also used to model the signal.

{\tolerance=800
The \MGvATNLO version 2.3.3 and 2.4.2 event generators~\cite{Alwall:2014hca} are used at leading order (LO) precision to generate simulated \wjets and DY+jets events with up to 4 additional partons for the analysis of 2016 and 2017 data, respectively. Exclusive event samples binned in jet multiplicity are used to enhance the statistical power of the simulation at higher values of jet multiplicity that are relevant to the phase space probed by this search. Production of top quark pairs, diboson and triboson events, and rare SM processes, such as single top quarks or top quark pairs associated with bosons, are generated at next-to-leading order (NLO) precision with \MGvATNLO and {\POWHEG}v2~\cite{Nason:2004rx,Frixione:2007vw,Alioli:2010xd,Re:2010bp}. Showering and hadronization of partons are carried out using the \PYTHIA~8.205 and 8.230 packages~\cite{Sjostrand:2014zea} for the 2016 and 2017 analyses, respectively, while a detailed simulation  of the CMS detector is based on the \GEANTfour~\cite{geant4} package. Finally, uncertainties in renormalization and factorization scale, and parton distribution functions (PDFs) have been obtained using the \textsc{SysCalc} package~\cite{Kalogeropoulos:2018cke}. Models of direct \PSGt pair production are generated with \MGvATNLO at LO precision up to the production of \PGt leptons, with their decay modeled by \PYTHIA~8.212 and 8.230 for the analysis of 2016 and 2017 data, respectively. The CUETP8M1~\cite{Khachatryan:2015pea} (CUETP8M2T4~\cite{CMS-PAS-TOP-16-021} for \ttbar) and CP5~\cite{Sirunyan:2019dfx} underlying-event tunes are used with \PYTHIA for the 2016 and 2017 analyses, respectively. The 2016 analysis uses the NNPDF3.0LO~\cite{Ball:2014uwa} set of PDFs in generating \wjets, DY+jets, and signal events, while the NNPDF3.0NLO PDFs are used for other processes. The NNPDF3.1NLO PDFs are used for all simulated events in the 2017 analysis.
\par}

Simulated events are reweighted to match the pileup profile observed in data. Differences between data and simulation in electron, muon, and \tauh identification and isolation efficiencies, jet, electron, muon, and \tauh energy scales, and {\cPqb} tagging efficiency are taken into account by applying scale factors to the simulation. We improve the modeling of initial-state radiation (ISR) in simulated signal events by reweighting the $\pt^{\mathrm{ISR}}$ distribution, where $\pt^{\mathrm{ISR}}$ corresponds to the total transverse momentum of the system of SUSY particles. This reweighting procedure is based on studies of the \pt of \PZ bosons~\cite{Chatrchyan:2013xna}. The signal production cross sections are calculated at NLO using next-to-leading logarithmic (NLL) soft-gluon resummations~\cite{Fuks:2013lya}. The most precise calculated cross sections available are used to normalize the simulated SM background samples, often corresponding to next-to-next-to-leading order accuracy.

\section{Event selection}
\label{sec:evtsel}
The search strategy in the \tautau final state relies on a cut-and-count analysis based on the SRs described below in Section~\ref{sec:tautausel}, while for the \leptau final states we make use of BDTs to discriminate between signal and background as described in Section~\ref{sec:leptausel}. The data used in this search are selected through triggers that require the presence of isolated electrons, muons, \tauh candidates, or \ptmiss. The data used for the \tautau analysis are collected with two sets of triggers. Events with $\ptmiss < 200$\GeV are selected using a trigger that requires the presence of two \tauh candidates, each with $\pt>35$ and $>$40\GeV in 2016 and 2017 data, respectively. We gain up to 7\% additional signal efficiency for events with $\ptmiss>200$\GeV with the help of a trigger that requires the presence of substantial \ptmiss, with a threshold varying between 100 and 140\GeV during the 2016 and 2017 data-taking periods. For the \etau final state, the trigger relies on the presence of an isolated electron satisfying stringent identification criteria and passing $\pt>25$ or $>$35\GeV in 2016 and 2017 data, respectively. For the \mutau final state, the trigger is based on the presence of an isolated muon with $\pt>24$ and $>$27\GeV in 2016 and 2017 data, respectively. Trigger efficiencies are measured in data and simulation. In addition to corrections mentioned in Section~\ref{sec:evtreco}, we apply scale factors to the simulation to account for any discrepancies in trigger efficiency with data. These scale factors are parameterized in the \pt and $\eta$ of the reconstructed electron, muon, or \tauh candidates, or the reconstructed \ptmiss for events selected using \ptmiss triggers.

\subsection{Event selection and search regions in the \texorpdfstring{\tautau}{tau-tau} final state}
\label{sec:tautausel}

Beyond the trigger selection, the baseline event selection for the \tautau analysis requires the presence of exactly two isolated \tauh candidates of opposite charge, satisfying the DeepPF selection described in Section~\ref{sec:evtreco}, with $\abs{\eta}<2.3$ and $\pt > 40$ and $>$45\GeV in the 2016 and 2017 analysis, respectively, as well as no additional \tauh candidates with $\pt > 30\GeV$ satisfying the very loose working point of the MVA-based discriminant. We veto events with additional electrons or muons with $\pt>20\GeV$ and $\abs{\eta} < 2.5$ or $<$2.4 for electrons and muons, respectively, and reject any events with a {\cPqb}-tagged jet to suppress top quark backgrounds. A requirement of $\abs{\Delta\phi(\tauh^{(1)},\tauh^{(2)})} > 1.5$ helps to suppress the DY+jets background, while retaining high signal efficiency. Finally, we require $\ptmiss>50\GeV$ to suppress the QCD multijet background.

{\tolerance=900
The removal of low-\pt jets in the forward ECAL region from the \ptvecmiss calculation in 2017 (see Section~\ref{sec:evtreco}) causes the background originating from DY+jets and other sources to increase in the SRs, since events with low-\pt jet activity in that region are assigned larger values of reconstructed \ptmiss. We recover some of the corresponding loss in sensitivity in the 2017 analysis by placing an upper bound of 50\GeV on the scalar \pt sum of low-\pt jets excluded from the \ptvecmiss calculation ($\HT^{\text{low}}$). This restriction reduces the impact of background events with significant low-\pt jet activity in the forward region, for which the \ptmiss would be overestimated. To ensure that the efficiency of this requirement is correctly estimated in simulation, a \zmumu CR is used to extract correction factors for the $\HT^{\text{low}}$ distribution in simulation that account for discrepancies with the distribution observed in data. The correction factors range from 0.8 for $\HT^{\text{low}}<10\GeV$ to 1.4 for $\HT^{\text{low}}>60\GeV$. In addition, to avoid effects related to jet mismeasurement that can contribute to spurious \ptmiss, we require the \ptvecmiss to have a minimum separation of 0.25 in $\abs{\Delta \phi}$ from jets with $\pt > 30\GeV$ and $\abs{\eta} < 2.4$, as well as from those with uncorrected $\pt > 50\GeV$ in the region $2.4 < \abs{\eta} < 3.14$.
\par}

Events satisfying the baseline selection criteria are subdivided into exclusive SRs using several discriminants. To improve the discrimination of signal from SM background, we take advantage of the expected presence of two \PSGczDo in the final state of signal events and their contribution to \ptmiss. Their presence skews the correlations between \ptvecmiss and the reconstructed leptons to be different from background processes,  even for those backgrounds with genuine \ptmiss. These differences can be exploited by mass observables calculated from the reconstructed lepton transverse momenta and \ptvecmiss to provide discrimination of signal from  background. For a particle decaying to a visible and an invisible particle, the transverse mass (\mT) calculated from the \ptvec of the visible decay products should have a kinematic endpoint at the mass of the parent particle. Assuming that the \ptmiss corresponds to the \pt of the invisible particle, we calculate the \mT observable for the visible particle q and the invisible particle as follows:
\begin{equation}
\mT(\mathrm{q}, \ptvecmiss) \equiv \sqrt{2 \pt^{\mathrm{q}} \ptmiss [1 - \cos \Delta\phi(\ptvec^{\mathrm{q}}, \ptvecmiss)]}.\label{eq:MT}
\end{equation}
We use as a discriminant the sum of the transverse masses calculated for each \tauh with \ptmiss, \sumMT, given by
\begin{equation}
\sumMT = \mT({\tauh^{(1)},\ptvecmiss}) + \mT({\tauh^{(2)},\ptvecmiss}).
\end{equation}
Another variable found to be useful in the discrimination of signal from  background is the ``stransverse mass" \mTii~\cite{MT2variable,MT2variable2, MT2variable3}. This mass variable is a generalization of \mT in the case of multiple invisible particles.  It serves as an estimator of the mass of pair-produced particles when both particles decay to a final state containing the same invisible particle. It is given by:
\begin{equation}
\mTii = \min_{\ptvec^{\mathrm{X}(1)} + \ptvec^{\mathrm{X}(2)} = \ptvecmiss}
  \left[ \max \left( \mT^{(1)} , \mT^{(2)} \right) \right],
\label{eq:MT2}
\end{equation}
where $\ptvec^{\mathrm{X}(i)}$  (with $i$=1, 2) are the unknown transverse momenta of the two undetected particles, X(1) and X(2), corresponding to the neutralinos in our signal models, and $\mT^{(i)}$ are the transverse masses obtained by pairing either of the two invisible particles with one of the two leptons.  The minimization ($\min$) is over the possible momenta of the invisible particles, taken to be massless, which are constrained to add up to the \ptvecmiss in the event.   For direct \PSGt pair production, with each \PSGt decaying to a \PGt lepton and a \PSGczDo, \mTii should be correlated with the mass difference between the \PSGt and \PSGczDo. A large value of \mTii is thus common in signal events for models with larger \PSGt masses and relatively rare in SM background events.

\begin{figure}[htb]
\centering
\includegraphics[width=0.48\textwidth]{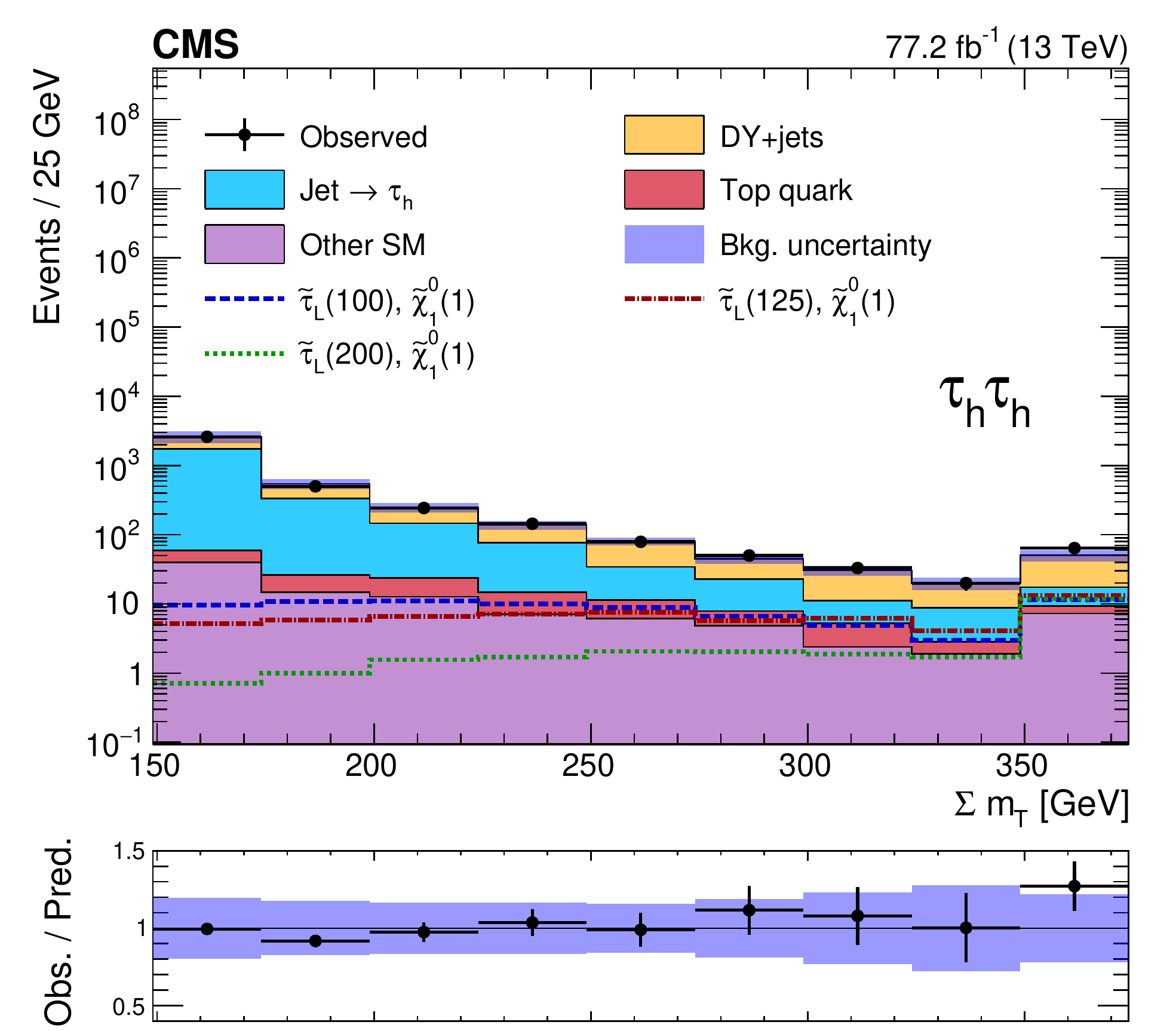}
\includegraphics[width=0.48\textwidth]{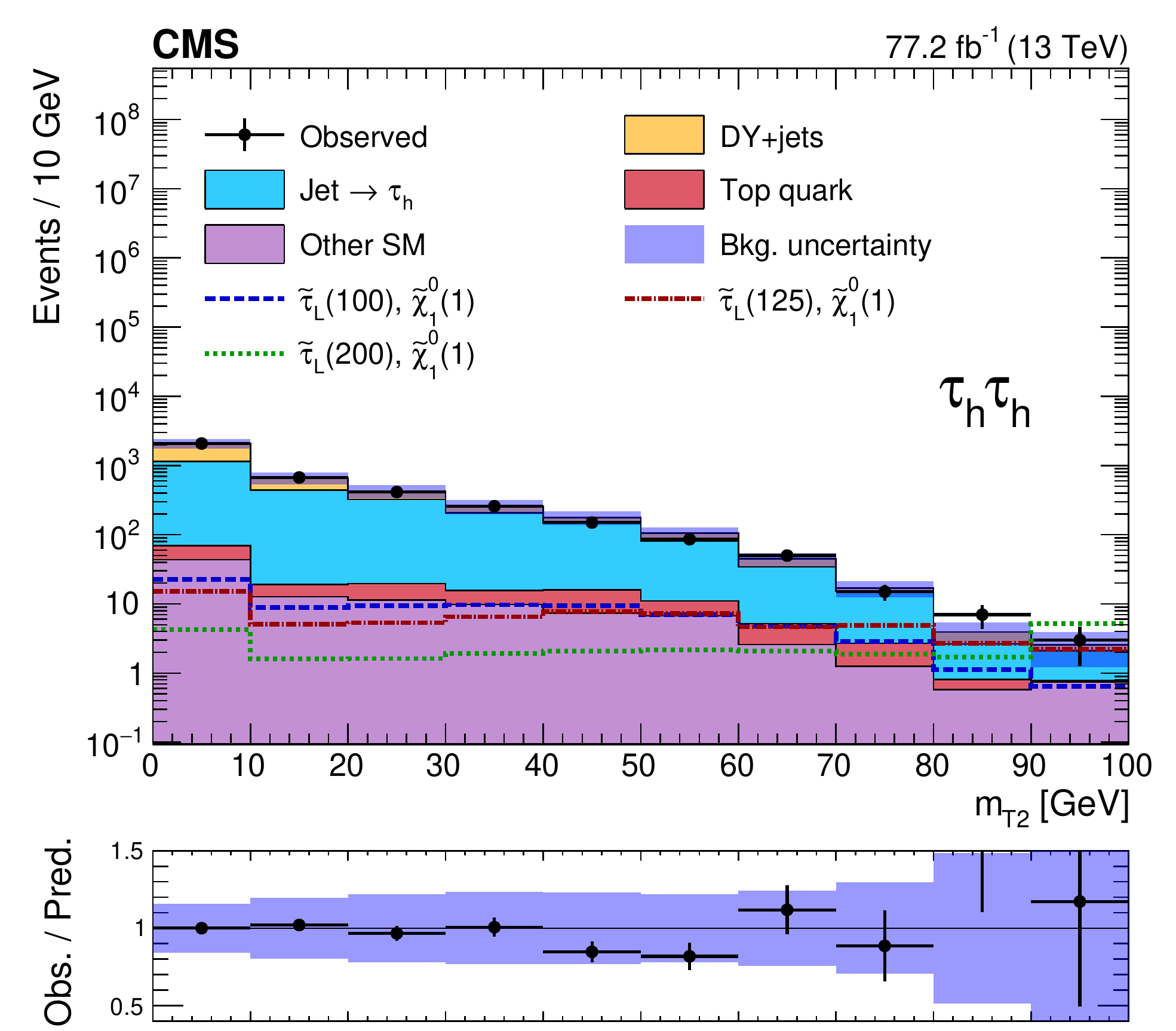}
\caption{\label{fig:srvariables_tautau} Distributions in \sumMT (\cmsLeft) and \mTii (\cmsRight) for events in the combined 2016 and 2017 data sets passing the baseline selection in the \tautau final state, along with the corresponding prediction for the SM background and three benchmark models for \staul pair production with $m(\staul)=100$, 125, and 200\GeV, $m(\PSGczDo)=1\GeV$. The numbers within parentheses in the legend correspond to the masses of the \staul and \PSGczDo in \GeV. The last bin includes overflow events in each case. The shaded uncertainty bands represent the combined statistical and systematic uncertainties in the background.}
\end{figure}

The SR definitions for the \tautau analysis, shown in Table~\ref{tab:sr}, are based on a cut-and-count analysis of the sample satisfying the baseline selections.  The regions are defined through criteria imposed on \mTii, \sumMT, and the number of reconstructed jets in an event, \nj. The \sumMT and \mTii distributions of events in the \tautau final state surviving the baseline selections are shown in Fig.~\ref{fig:srvariables_tautau}. The distributions obtained for 2016 and 2017 data are combined. Separate sets of simulated events are used to model signal and background events in 2016 and 2017 data using the methods described in Section~\ref{sec:evtreco}. In all distributions, the last bin includes overflow events. After applying a minimum requirement of $\mTii>25\GeV$ in all SRs, we subdivide events into low (25--50\GeV) and high (${>}50\GeV$) \mTii regions, to improve the sensitivity to lower and higher \PSGt mass signals, respectively. For each \mTii region, the \sumMT distribution is exploited to provide sensitivity for a large range of \PSGt mass signals. We define three bins in \sumMT: 200--250, 250--300, and $>$300\GeV. Finally, we subdivide events in each \mTii and \sumMT region into the categories $\nj = 0$ and $\nj \geq 1$.  This binning is beneficial as background events passing the SR kinematic selections are largely characterized by additional jet activity, while signal contains very few additional jets. The 0-jet category therefore provides nearly background-free SRs. However, we retain the SRs with $\nj \geq 1$ that are also expected to contain signal events with ISR or pileup jets.

\begin{table*}[!h]
\centering
\topcaption{Ranges in \mTii, \sumMT, and \nj used to define the SRs used in the \tautau analysis.}
\label{tab:sr}
\begin{tabular}{l c c c c c c c c c c c c c c c}
\hline
\mTii [\GeVns{}] & & \multicolumn{6}{c}{25--50} & & \multicolumn{6}{c}{$>$50} \\
[\cmsTabSkip]
\sumMT [\GeVns{}] & & \multicolumn{2}{c}{200--250} & \multicolumn{2}{c}{250--300} & \multicolumn{2}{c}{$>$300} & & \multicolumn{2}{c}{200--250} & \multicolumn{2}{c}{250--300} & \multicolumn{2}{c}{$>$300} \\
[\cmsTabSkip]
\nj & & 0 & $\geq$1 & 0 & $\geq$1 & 0 & $\geq$1 & & 0 & $\geq$1 & 0 & $\geq$1 & 0 & $\geq$1 \\
\hline
\end{tabular}
\end{table*}

\subsection{Event selection in the \texorpdfstring{\leptau}{lep-tau} final states}
\label{sec:leptausel}
The baseline event selections for the \leptau analyses require either an electron with $\pt > 26\,(35)$\GeV and $\abs{\eta} < 2.1$ or a muon with $\pt > 25\,(28)$\GeV and $\abs{\eta}<2.4$ for the 2016 (2017) data, and a \tauh candidate with $\pt > 30$\GeV and $\abs{\eta} < 2.3$. Electrons, muons, and \tauh candidates are required to have $\abs{\dz} < 0.2\cm$, and electrons and muons are also required to have $\abs{\dxy} < 0.045\cm$. Electrons and muons have to satisfy $\irel<0.15$ and $<$0.1, respectively. Backgrounds from \ttbar and \wjets are greatly reduced by vetoing events that contain jets with $\pt > 20\GeV$. Events from the \wjets background are further reduced by requiring the transverse mass $\mT(\ell, \ptvecmiss)$, calculated using the electron or muon momentum vector and \ptvecmiss, to be between 20 and 60\GeV or above 120\GeV. A  significant background from DY+jets events is reduced by requiring the invariant mass of the electron or muon and the \tauh, $m_{\ell \tauh}$  to be above 50\GeV. To reduce background from QCD multijet events, we require $2.0< \Delta R(\ell, \tauh) <3.5$.

\begin{figure}[htb]
\centering
\includegraphics[width=0.48\textwidth]{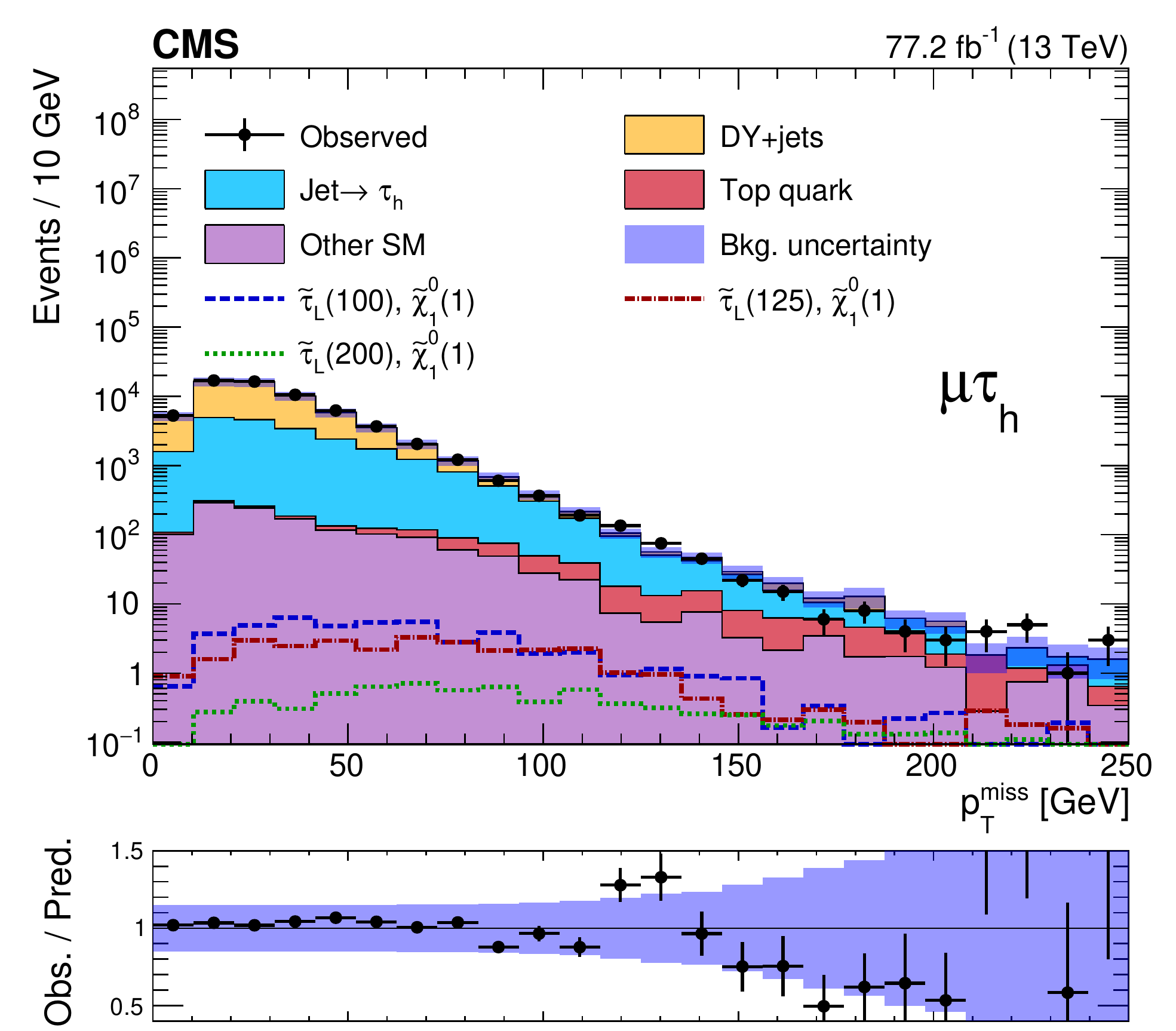}
\includegraphics[width=0.48\textwidth]{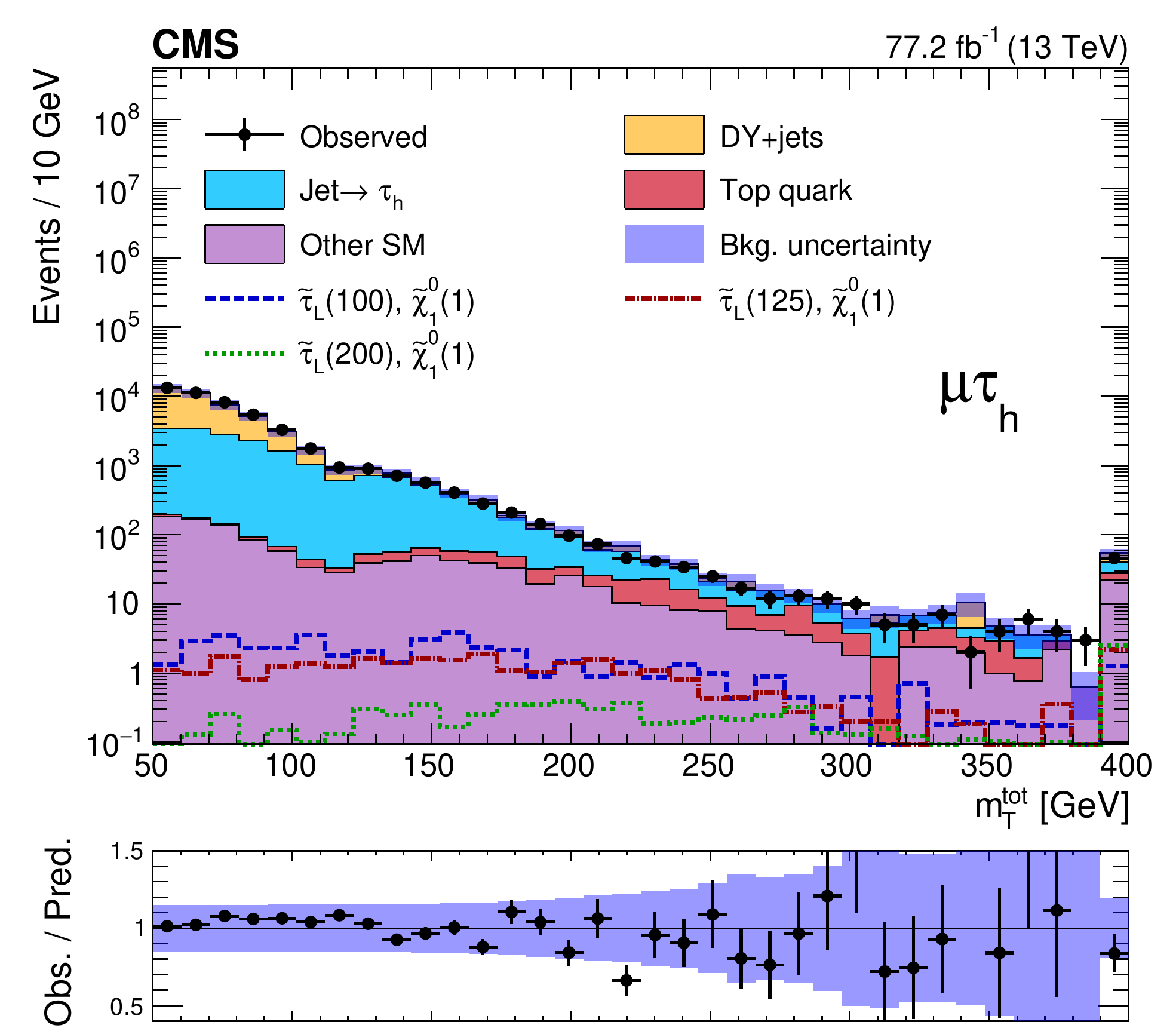}
\caption{\label{fig:srvariables_leptau} Distributions in \ptmiss (\cmsLeft) and $\mT^{\text{tot}}$ (\cmsRight) for events in the combined 2016 and 2017 data passing the baseline selections in the \mutau final state, along with the corresponding prediction for SM background and three benchmark models of \staul pair production with $m(\staul)=100$, 125, and 200\GeV and $m(\PSGczDo)=1\GeV$. The numbers within parentheses in the legend correspond to the masses of the \staul and \PSGczDo in \GeV. The last bin includes overflow events in each case. The shaded uncertainty bands represent the combined statistical and average systematic uncertainties in the background.}
\end{figure}

{\tolerance=800
With these preselection criteria in place, we train several BDTs corresponding to different signal hypotheses to classify signal and background events. The input variables are the \pt of the electron or muon, the \pt of the \tauh candidate, \ptmiss, $\mT(\ell, \ptvecmiss)$, $\Delta \eta(\ell,\tauh)$, $\Delta \phi(\ell, \ptvecmiss)$, $\Delta \phi(\tauh, \ptvecmiss)$, $\Delta R(\ell, \tauh)$, $m(\ell \tauh)$, and $\mT^{\text{tot}}\equiv\sqrt{\smash[b]{\mT^2(\ell, \ptvecmiss)+\mT^2(\tauh, \ptvecmiss)}}$. We also include \mTii and the contransverse mass ($m_{\mathrm{CT}}$)~\cite{Tovey:2008ui,Polesello:2009rn}, computed from the visible decay products and defined as
\begin{equation}
m_{\mathrm{CT}} \equiv \sqrt{2 \pt^{\ell} \pt^{\tauh} [1 + \cos\Delta\phi(\ell,\tauh)]}.
\end{equation}
For signal events, $m_{\mathrm{CT}}$ is expected to have an endpoint near ${(m(\PSGt)^2-m(\PSGczDo)^2)}/m(\PSGt)$. Finally, we include the variable $\DZ=\ptvecmiss \cdot \vec \zeta - 0.85 (\ptvec^{\ell}+\ptvec^{\tauh}) \cdot \vec \zeta$, with $\vec{\zeta}$ being the bisector of the directions of the transverse momenta of the electron or muon and the \tauh candidate~\cite{CuencaAlmenar:2008zza,Khachatryan:2014wca}. The value of 0.85 reflects an optimization to efficiently distinguish DY+jets events from other backgrounds and the signal. Figure~\ref{fig:srvariables_leptau} shows the distributions of events passing the baseline selections in the \mutau final state in two of the BDT input variables that provide the highest discriminating power, \ptmiss and $\mT^{\text{tot}}$. The distributions observed in the \etau final state are similar.
\par}

Since the signal kinematics depend on mass, we train BDTs for signals with \PSGt masses of 100, 150, and 200\GeV. In all cases we use a \PSGczDo mass of 1\GeV. As the results of the training depend critically on the number of input events, we relax the \tauh MVA-based isolation criteria and reduce the  \pt threshold for the \tauh to 20\GeV for the training sample in order to increase the number of training and test events. The ``very tight" isolation and a \pt threshold of 30\GeV for the \tauh are applied in the final analysis. For a given signal hypothesis, we choose the BDT trained with the same \PSGt mass for models with \PSGt masses of 100, 150, and 200\GeV, or the one that provides optimal sensitivity for models with other \PSGt mass values. For signal models with \PSGt masses of 90 and 125\GeV, we use the BDT trained for $m(\PSGt)=100\GeV$, while for those with a \PSGt mass of 175\GeV, we use the BDT trained for $m(\PSGt)=200\GeV$.  While signal events are largely expected to have high BDT output values, we include the full BDT distribution in a binned fit for the statistical interpretation of the analysis as described in Section~\ref{sec:results}. The binning is chosen to optimize signal significance.

\section {Background estimation}
\label{sec:bkgest}
{\tolerance=800
Our most significant backgrounds are from DY+jets, \wjets, QCD multijet, \ttbar, and diboson processes. They have relative contributions that vary with final state. For the \tautau final state, the dominant background arises from the misidentification of jets as \tauh candidates in QCD multijet and \wjets events, constituting $\approx$65\% of background after the baseline selection. For the \leptau final states after the baseline selection, the main backgrounds are from DY+jets ($\approx$50\%), \wjets ($\approx$30\%), and QCD multijet ($\approx$10\%) events. The DY+jets contribution, which is also a major background in the \tautau final state ($\approx$20\%), usually consists of events with two prompt \PGt leptons. This background is determined with simulation samples after applying corrections to match the normalization and to be consistent with variable distributions in collider data. The \wjets and QCD multijet backgrounds usually contain one or more jets misidentified as \tauh and their contributions are determined via  methods that rely on data. Finally, we have smaller contributions from other SM processes such as the production of Higgs bosons, dibosons, and top quark pairs with or without vector bosons.  These are estimated via MC simulation with appropriate correction factors applied as described in Section~\ref{sec:evtreco}. For the \leptau analyses, dedicated CRs that are each enriched in one of the major background processes are used to validate the modeling of the BDT distribution and to extract uncertainties that are used to account for any potential mismodeling of the distributions in simulation. These CRs are described in the following subsections below.
\par}

\subsection{Estimation of background from misidentified jets}
\label{sec:fakeestimation}

\subsubsection{Misidentified jets in the \texorpdfstring{\tautau}{tau-tau} final state}

{\tolerance=800
After requiring two \tauh candidates with high \pt, events with misidentified \tauh candidates are the dominant background in the \tautau final state.  This background, which originates predominantly from QCD multijet and \wjets production, is predicted by extrapolating the event count in a data sample selected with a relaxed isolation requirement into the SR. The fraction of non-prompt or misidentified \tauh candidates selected with the very loose MVA-based isolation working point that also pass the tight DeepPF isolation requirement is measured in a QCD multijet-enriched sample of same-charge \tautau events. The same-charge \tautau events are collected with the same \tautau trigger as opposite-charge \tautau events to avoid additional trigger-related biases.  We also require \mTii to be low ($<$40\GeV) to reduce potential contributions from signal events.  We find that roughly 20\% of the same-charge events with misidentified \tauh candidates selected with very loose isolation also pass the tight isolation requirement. However, the rate depends on the \pt and decay mode (one- or three-prongs) of the \tauh candidate, as well as the jet flavor, \ie, whether the misidentified jet originates from the hadronization of light-flavor quarks, heavy-flavor quarks, or gluons.  The \tauh misidentification rate is therefore measured in bins of \pt and decay mode to mitigate the dependence on these factors. The measurement is also binned in the number of primary vertices (\npv) to capture the effects of pileup. From studies performed with MC simulation samples, a systematic uncertainty of $\approx$30\%  is assigned to account for the dependence of the misidentification rate on jet flavor.
\par}

Since the isolation efficiency for prompt \tauh candidates is only around 70--80\%, processes containing genuine \tauh candidates can enter the sideband regions in events that are selected with the relaxed isolation requirement. To take this into account when calculating the final background estimate, we define three categories of events with at least two loosely isolated \tauh candidates: (i) events in which both \tauh candidates pass the tight DeepPF isolation requirement, (ii) events in which one passes and one fails the tight isolation requirement, and (iii) events in which both \tauh candidates fail the tight isolation requirement. We then equate the count of events in each of these three event categories to the sum of expected counts for the events with two prompt \tauh candidates, two jets misidentified as \tauh candidates, or one prompt \tauh candidate and one jet misidentified as a \tauh candidate, that contribute to each category.  The contributions from backgrounds with one or two jets misidentified as \tauh candidates in the SRs are then determined analytically by solving a set of linear equations.

\subsubsection{Misidentified jets in the \texorpdfstring{\etau and \mutau}{e-tau and mu-tau} final states}

The misidentification of jets as \tauh candidates also gives rise to a major source of background in the \etau and \mutau final states that arises mainly from \wjets events with leptonic {\PW} boson decays. We estimate this background from a sideband region in data selected using the SR selection criteria, with the exception that the \tauh candidates are required to satisfy the loose isolation working point and not the very tight working point. A transfer factor for the extrapolation of event counts from this \tauh-isolation range into the tight isolation range of the SR is determined with a \wjets CR selected from events with one muon and at least one \tauh candidate that passes the loose isolation requirement. In events with more than one \tauh candidate, the candidate with the highest value of the MVA-based isolation discriminant is used.  To increase the purity of \wjets events in this region, we reduce the contribution from \ttbar and QCD multijet events by requiring $60 < \mT(\ell,\ptvecmiss) < 120\GeV$, $\ptmiss > 40\GeV$, no more than two jets, and an azimuthal separation of at least 2.5 radians between any jet and the {\PW} boson reconstructed from the muon and \ptvecmiss ($\Delta \phi(\PW,\text{jet})>2.5$). We also reject events with additional electrons or muons satisfying looser identification criteria. The remaining sample has an expected purity of $\approx$85\% for \wjets events. The transfer factor, $R$, is then determined from this control sample after subtracting the remaining non-\wjets background contributions estimated from simulation, as follows:
\begin{equation}
\label{eq:tranferFactor}
R = \frac{N^{\mathrm{CR}}_{\text{data}}({\mathrm{VT}})-N^{\mathrm{CR}}_{\text{MC  no \PW}}({\mathrm{VT}})}{N^{\mathrm{CR}}_{\text{data}} (\mathrm{L}\overline{\mathrm{VT}})-N^{\mathrm{CR}}_{\text{MC  no \PW}} (\mathrm{L}\overline{\mathrm{VT}})},
\end{equation}
where $N^{\mathrm{CR}}_{\text{data}}$ corresponds to the number of events in the CR in data. The parenthetical argument $\mathrm{VT}$ denotes events in which the \tauh candidate satisfies the very tight isolation working point, while $\mathrm{L}\overline{\mathrm{VT}}$ denotes those that satisfy the loose, but not the very tight requirement. Transfer factors are determined separately in bins of \pt and $\eta$ of \tauh candidates in order to achieve an accurate description of the background.

{\tolerance=800
The contribution of the background originating from a jet misidentified as a \tauh candidate in the SR is then determined from the corresponding sideband in data:
\begin{equation}
N(\text{jet} \to \tauh) = R \, (N^{\text{sideband}}_{\text{data}} - N^{\text{sideband}}_{\mathrm{MC}, \PGt}),
\end{equation}
where $N^{\text{sideband}}_{\text{data}}$ is the number of events in the sideband in data, from which $N^{\text{sideband}}_{\mathrm{MC}, \PGt}$, the number with genuine \PGt leptons as estimated with   MC simulation by generator-level matching, is subtracted. We validate the estimation of jets misidentified as  \tauh  in a CR requiring $60 < \mT(\ell,\ptvecmiss) < 120\GeV$ and  $\Delta \phi(\PW,\text{jet})<2.5$ to ensure that the region is independent of the region described above that is used to estimate the background.
\par}

\subsection{Estimation of background from Drell--Yan+jets}
\label{sec:dyestimation}
\begin{figure}[htb]
\centering
\includegraphics[width=0.48\textwidth]{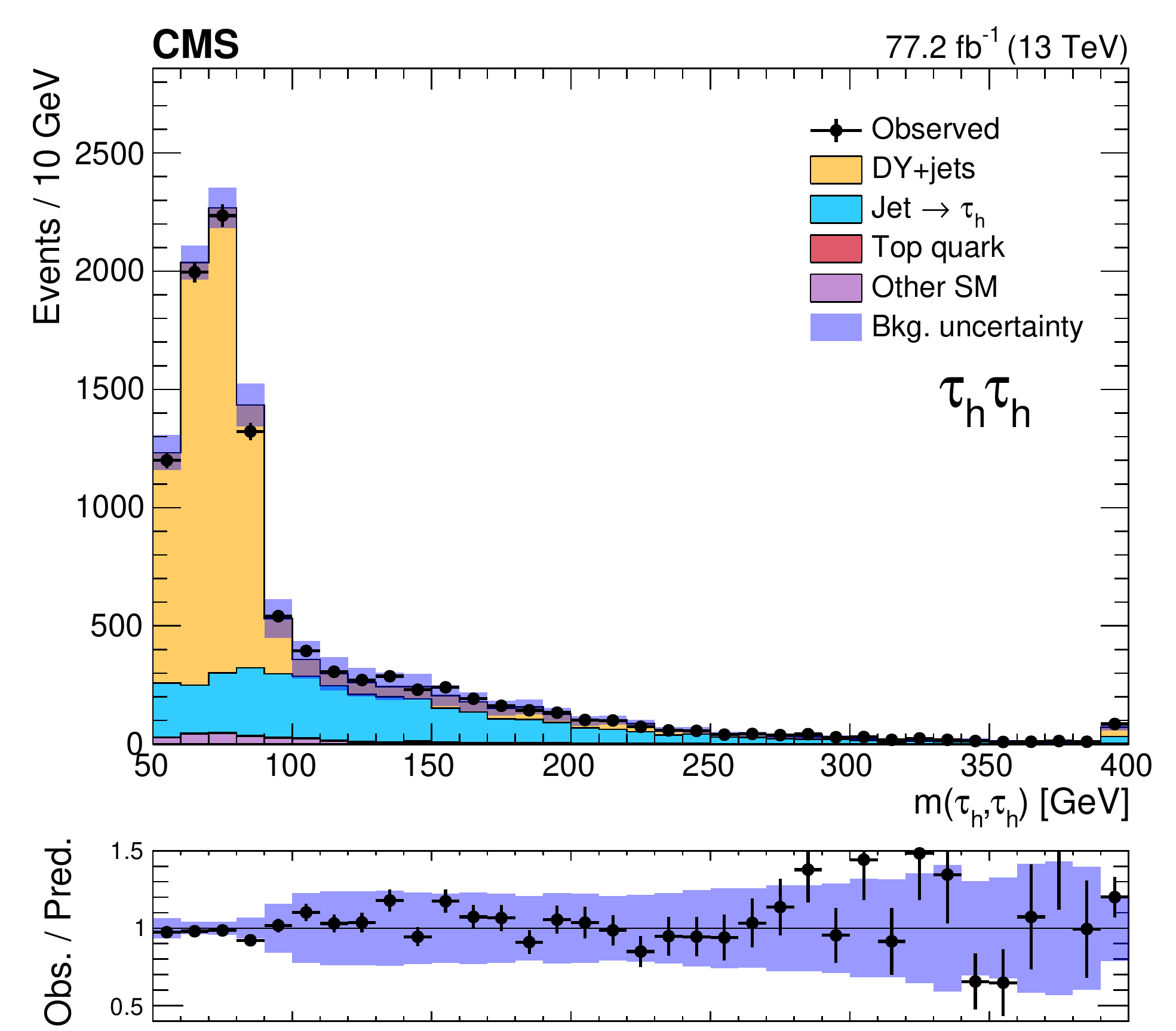}
\includegraphics[width=0.48\textwidth]{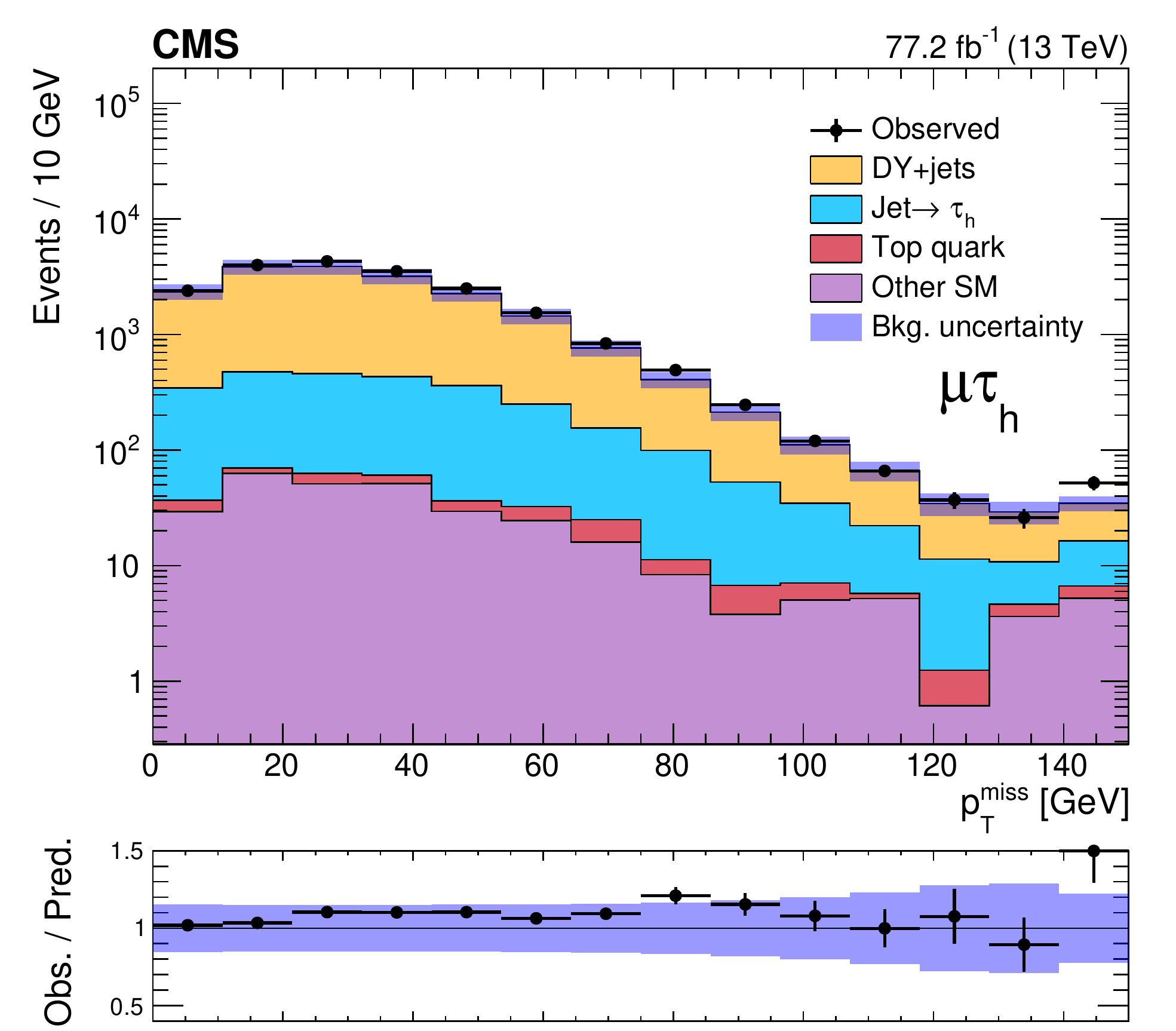}
\caption{\label{fig:ztautau} Visible-mass spectra of \PGt lepton pairs in \tautau events (\cmsLeft) and \ptmiss distribution in \mutau events (\cmsRight) in data and the corresponding prediction for SM background in the combined 2016 and 2017 DY+jets validation regions. The last bin includes overflow events in each case. The shaded uncertainty band represents the statistical and systematic uncertainties in the background prediction. For the \mutau distribution, the systematic uncertainty included in each bin corresponds to a single common average value.}
\end{figure}

The DY+jets background comes primarily from \ztautau decays. We estimate this contribution via simulation, after applying corrections based on CRs in data. Mismodeling of the \PZ boson mass or \pt distribution in simulation can lead to significant differences between data and simulation in kinematic discriminant distributions, especially when considering the large values of these variables that are relevant for the \tautau SRs. We therefore use a high-purity \zmumu CR to compare the dimuon mass and \pt spectra between data and simulation and use the observed differences to correct the simulation in the SRs with weights parameterized by generator-level \PZ boson mass and \pt. The correction factors range up to 30\% for high-mass and high-\pt values. Because these factors are intended to compensate for missing higher-order effects in the simulation, we assign the differences between the generator-level \PZ boson mass and \pt distributions in LO and NLO simulated events as systematic uncertainties. The differences between data and simulation are taken into account through the use of scale factors, as described in Section~\ref{sec:evtreco}.  The uncertainties in these corrections are propagated to the final background estimate. The corrected simulation is validated in the \tautau final state using a \ztautau CR selected by inverting the \mTii and \sumMT requirements used to define the SRs. In addition, requiring a \pt of at least 50\GeV for the \tautau system reduces the QCD multijet background and improves the purity of this CR. This choice makes it possible to increase the statistical power of this region by removing the $\ptmiss>50\GeV$ requirement. The visible mass distribution of the \tautau system shown in Fig.~\ref{fig:ztautau} (\cmsLeft) demonstrates that the corrected simulation agrees  with the data within experimental uncertainties.

For the analysis in the \leptau final states, a normalization scale factor, as well as corrections to the \pt distribution of the \PZ boson in simulation are obtained from a very pure \zmumu CR in data. These events are selected by requiring two isolated muons and no additional leptons, at most one jet, no {\cPqb}-tagged jets, and a dimuon mass in a window of 75--105\GeV, to increase the probability to $>$99\% that they originate from \zmumu decays. After subtracting all other contributions estimated from simulation, a normalization scale factor of $0.96\pm0.05$, which is compatible with unity, is extracted from the ratio of data to simulated events. The uncertainty in the scale factor is determined by varying systematic uncertainties associated with objects such as the muon efficiency and jet energy uncertainties.

{\tolerance=800
To validate the DY+jets background prediction in the \leptau analyses, we construct a CR in \mutau events with $ \mT(\PGm,\ptvecmiss) < 20\GeV$, $ 50 < m(\mutau) < 80\GeV$, and $\nj = 0$. These requirements are chosen to obtain a \ztautau  sample with good purity. The $m(\mutau)$ range is chosen to select the \PZ boson peak, low $ \mT(\PGm,\ptvecmiss)$ helps to remove \wjets and potential signal contamination while the 0-jet requirement helps remove other backgrounds. The \ptmiss distribution of these events is shown in Fig.~\ref{fig:ztautau} (\cmsRight). We observe good agreement between data and the predicted background.
\par}

\subsection{Estimation of other backgrounds}
Smaller contributions are expected from other SM backgrounds, including diboson, triboson, and Higgs boson production. There are also contributions from \ttbar and single top quark production, or top quark pair production in association with a vector boson. These are estimated via MC simulation after application of efficiency and energy-scale corrections. Experimental and theoretical uncertainties are evaluated as described below in Section~\ref{sec:sysunc}.

For the \leptau analyses, we check the BDT distribution in a \ttbar-enriched CR that is defined by requiring the event selection to be the same as in the SR, except for a requirement of one or two {\cPqb}-tagged jets. To validate the {\PW}{\PW} background prediction, we construct a CR of events with  oppositely charged muon-electron pairs that have $m_{\PGm \Pe} > 90\GeV$ and $\nj = 0$. We obtain systematic uncertainties for the normalization of the corresponding backgrounds and any potential mismodeling of the BDT distribution in these CRs. The latter is done by constructing a $\chi^2$ test for all CRs with the BDT modeling taken into account by including an additional floating uncertainty that is determined by requiring a $p$-value~\cite{pvalue} of at least 68\% in all CRs. In this way, the BDT shape uncertainty is estimated to be 9\%.

\section{Systematic uncertainties}
\label{sec:sysunc}
The dominant uncertainties in this analysis are the statistical uncertainties resulting from limited event counts in data sidebands or in simulated event samples used to obtain background estimates and the systematic uncertainties in the estimated rates for jets to be misidentified as  \tauh candidates. We rely on an extrapolation in \tauh isolation to obtain an estimate of the background originating from jets misidentified as \tauh candidates. In the \tautau analysis, the uncertainty in this extrapolation is dominated by the dependence of isolation on jet flavor. It also includes the statistical uncertainty associated with the CR samples from which the extrapolation factors are obtained, which can be significant in the case of search regions with limited event counts that are defined with stringent kinematic requirements. The uncertainty in the combined identification and isolation efficiency for prompt \tauh candidates is also propagated to the final estimated uncertainty. In the \leptau analyses, we estimate a transfer factor for the extrapolation in \tauh isolation from a \wjets-enriched CR. The purity of \wjets events this region is $\approx$85\% as determined from simulation. We therefore propagate a relative uncertainty of 15\% to account for contamination from other sources.

{\tolerance=800
We use simulation to obtain estimates of the yields from other background contributions and to estimate the potential signal contributions. We propagate uncertainties related to the {\cPqb} tagging, trigger, and selection efficiencies, the renormalization and factorization scales, PDFs, jet energy scale and resolution, unclustered energy contributing to \ptmiss, and the energy scales of electrons, muons, and \tauh candidates. The correction factors and the corresponding uncertainties for the \tauh energy scale in simulation are derived from \ztautau events in the \leptau final states by fits to distributions of the reconstructed \tauh mass and the visible mass of the \leptau system~\cite{Sirunyan:2018pgf}. The systematic uncertainties corresponding to energy scale variations can be significant in the \tautau search regions defined with stringent kinematic requirements, which are affected by large statistical uncertainties, because of potentially large event migrations. For the DY+jets background, we have an additional uncertainty associated with the corrections applied to the mass and \pt distributions. We assign a 15\% normalization uncertainty in the \tautau final state for the cross sections of processes estimated from simulation, namely DY+jets, \ttbar, diboson, and rare SM processes, based on the results of CMS differential cross section measurements~\cite{Sirunyan:2018owv,Sirunyan:2018ucr}. For the \leptau analyses, we extract normalization uncertainties of 5, 5, and 20\% for the DY+jets, \ttbar, and {\PW}{\PW} backgrounds, respectively, based on the estimated impurity of the corresponding process-enriched CRs. An additional uncertainty of 9\% is assigned to cover potential mismodeling of the BDT distribution in simulation that is based on studies in CRs.
\par}

The categorization of events in the \tautau final state by the number of reconstructed jets induces sensitivity to the modeling of ISR in the signal simulation. The $\pt^{\mathrm{ISR}}$ distribution of simulated signal events is reweighted to improve the ISR modeling. The  reweighting factors are obtained from studies of \PZ boson events. We take the deviation of the reweighting factors from unity as a systematic uncertainty.

The uncertainty in the integrated luminosity is taken into account in all background estimates for which we do not extract normalization scale factors in dedicated data CRs, as well as for signal estimates. This uncertainty corresponds to 2.5\%~\cite{CMS-PAS-LUM-17-001} and 2.3\%~\cite{CMS-PAS-LUM-17-004} for the 2016 and 2017 data, respectively. With the exception of statistical uncertainties, most other uncertainties are of similar size between the 2016 and 2017 analyses. The main systematic uncertainties for signal and background are summarized in Table~\ref{tab:systematics}.

In general, we treat all statistical uncertainties as uncorrelated. In addition, all systematic uncertainties arising from statistical limitations in the 2016 and 2017 data are assumed to be uncorrelated while systematic uncertainties from similar sources are treated as correlated or partially correlated across the various background and signal predictions. For the combination of the \tautau and \leptau analyses, we correlate uncertainties related to object reconstruction, with the exception of the \tauh selection efficiency, which is treated as uncorrelated  because of the use of different isolation algorithms.

\begin{table*}[tbh]
\centering
\topcaption{\label{tab:systematics} Systematic uncertainties of SM background predictions and a representative signal model, corresponding to a left-handed \PSGt, with $m(\PSGt) = 100\GeV$ and $m(\PSGczDo) = 1\GeV$. The uncertainty ranges are given in percent. The spread of values reflects uncertainties in different SRs.}
\cmsTableNoResize{
\begin{tabular}{lccccc}
\hline
Uncertainty (\%)          & Signal & Misidentified \tauh & DY+jets & Top quark & Other SM    \\
\hline
\tauh efficiency  & 5--13 & \NA & 5--15 & 1--14 & 10--51 \\
$\Pe/\PGm$ efficiency (\leptau)  & 2--3 & \NA & 2--3 & 2--3 & 2--3 \\
\tauh energy scale  & 0.5--12 & \NA & 2.6--27 & 1.2--11 & 4.1--13 \\
$\Pe/\PGm$ energy scale (\leptau) & 0.1--25 & 0.1--5  & 0.1--30  & 0.1--20 & 0.1--10 \\
Jet energy scale  & 0.5--38 & \NA & 1.1--19 & 0.6--13 & 2.4--14 \\
Jet energy resolution  & 0.3--22 & \NA & 1.9--10 & 0.7--22 & 0.2--11 \\
Unclustered energy  & 0.3--21 & \NA & 2.6--30 & 0.2--6.4 & 1.7--14 \\
{\cPqb} tagging  & 0.2--0.9 & \NA & 0.2--23 & 1.7--25 & 0.2--1.2 \\
Pileup  & 0.9--9.1 & \NA & 2--22 & 0.1--24 & 0.3--25 \\
BDT distribution (\leptau)  & 9 & \NA & 9 & 9 & 9 \\
$\ell\to\tauh$ misidentification rate (\leptau)  & \NA & \NA & \NA & 1 & 1 \\
Integrated luminosity  & 2.3--2.5 & \NA & 2.3--2.5 & 2.3--2.5 & 2.3--2.5 \\
Background normalization  & \NA & 10 & 5--15 & 2.5--15 & 15--25 \\
DY+jets mass and \pt  & \NA & \NA & 0.2--11 & \NA & \NA \\
\tauh misidentification rate  & \NA & 4.6--51 & \NA & \NA & \NA \\
Signal ISR  & 0.2--8.2 & \NA & \NA & \NA & \NA \\
Renormalization and factorization scales  & 1.6--7 & \NA & 0.7--14 & 0.7--30 & 6.7--16 \\
PDFs  & \NA & \NA & 0.1--1.2 & 0.1--0.4 & 0.1--0.6 \\
\hline
\end{tabular}
}
\end{table*}

\section{Results and interpretation}
\label{sec:results}
\begin{figure*}[!h]
\centering
\includegraphics[width=0.48\textwidth]{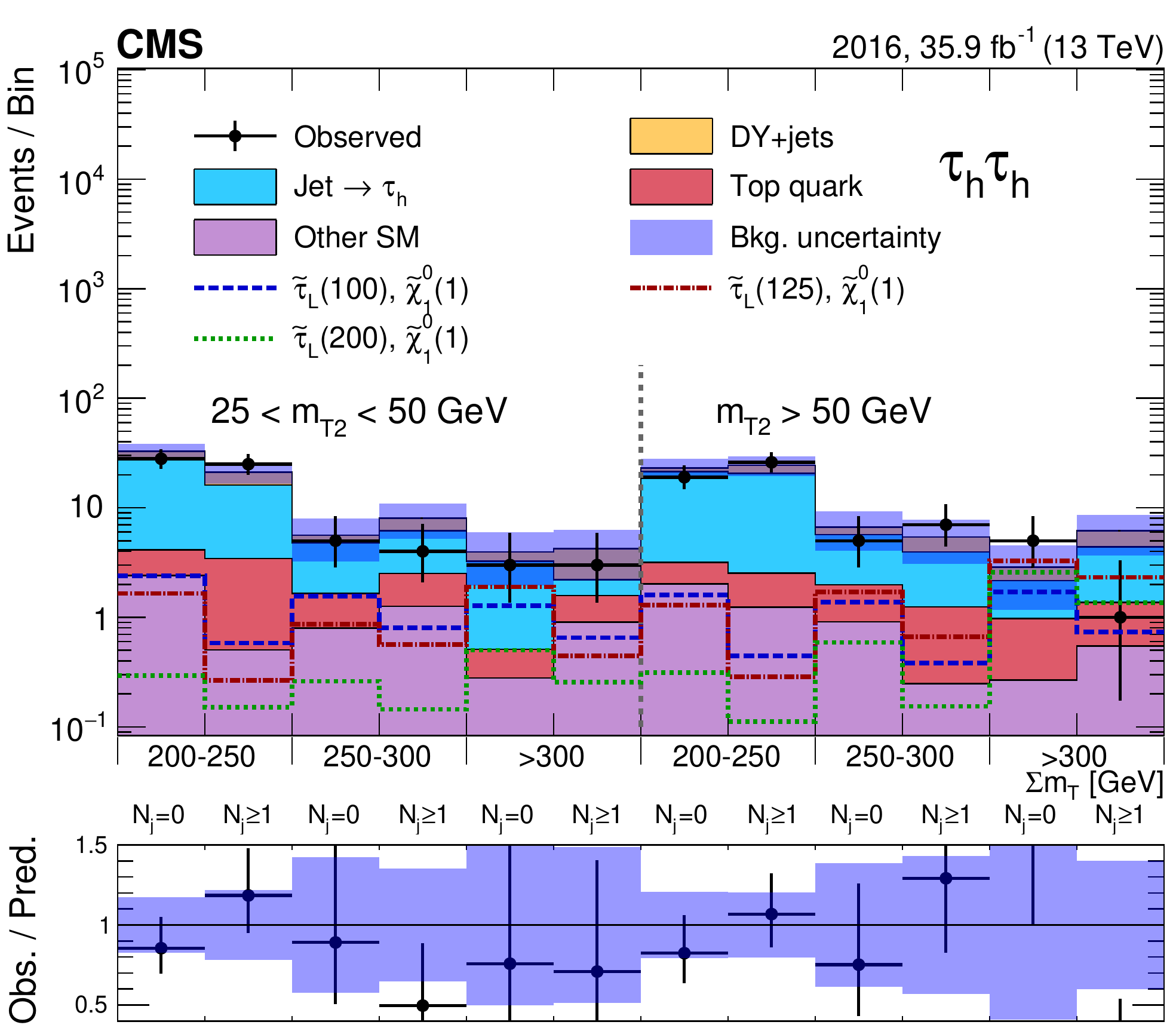}
\includegraphics[width=0.48\textwidth]{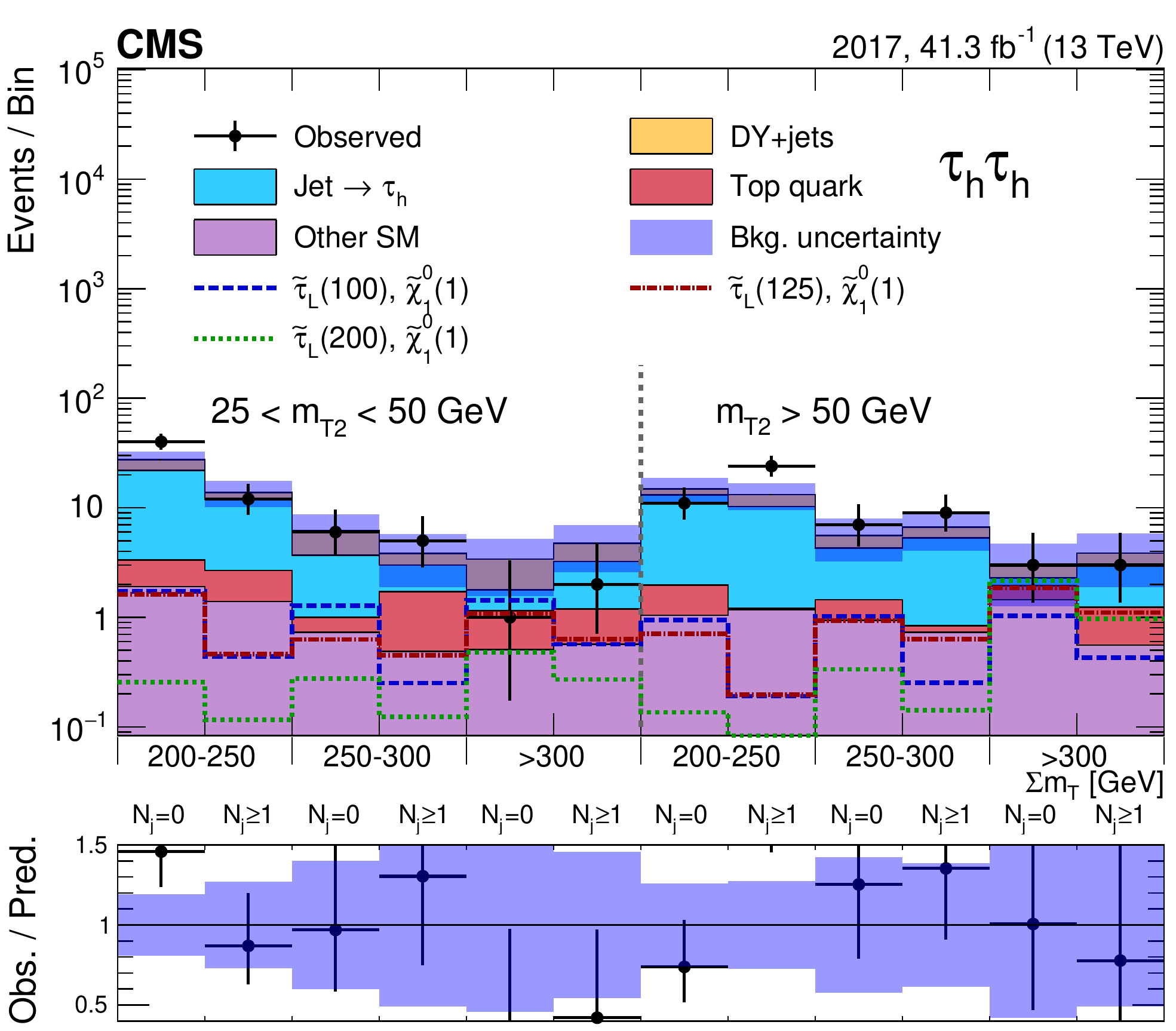} \\
\includegraphics[width=0.48\textwidth]{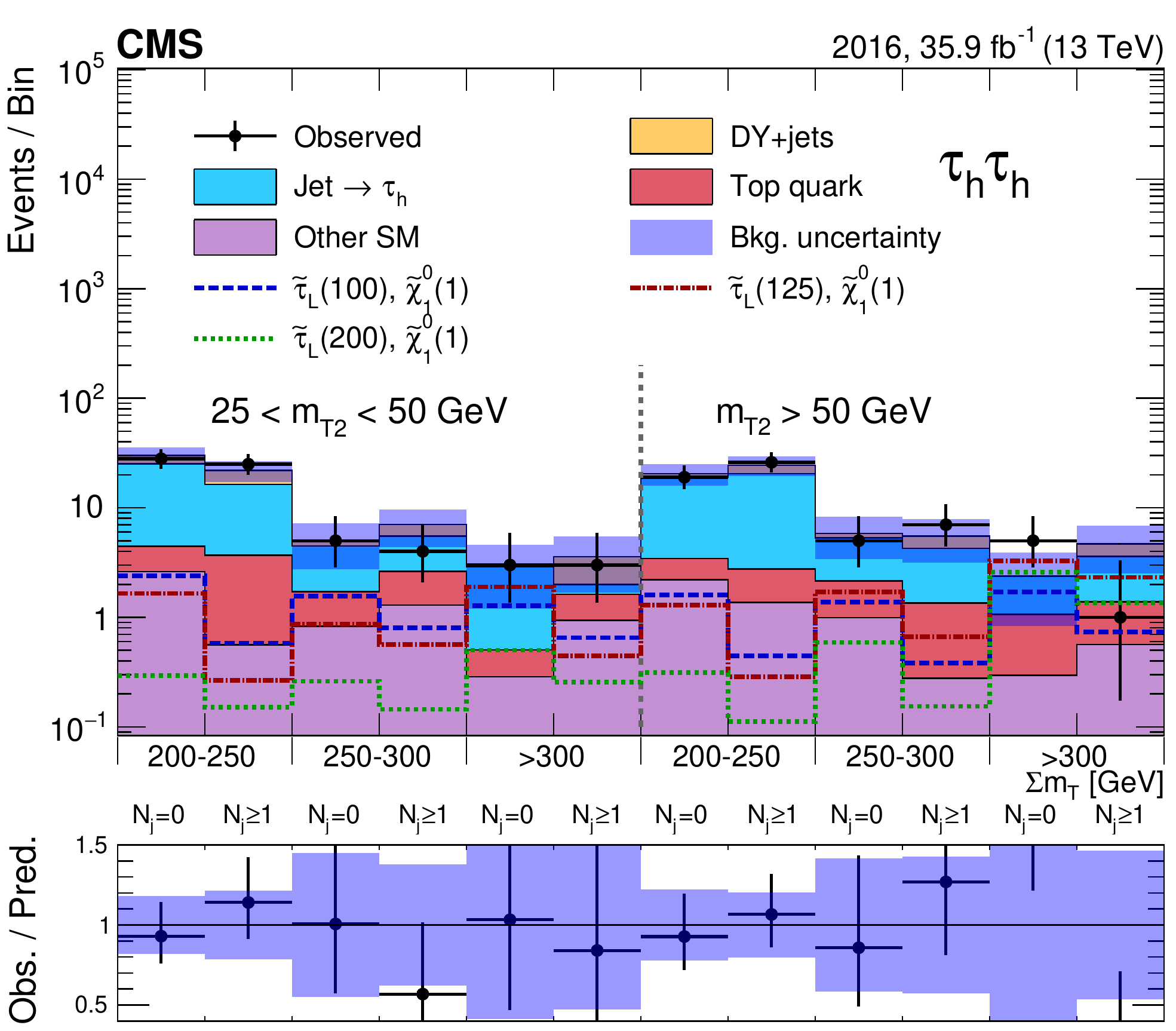}
\includegraphics[width=0.48\textwidth]{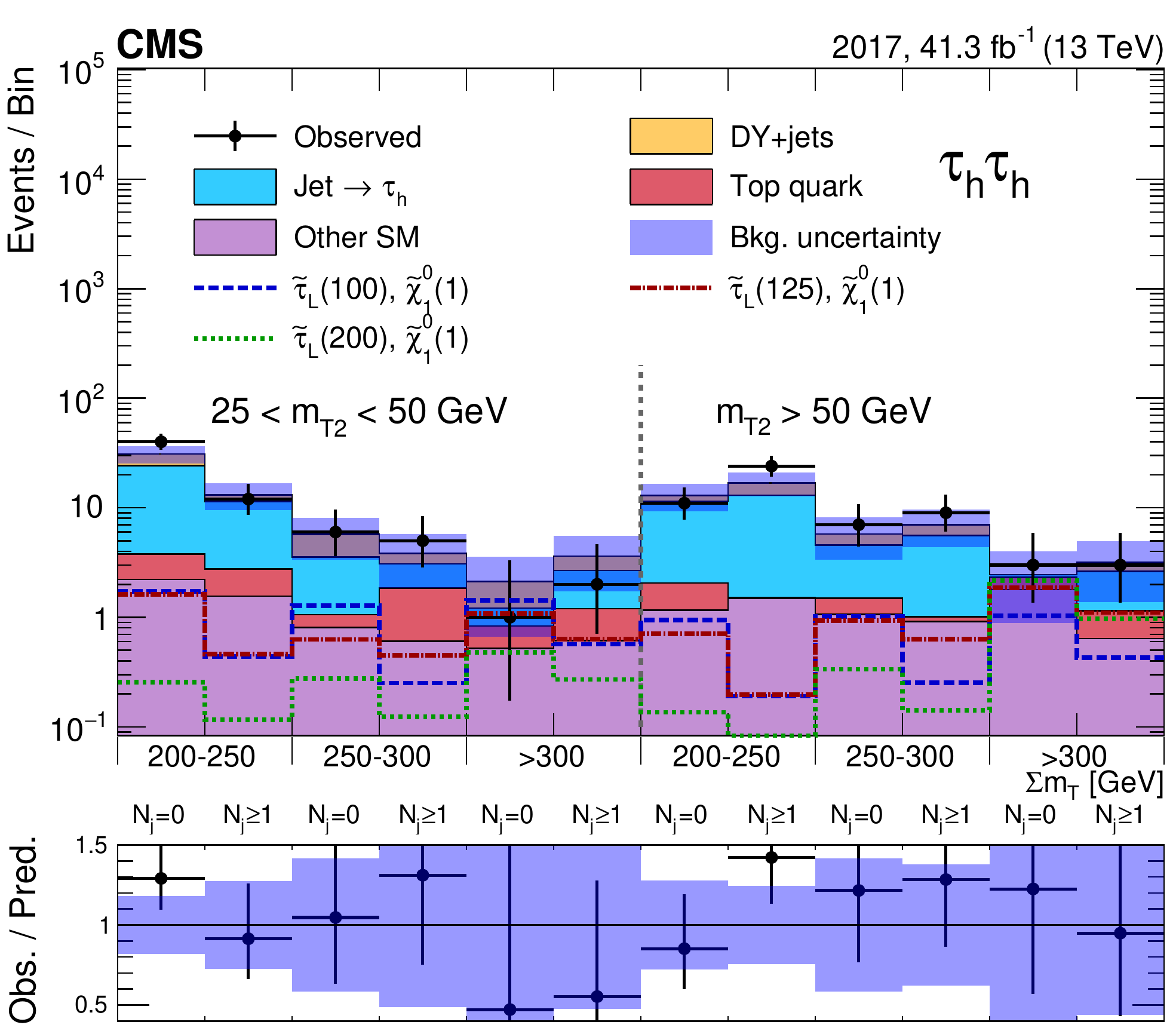}
\caption{Event counts and predicted yields for the SM background in the \tautau analysis for the 2016 (left) and 2017 (right) data, before (upper) and after (lower) a maximum-likelihood fit to the data. Predicted signal yields are also shown for benchmark signal models of \staul pair production with $m(\staul)=100$, 125, and 200\GeV and $m(\PSGczDo)=1\GeV$.}
\label{fig:tautauresults}
\end{figure*}

\begin{figure*}[!h]
\centering
\includegraphics[width=0.48\textwidth]{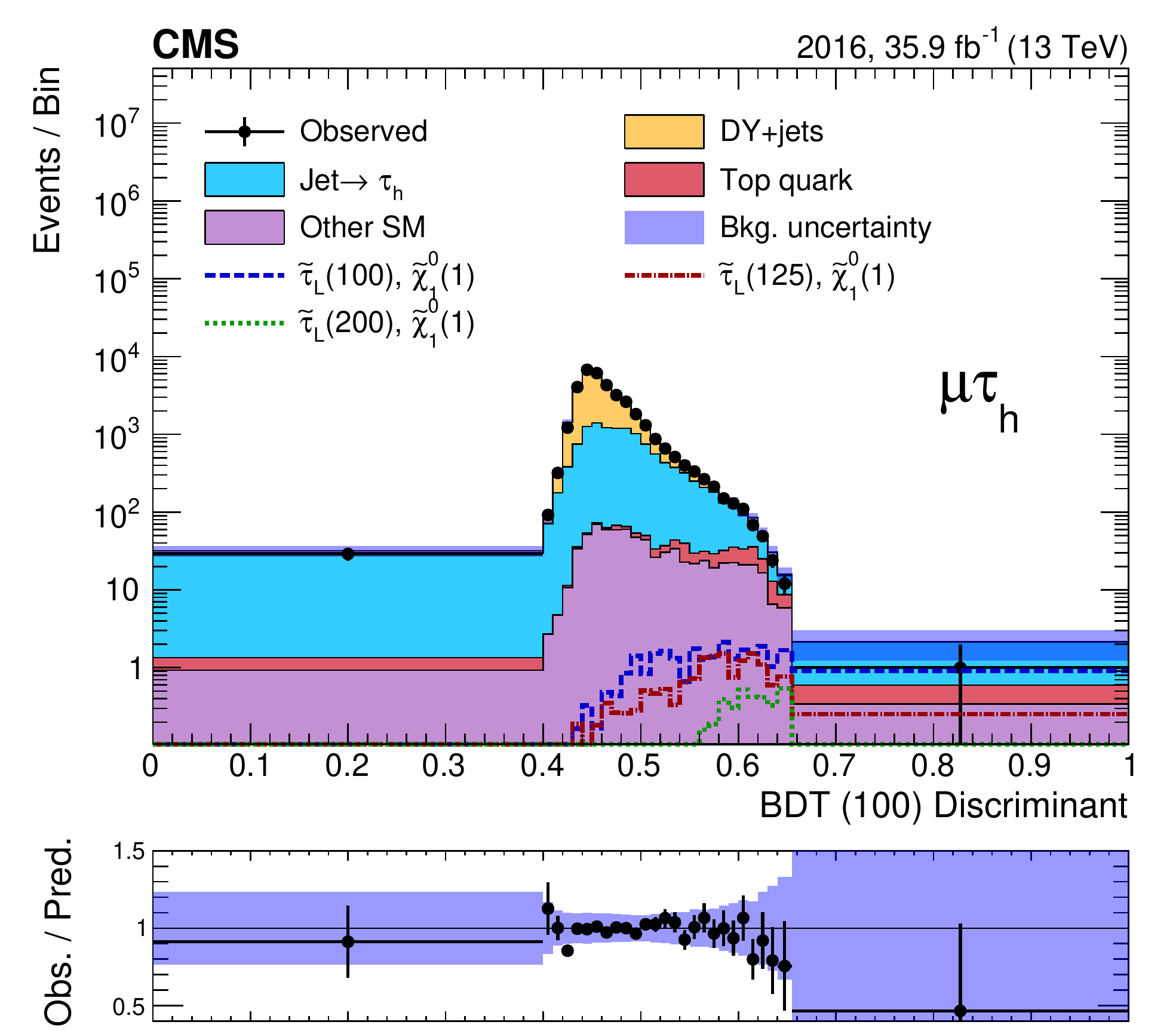}
\includegraphics[width=0.48\textwidth]{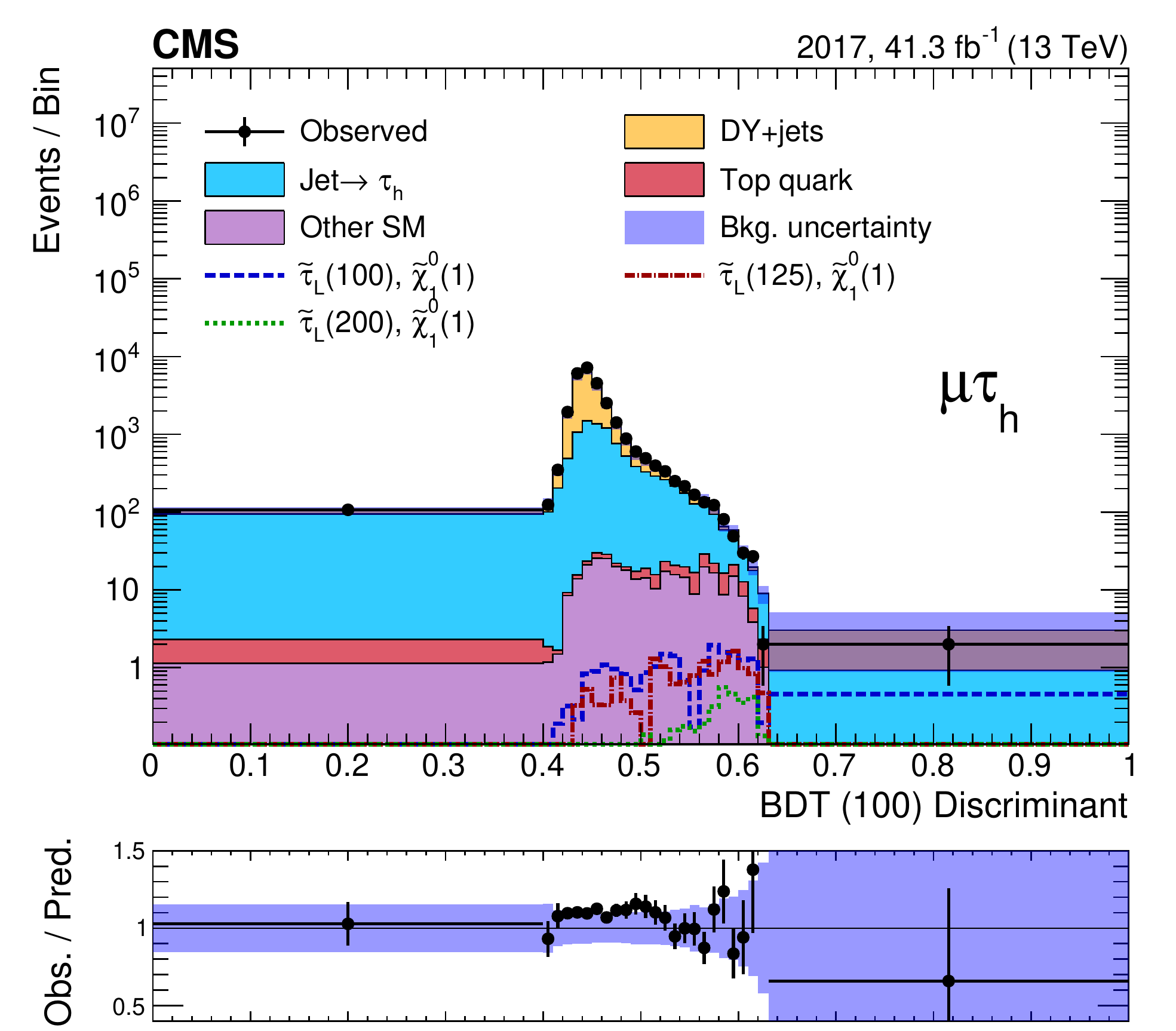} \\
\includegraphics[width=0.48\textwidth]{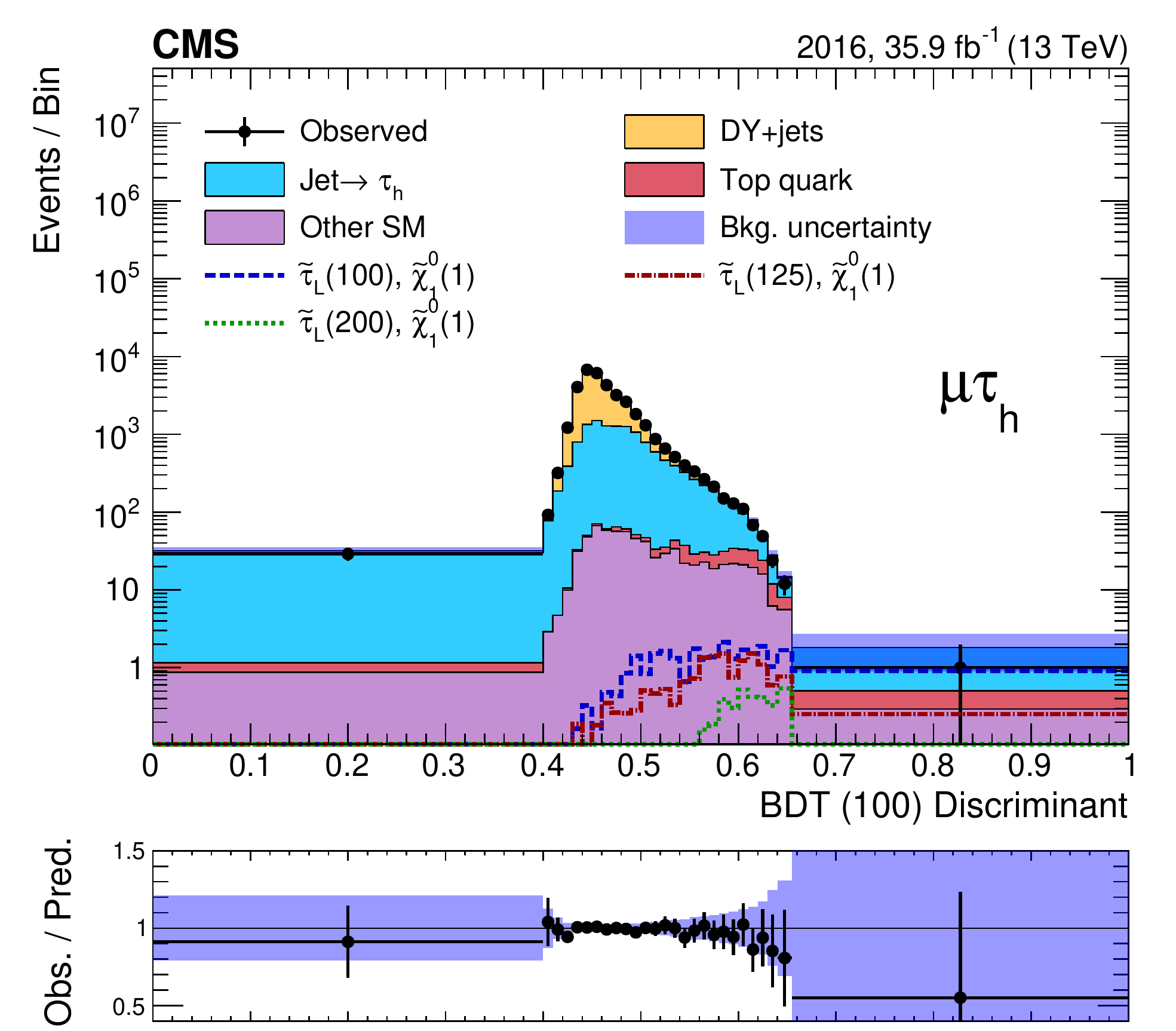}
\includegraphics[width=0.48\textwidth]{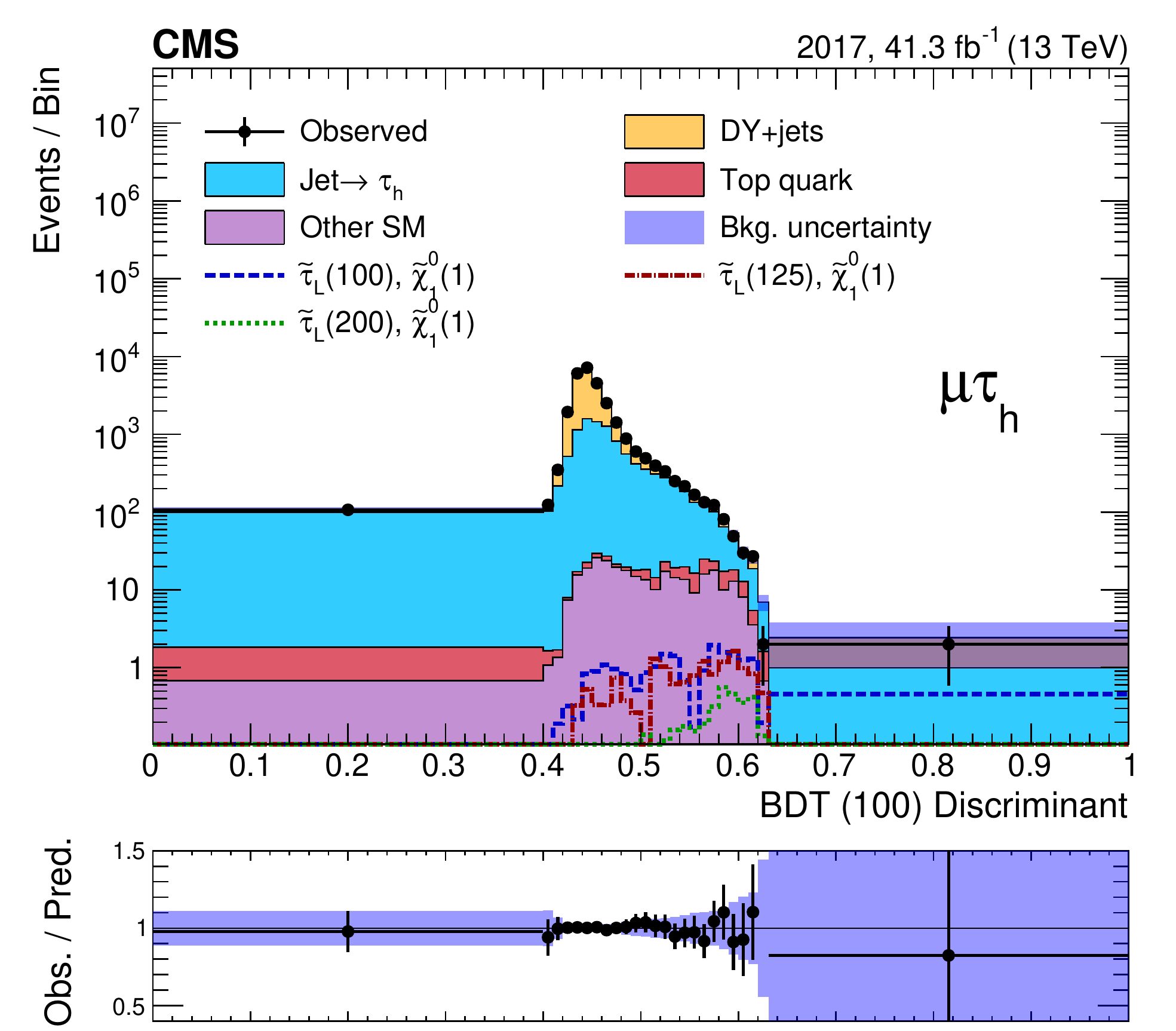} \\
\caption{Discriminant distributions for the BDT trained for a \PSGt mass of 100\GeV and a \PSGczDo mass of 1\GeV (BDT (100)) in the \mutau final state for the 2016 (left) and 2017 (right) data, before (upper) and after (lower) a maximum-likelihood fit to the data. Predicted signal yields are also shown for benchmark models of \staul pair production with $m(\staul)=100$, 125, and 200\GeV and $m(\PSGczDo)=1\GeV$.}
\label{fig:fitplots_mutau}
\end{figure*}

\begin{figure*}[!h]
\centering
\includegraphics[width=0.48\textwidth]{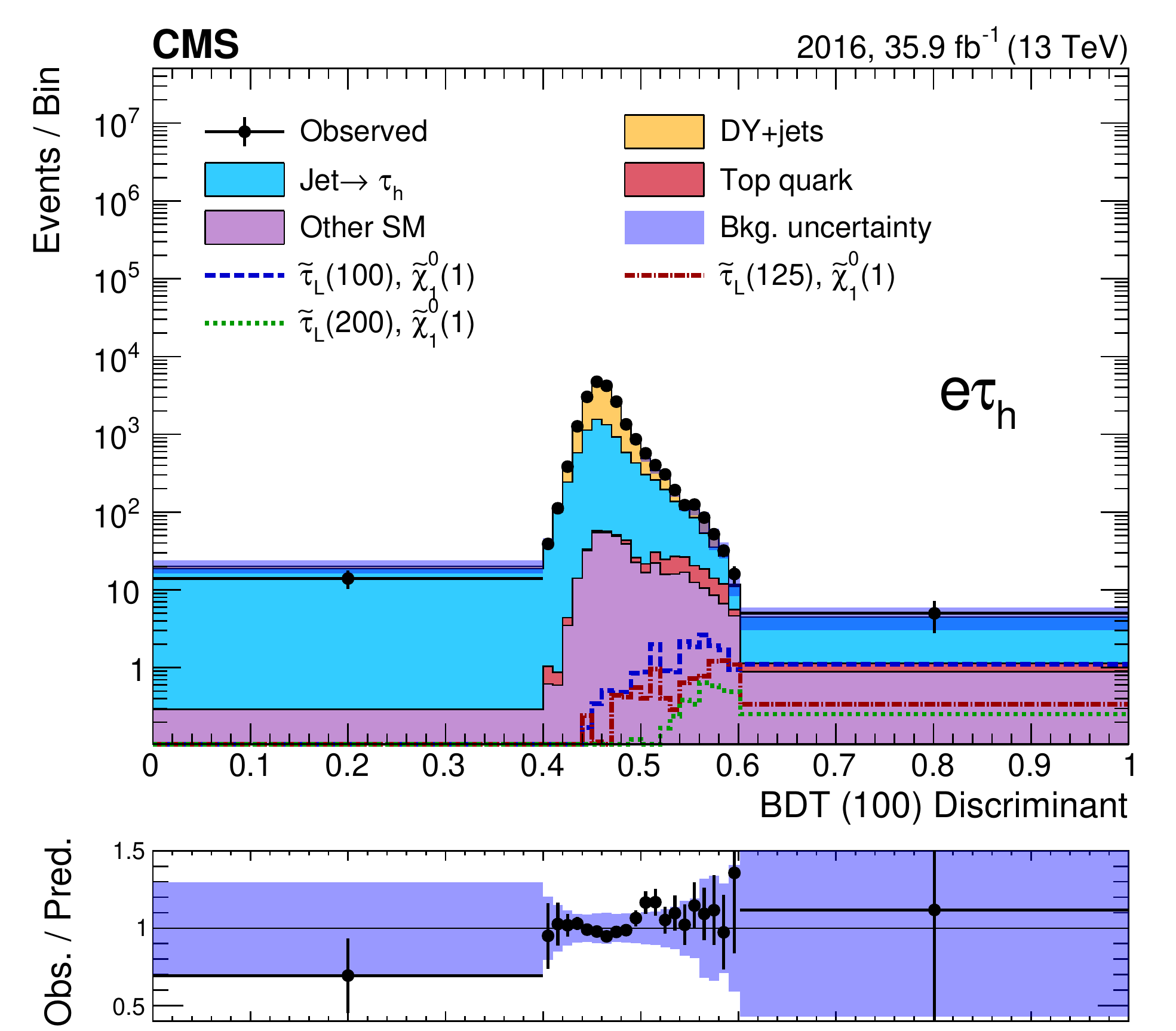}
\includegraphics[width=0.48\textwidth]{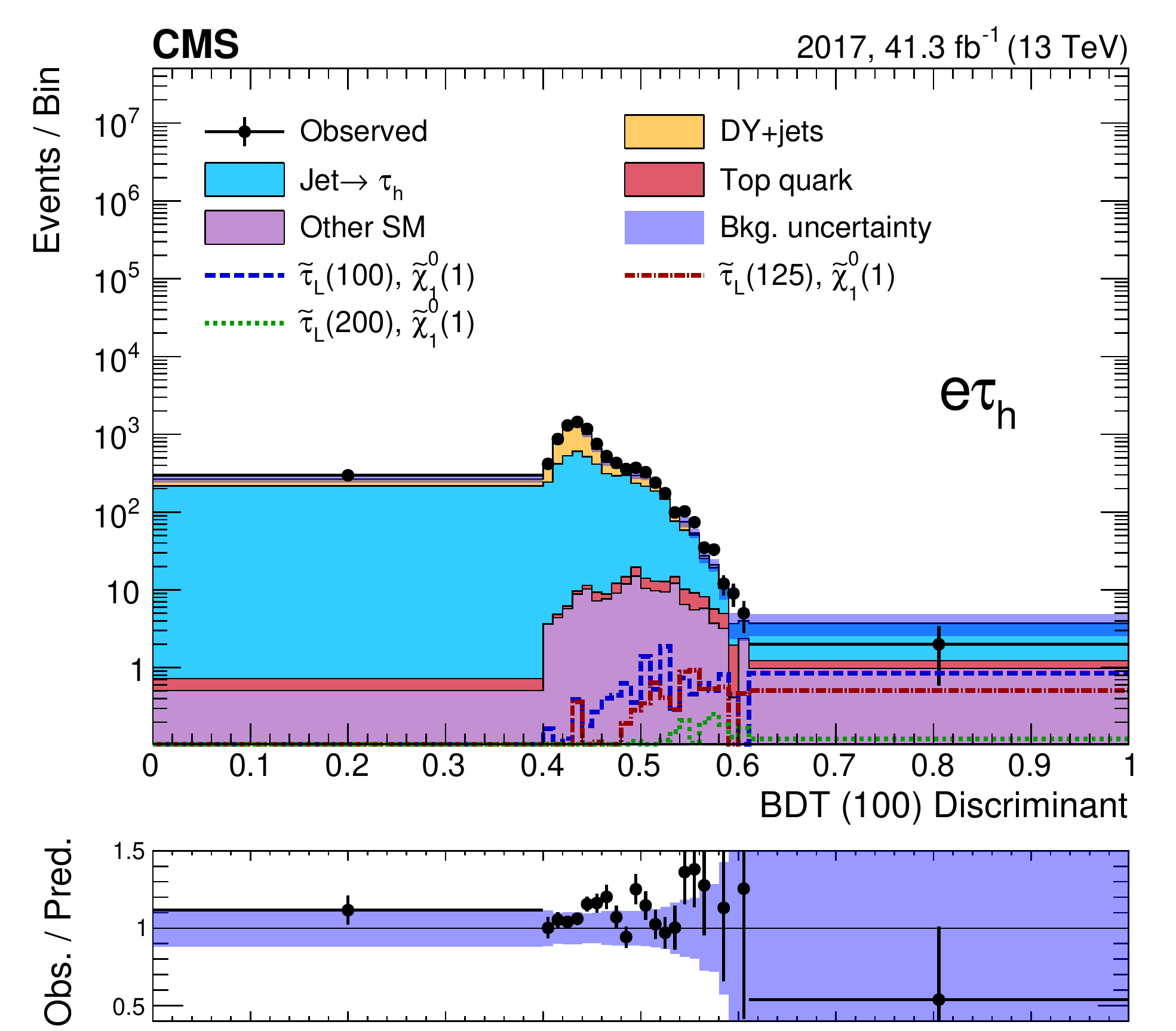} \\
\includegraphics[width=0.48\textwidth]{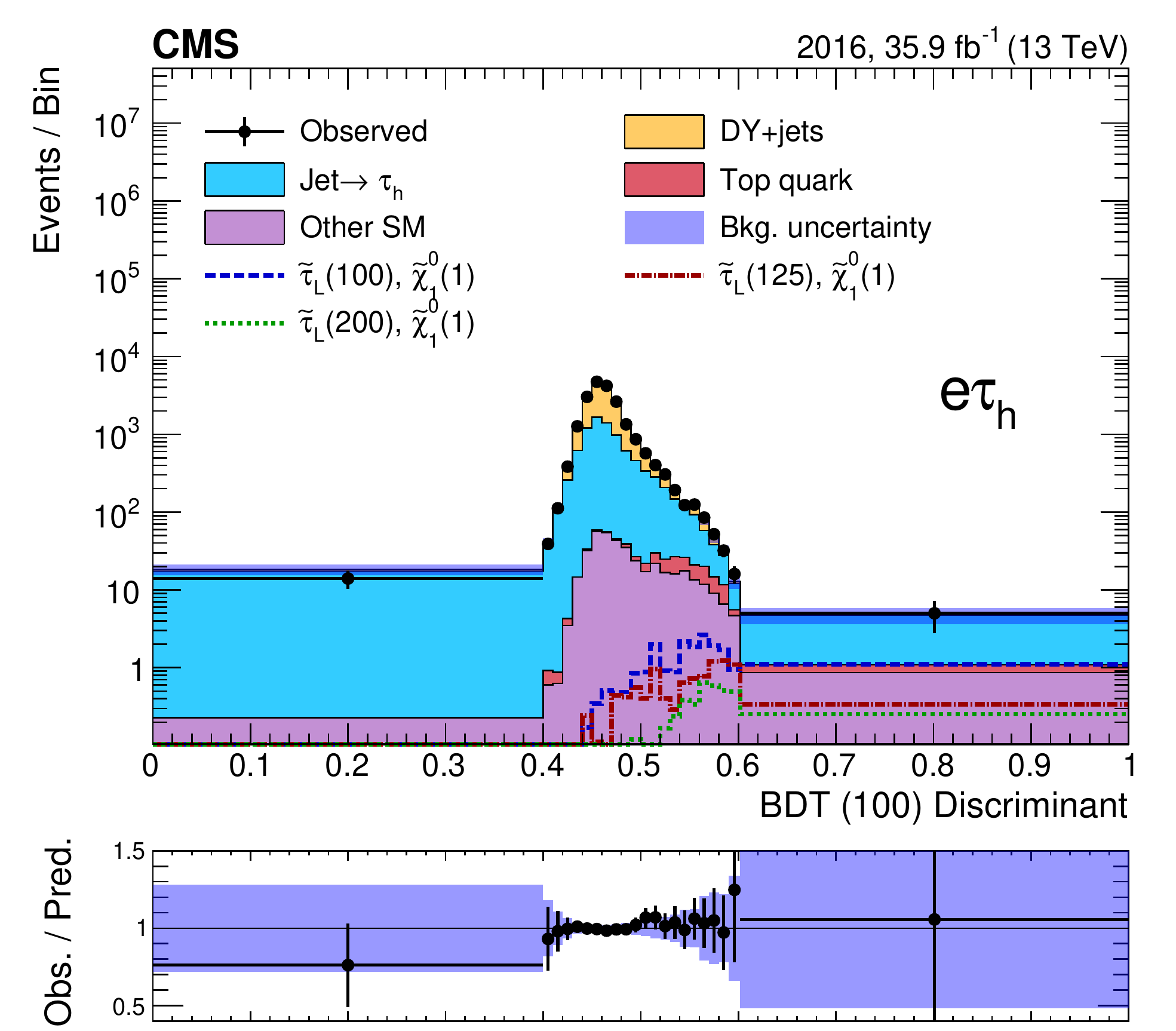}
\includegraphics[width=0.48\textwidth]{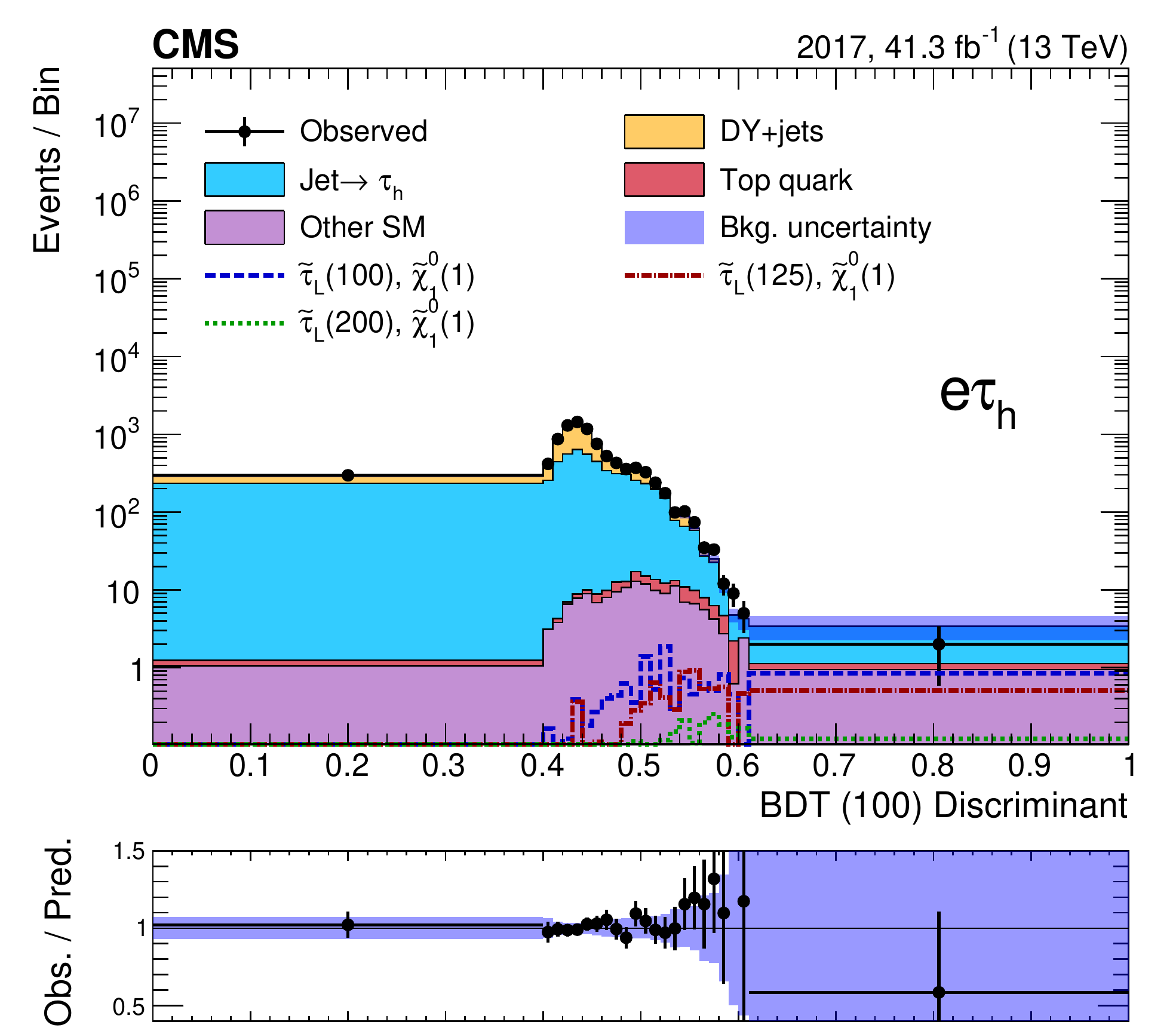} \\
\caption{Discriminant distributions for the BDT trained for a \PSGt mass of 100\GeV and a \PSGczDo mass of 1\GeV (BDT (100)) in the \etau final state for the 2016 (left) and 2017 (right) data, before (upper) and after (lower) a maximum-likelihood fit to the data. Predicted signal yields are also shown for benchmark models of \staul pair production with $m(\staul)=100$, 125, and 200\GeV and $m(\PSGczDo)=1\GeV$.}
\label{fig:fitplots_etau}
\end{figure*}

The results of the search in the \tautau final state are presented in Fig.~\ref{fig:tautauresults} and summarized in Tables~\ref{tab:results2016} and~\ref{tab:results2017}. The background predictions resulting from a maximum likelihood fit to the data under the background-only hypothesis are shown in the lower row of Fig.~\ref{fig:tautauresults}. The BDT distributions corresponding to a training for a \PSGt mass of 100\GeV and a \PSGczDo mass of 1\GeV are shown before and after the maximum-likelihood fit to the data in Figs.~\ref{fig:fitplots_mutau} and~\ref{fig:fitplots_etau} for the \mutau and \etau final states, respectively. The data are consistent with the prediction for SM background. The predicted and observed event yields in the last, most sensitive BDT bins are summarized in Tables~\ref{tab:resultsSemilep2016} and~\ref{tab:resultsSemilep2017} for \leptau final states. For the statistical interpretation of these results, the normalization uncertainties affecting background and signal predictions are generally assumed to be log-normally distributed. For statistical uncertainties limited by small event counts in data or simulation, we use a $\Gamma$ distribution.

\begin{figure*}[!h]
\centering
\includegraphics[width=0.48\textwidth]{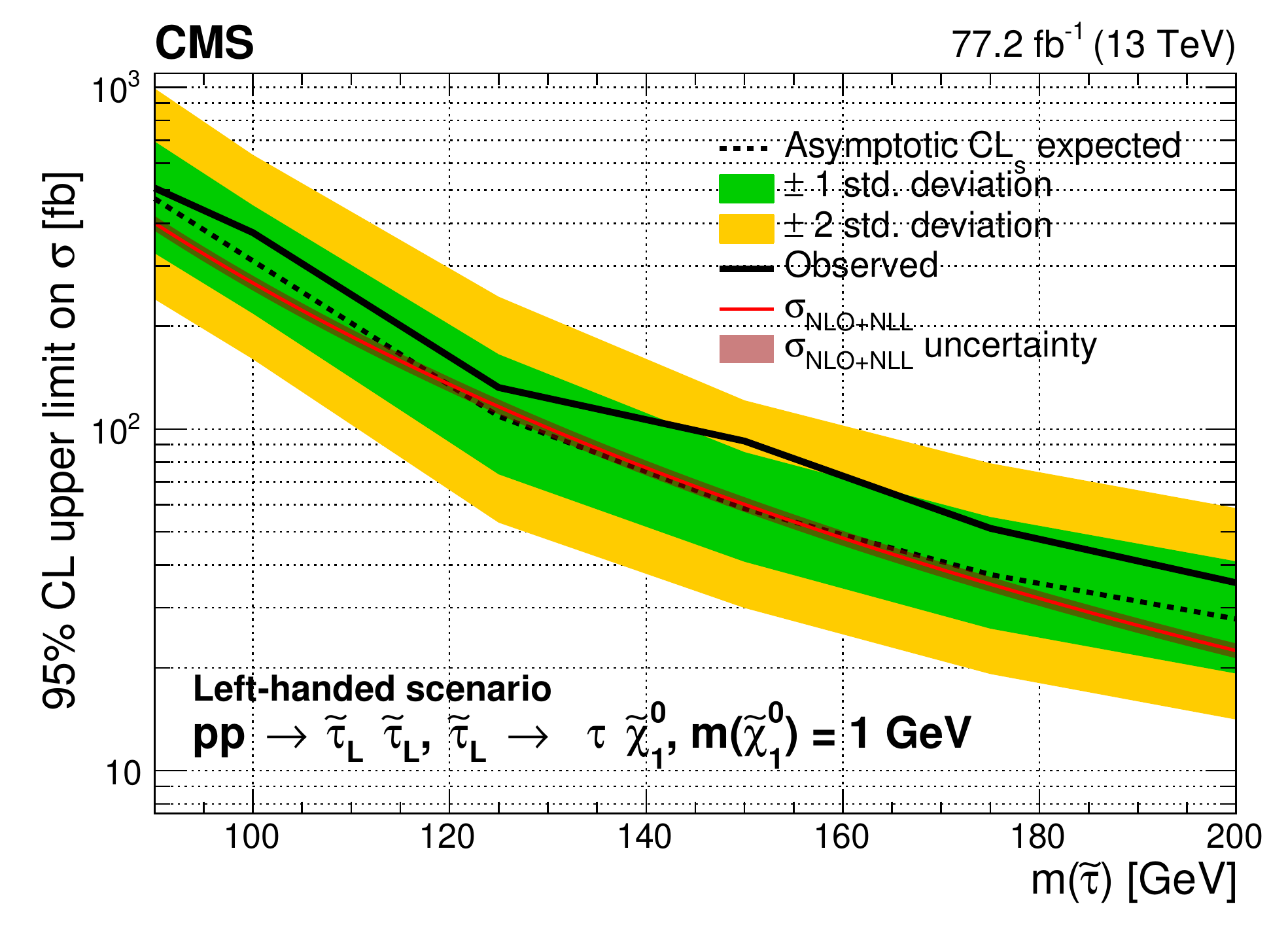}
\includegraphics[width=0.48\textwidth]{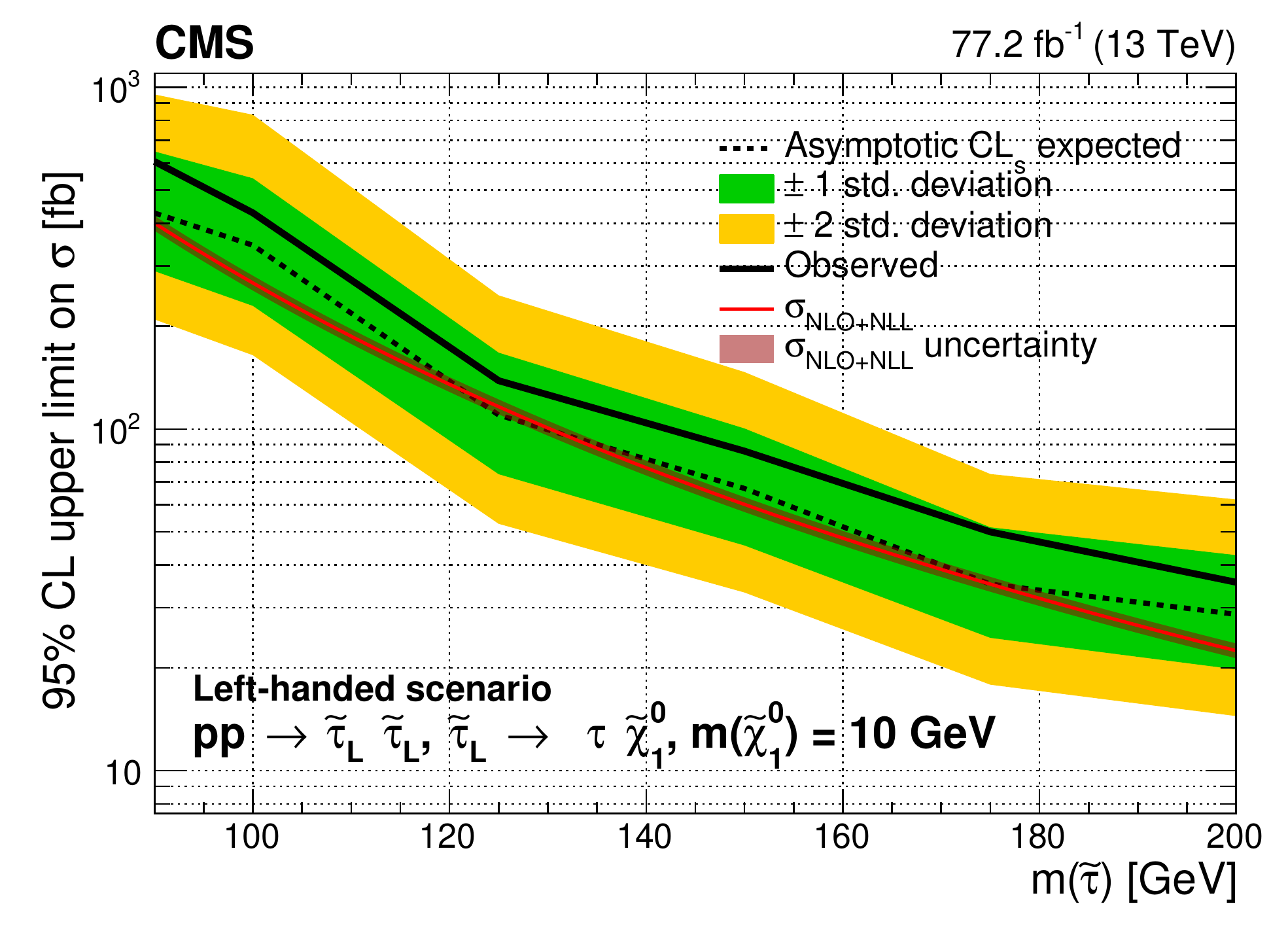}
\includegraphics[width=0.48\textwidth]{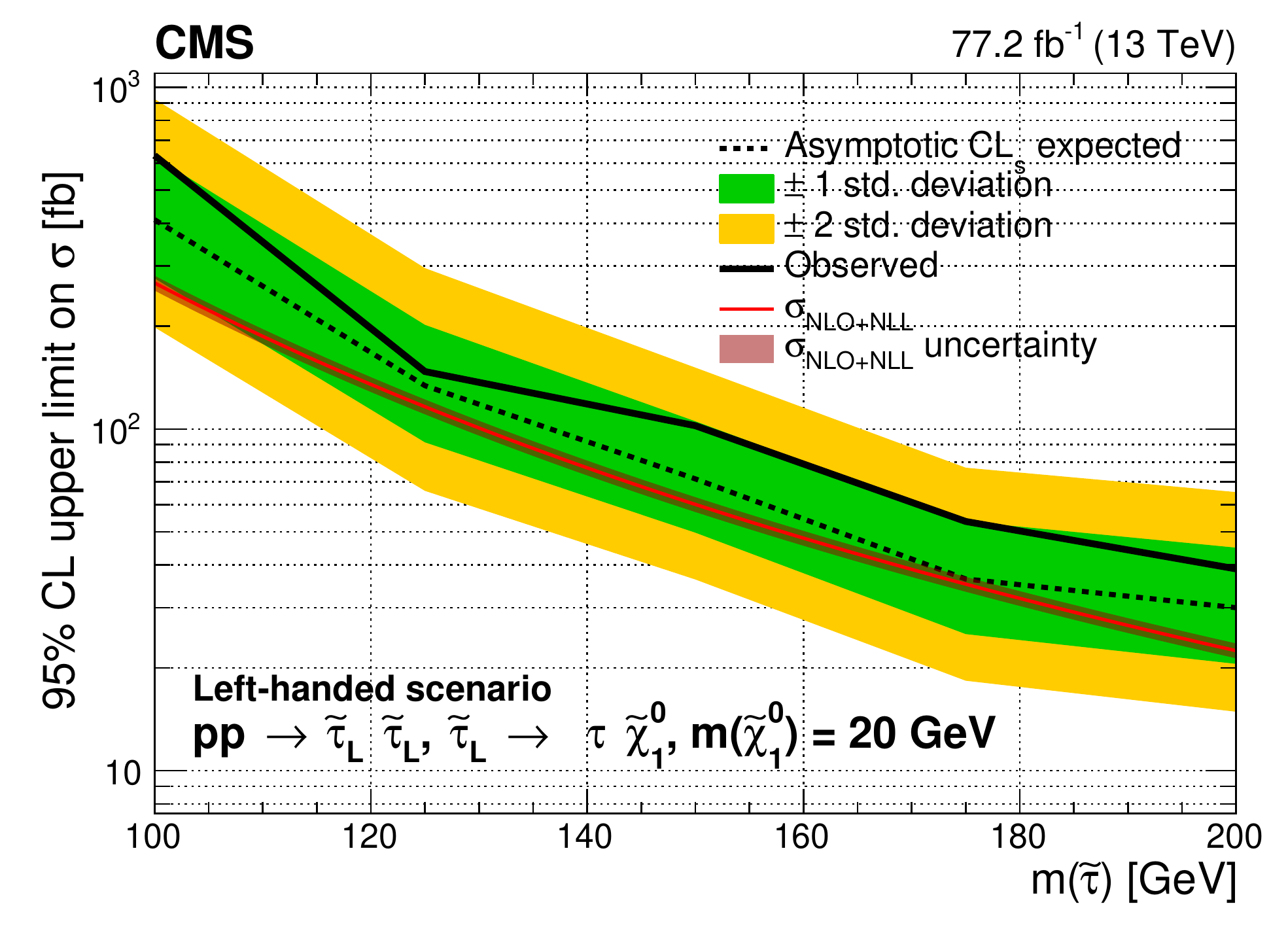}
\caption{Upper limit on the cross section ($\sigma$) of \PSGt pair production excluded at 95\% \CL as a function of the \PSGt mass in the purely left-handed \PSGt models for a \PSGczDo mass of 1\GeV (upper left), 10\GeV (upper right) and 20\GeV (lower). The results shown are for the statistical combination of the 2016 and 2017 data in the \tautau and \leptau analyses. The inner (green) and outer (yellow) bands indicate the respective regions containing 68 and 95\% of the distribution of limits expected under the background-only hypothesis. The solid red line indicates the NLO+NLL prediction for the signal production cross section calculated with \textsc{Resummino}~\cite{Fuks:2013lya}, while the red shaded band represents the uncertainty in the prediction.}
\label{fig:limitsleft}
\end{figure*}

\begin{figure*}[!h]
\centering
\includegraphics[width=0.48\textwidth]{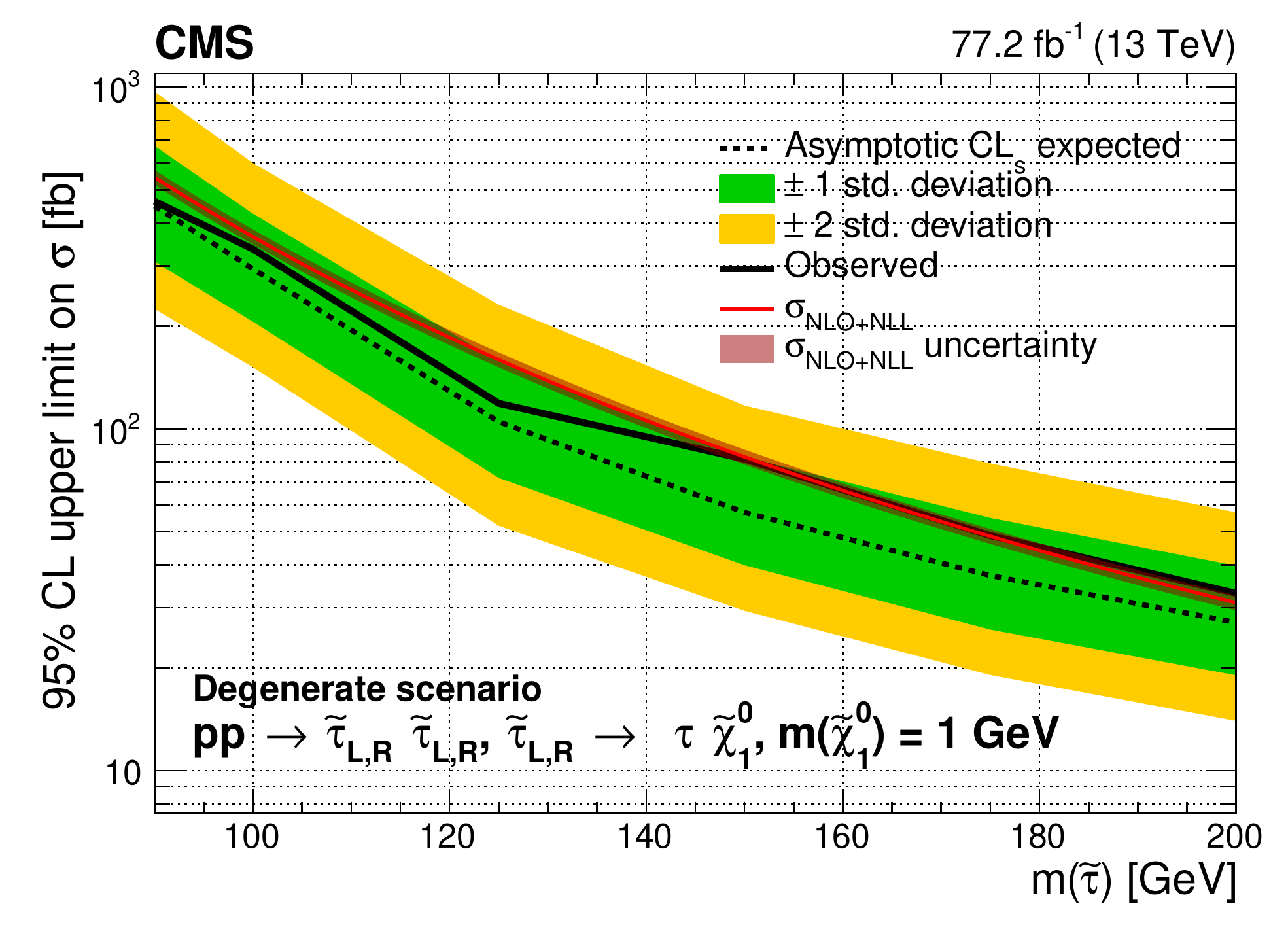}
\includegraphics[width=0.48\textwidth]{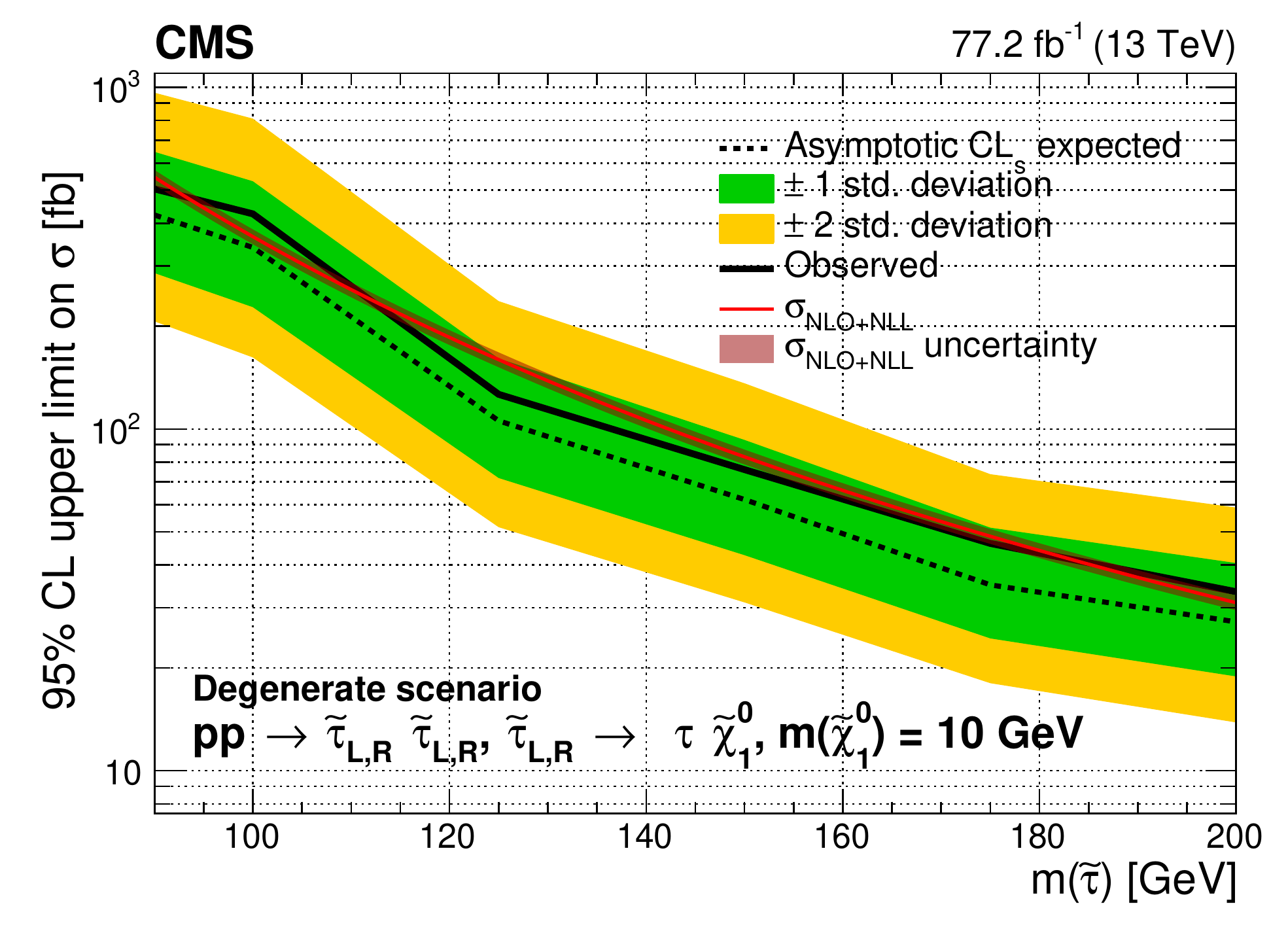}
\includegraphics[width=0.48\textwidth]{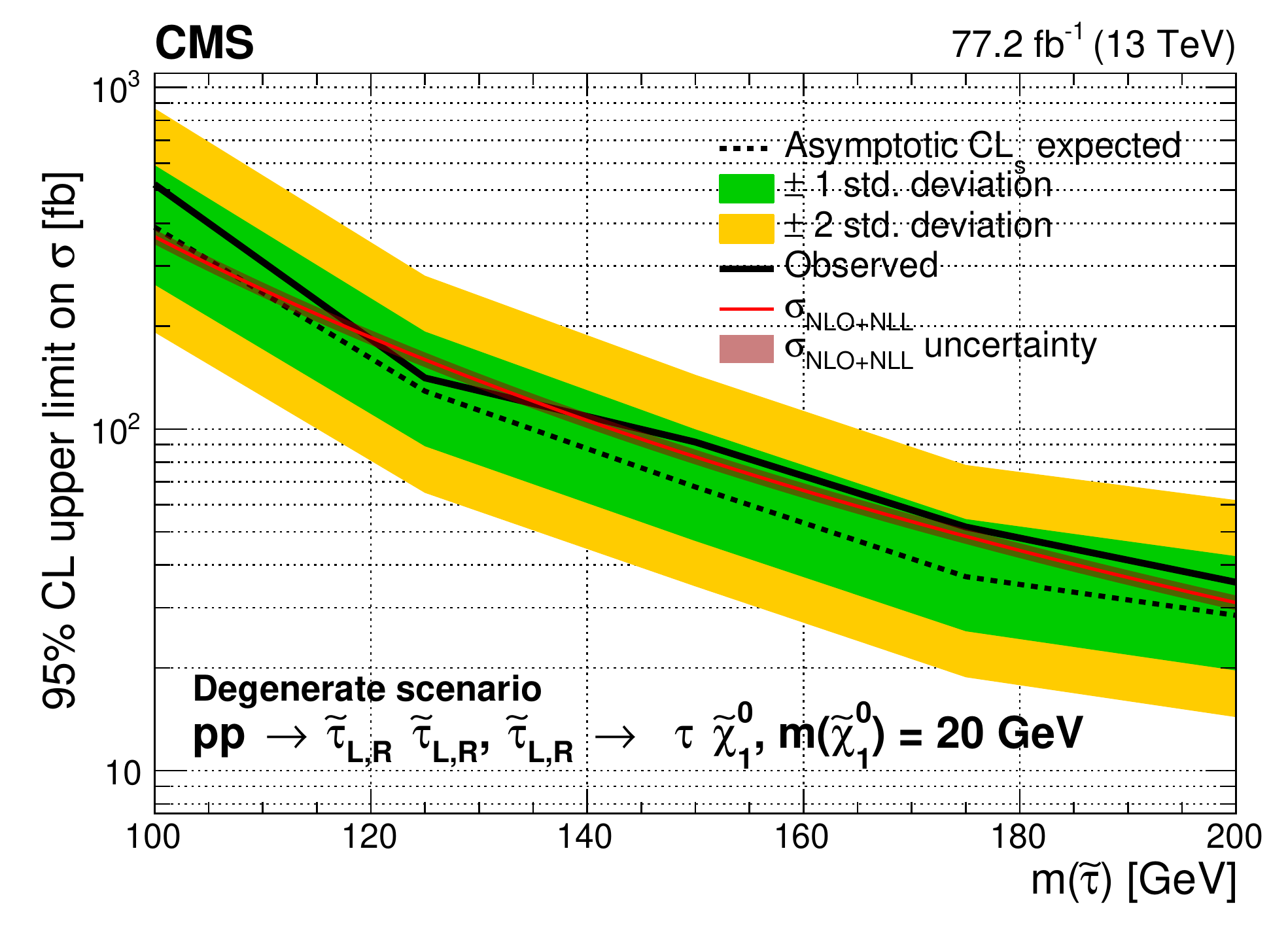}
\caption{Upper limit on the cross section ($\sigma$) of \PSGt pair production excluded at 95\% \CL as a function of the \PSGt mass in the degenerate \PSGt models for a \PSGczDo mass of 1\GeV (upper left), 10\GeV (upper right) and 20\GeV (lower). The results shown are for the statistical combination of the 2016 and 2017 data in the \tautau and \leptau analyses. The inner (green) and outer (yellow) bands indicate the respective regions containing 68 and 95\% of the distribution of limits expected under the background-only hypothesis. The solid red line indicates the NLO+NLL prediction for the signal production cross section calculated with \textsc{Resummino}~\cite{Fuks:2013lya}, while the red shaded band represents the uncertainty in the prediction.}
\label{fig:limitsdegenerate}
\end{figure*}

The results are used to set upper limits on the cross section for the production of \PSGt pairs in the context of simplified models~\cite{Simp,Alwall:2008ag,Alwall:2008va,Alves:2011wf} using all of the exclusive \tautau SRs and the \leptau BDT distributions in a full statistical combination. The limits are evaluated using likelihood fits with the signal strength, background event yields, and nuisance parameters corresponding to the uncertainties in the signal and background estimates as fitted parameters. The nuisance parameters are constrained within their uncertainties in the fit. We assume that the \PSGt decays with 100\% branching fraction to a \PGt lepton and a \PSGczDo. The 95\% \CL upper limits on SUSY production cross sections are calculated using a modified frequentist approach with the \CLs criterion~\cite{Junk:1999kv,Read:2002hq}. An asymptotic approximation is used for the test statistic~\cite{CMS-NOTE-2011-005,Cowan:2010js}, $q_{\mu} = -2\ln \mathcal{L}_{\mu}/\mathcal{L}_{\text{max}}$, where $\mathcal{L}_{\text{max}}$ is the maximum likelihood determined by allowing all fitted parameters, including the signal strength $\mu$, to vary, and $\mathcal{L}_{\mu}$ is the maximum likelihood for a fixed signal strength. Figure~\ref{fig:limitsleft} shows the limits obtained for purely left-handed \PSGt pair production, while Fig.~\ref{fig:limitsdegenerate} shows the limits obtained for the degenerate \PSGt model in which both left- and right-handed \PSGt pairs are produced. The \tautau analysis makes the dominant contribution to the search sensitivity. A slight excess of events over the background expectation in the \tautau SRs results in an observed limit that is weaker than the expected limit. The strongest limits are observed in the case of a nearly massless \PSGczDo. In general, the constraints are weaker for higher values of the \PSGczDo mass because of smaller experimental acceptances. For \PSGt masses above $\approx$150\GeV, however, the sensitivity does not degrade significantly when the \PSGczDo mass increases up to 20\GeV. In the purely left-handed model, the strongest limits are observed for a \PSGt mass of 125\GeV where we exclude a \PSGt pair production cross section of 132\unit{fb}. This value is a factor of 1.14 larger than the theoretical cross section. In the degenerate \PSGt model we exclude \PSGt masses between 90 and 150\GeV under the assumption of a nearly massless \PSGczDo.

\begin{table*}[tbh]
\centering
\topcaption{Predicted background yields and observed event counts in \tautau SRs in 2016 data. For the background estimates with no events in the sideband or in the simulated sample, we calculate the 68\% \CL upper limit on the yield. The first and second uncertainties given are statistical and systematic, respectively. We also list the predicted signal yields corresponding to the purely left-handed model for a \PSGt mass of 100\GeV and a \PSGczDo mass of 1\GeV.}
\cmsTable{
\begin{tabular}{l  c  c  c  c  c  c }
\hline
\mTii [\GeVns{}] & \multicolumn{6}{c}{25--50} \\
[\cmsTabSkip]
$\sumMT$ [\GeVns{}] & \multicolumn{2}{c}{200--250} & \multicolumn{2}{c}{250--300} & \multicolumn{2}{c}{$>$300} \\
[\cmsTabSkip]
\nj & 0 & $\geq$1 & 0 & $\geq$1 & 0 & $\geq$1 \\
\hline
Misidentified \tauh & 23.5 $\pm$ 2.9 $\pm$ 9.8 & 12.7 $\pm$ 2.4 $\pm$ 4.2 & 3.1 $\pm$ 1.0 $\pm$ 1.7 & 3.6 $\pm$ 1.1 $\pm$ 2.0 & 2.8 $\pm$ 0.8 $\pm$ 1.8 & 0.5 $\pm$ 0.5 $\pm$ 0.2 \\
DY+jets & 4.3 $\pm$ 2.1 $\pm$ 0.7 & 4.5 $\pm$ 1.5 $\pm$ 0.9 & 0.4 $\pm$ 0.4 $\pm$ 0.1 & 1.6 $\pm$ 0.9 $\pm$ 0.3 & $<$0.7 & 1.5 $\pm$ 0.9 $\pm$ 0.5 \\
Top quark & 1.7 $\pm$ 0.3 $\pm$ 0.3 & 2.9 $\pm$ 0.4 $\pm$ 0.3 & 0.8 $\pm$ 0.2 $\pm$ 0.1 & 1.3 $\pm$ 0.2 $\pm$ 0.2 & 0.2 $\pm$ 0.1 $\pm$ 0.1 & 0.6 $\pm$ 0.2 $\pm$ 0.2 \\
Other SM & 2.4 $\pm$ 0.7 $\pm$ 0.4 & 0.5 $\pm$ 0.2 $\pm$ 0.1 & 0.7 $\pm$ 0.4 $\pm$ 0.1 & 1.2 $\pm$ 0.5 $\pm$ 0.3 & 0.3 $\pm$ 0.2 $\pm$ 0.1 & 0.9 $\pm$ 0.4 $\pm$ 0.2 \\
Total prediction & 31.9 $\pm$ 3.7 $\pm$ 9.8 & 20.6 $\pm$ 2.9 $\pm$ 4.3 & 5.1 $\pm$ 1.2 $\pm$ 1.7 & 7.7 $\pm$ 1.5 $\pm$ 2.1 & 3.2 $\pm$ 0.9 $\pm$ 1.8 & 3.5 $\pm$ 1.1 $\pm$ 0.6 \\
Observed & 28 & 25 & 5 & 4 & 3 & 3 \\
$m(\staul)=100 \GeV$ & 2.4  $\pm$ 0.3  $\pm$ 0.4 & 0.6  $\pm$ 0.2  $\pm$ 0.1 & 1.6  $\pm$ 0.2  $\pm$ 0.2 & 0.8  $\pm$ 0.2  $\pm$ 0.1 & 1.3  $\pm$ 0.2  $\pm$ 0.4 & 0.7  $\pm$ 0.2  $\pm$ 0.2 \\
\hline
\mTii [\GeVns{}] & \multicolumn{6}{c}{$>$50} \\
[\cmsTabSkip]
$\sumMT$ [\GeVns{}] & \multicolumn{2}{c}{200--250} & \multicolumn{2}{c}{250--300} & \multicolumn{2}{c}{$>$300} \\
[\cmsTabSkip]
$\nj$ & 0 & $\geq$1 & 0 & $\geq$1 & 0 & $\geq$1 \\
\hline
Misidentified \tauh & 18.2 $\pm$ 2.8 $\pm$ 9.5 & 18.1 $\pm$ 2.9 $\pm$ 6.0 & 3.7 $\pm$ 1.0 $\pm$ 2.2 & 2.7 $\pm$ 1.1 $\pm$ 0.5 & 1.1 $\pm$ 0.6 $\pm$ 0.6 & 2.9 $\pm$ 0.8 $\pm$ 1.6 \\
DY+jets & 1.1 $\pm$ 0.8 $\pm$ 0.2 & 3.3 $\pm$ 1.3 $\pm$ 0.7 & 0.5 $\pm$ 0.5 $\pm$ 0.1 & 1.0 $\pm$ 0.7 $\pm$ 0.1 & $<$0.7 & 1.3 $\pm$ 0.8 $\pm$ 0.5 \\
Top quark & 1.1 $\pm$ 0.3 $\pm$ 0.1 & 1.3 $\pm$ 0.2 $\pm$ 0.3 & 1.1 $\pm$ 0.2 $\pm$ 0.2 & 1.0 $\pm$ 0.2 $\pm$ 0.1 & 0.7 $\pm$ 0.2 $\pm$ 0.1 & 0.8 $\pm$ 0.2 $\pm$ 0.1 \\
Other SM & 2.0 $\pm$ 0.6 $\pm$ 0.3 & 1.2 $\pm$ 0.4 $\pm$ 0.2 & 0.9 $\pm$ 0.4 $\pm$ 0.1 & 0.2 $\pm$ 0.1 $\pm$ 0.1 & 0.3 $\pm$ 0.1 $\pm$ 0.1 & 0.5 $\pm$ 0.2 $\pm$ 0.2 \\
Total prediction & 22.5 $\pm$ 3.0 $\pm$ 9.5 & 23.9 $\pm$ 3.3 $\pm$ 6.0 & 6.2 $\pm$ 1.2 $\pm$ 2.2 & 4.9 $\pm$ 1.3 $\pm$ 0.5 & 2.1$\pm$ 0.6 $\pm$ 0.6 & 5.5 $\pm$ 1.2 $\pm$ 1.7 \\
Observed & 19 & 26 & 5 & 7 & 5 & 1 \\
$m(\staul)=100 \GeV$ & 1.6  $\pm$ 0.2  $\pm$ 0.3 & 0.4  $\pm$ 0.1  $\pm$ 0.1 & 1.4  $\pm$ 0.2  $\pm$ 0.2 & 0.4  $\pm$ 0.1  $\pm$ 0.1 & 1.7  $\pm$ 0.2  $\pm$ 0.4 & 0.7  $\pm$ 0.2  $\pm$ 0.2 \\
\hline
\end{tabular}
}
\label{tab:results2016}
\end{table*}

\begin{table*}[tbh]
\centering
\topcaption{Predicted background yields and observed event counts in \tautau SRs in 2017 data. For the background estimates with no events in the sideband or in the simulated sample, we calculate the 68\% \CL upper limit on the yield. The first and second uncertainties given are statistical and systematic, respectively. We also list the predicted signal yields corresponding to the purely left-handed model for a \PSGt mass of 100\GeV and a \PSGczDo mass of 1\GeV.}
\cmsTable{
\begin{tabular}{l  c  c  c  c  c  c }
\hline
\mTii [\GeVns{}] & \multicolumn{6}{c}{25--50} \\
[\cmsTabSkip]
$\sumMT$ [\GeVns{}] & \multicolumn{2}{c}{200--250} & \multicolumn{2}{c}{250--300} & \multicolumn{2}{c}{$>$300} \\
[\cmsTabSkip]
\nj & 0 & $\geq$1 & 0 & $\geq$1 & 0 & $\geq$1 \\
\hline
Misidentified \tauh & 18.6 $\pm$ 3.1 $\pm$ 3.6  & 9.4 $\pm$ 2.1 $\pm$ 1.7 & 2.7 $\pm$ 0.9 $\pm$ 1.0  & 1.1 $\pm$ 0.8 $\pm$ 0.3  & 0.5 $\pm$ 0.5 $\pm$ 0.1  & 1.9 $\pm$ 0.8 $\pm$ 1.3  \\
DY+jets & 5.0 $\pm$ 2.0 $\pm$ 0.7  & 1.5 $\pm$ 0.7 $\pm$ 0.2 & 1.9 $\pm$ 1.4 $\pm$ 0.5  & 0.6 $\pm$ 0.4 $\pm$ 0.2  & 1.1 $\pm$ 0.8 $\pm$ 0.3  & 1.0 $\pm$ 0.8 $\pm$ 0.1  \\
Top quark & 1.2 $\pm$ 0.6 $\pm$ 0.2   & 1.1 $\pm$ 0.5 $\pm$ 0.2  & 0.2 $\pm$ 0.1 $\pm$ 0.1   & 1.0 $\pm$ 0.6 $\pm$ 0.1   & 0.3 $\pm$ 0.3 $\pm$ 0.1   & 0.5 $\pm$ 0.2 $\pm$ 0.1   \\
Other SM & 1.9 $\pm$ 0.7 $\pm$ 0.4  & 1.4 $\pm$ 0.6 $\pm$ 0.4 & 0.7 $\pm$ 0.5 $\pm$ 0.1  & 0.5 $\pm$ 0.5 $\pm$ 0.1  & 0.5 $\pm$ 0.3 $\pm$ 0.1  & 0.6 $\pm$ 0.4 $\pm$ 0.3  \\
Total prediction & 26.7 $\pm$ 3.8 $\pm$ 3.7  & 13.3 $\pm$ 2.3 $\pm$ 1.8 & 5.5 $\pm$ 1.8 $\pm$ 1.1  & 3.2 $\pm$ 1.2 $\pm$ 0.4  & 2.4 $\pm$ 1.0 $\pm$ 0.4  & 4.0 $\pm$ 1.2 $\pm$ 1.4  \\
Observed & 40  & 12 & 6  & 5  & 1  & 2  \\
$m(\staul)=100\GeV$ & 1.7  $\pm$ 0.2  $\pm$ 0.2 & 0.4  $\pm$ 0.1  $\pm$ 0.1 & 1.3  $\pm$ 0.2  $\pm$ 0.2 & 0.3  $\pm$ 0.1  $\pm$ 0.1 & 1.4  $\pm$ 0.2  $\pm$ 0.4  & 0.6  $\pm$ 0.1  $\pm$ 0.2 \\
\hline
\mTii [\GeVns{}] & \multicolumn{6}{c}{$>$50} \\
[\cmsTabSkip]
$\sumMT$ [\GeVns{}] & \multicolumn{2}{c}{200--250} & \multicolumn{2}{c}{250--300} & \multicolumn{2}{c}{$>$300} \\
[\cmsTabSkip]
$\nj$ & 0 & $\geq$1 & 0 & $\geq$1 & 0 & $\geq$1 \\
\hline
Misidentified \tauh & 11.2 $\pm$ 2.3 $\pm$ 4.7  & 9.0 $\pm$ 2.6 $\pm$ 1.1 & 2.8 $\pm$ 1.3 $\pm$ 0.3  & 4.5 $\pm$ 1.4 $\pm$ 1.8  & 0.2 $\pm$ 0.7 $\pm$ 0.5  & 1.6 $\pm$ 0.8 $\pm$ 0.2  \\
DY+jets & 1.3 $\pm$ 0.8 $\pm$ 0.2  & 2.6 $\pm$ 1.0 $\pm$ 0.4 & 1.0 $\pm$ 0.6 $\pm$ 0.1  & 1.0 $\pm$ 0.6 $\pm$ 0.1  & $<$0.7  & 0.5 $\pm$ 0.5 $\pm$ 0.1  \\
Top quark & 0.8 $\pm$ 0.4 $\pm$ 0.1   & $<$0.2  & 0.3 $\pm$ 0.3 $\pm$ 0.1   & 0.1 $\pm$ 0.1 $\pm$ 0.1   & 0.4 $\pm$ 0.3 $\pm$ 0.1   & 0.6 $\pm$ 0.5 $\pm$ 0.2   \\
Other SM & 1.0 $\pm$ 0.4 $\pm$ 0.2  & 1.2 $\pm$ 0.6 $\pm$ 0.2 & 0.9 $\pm$ 0.5 $\pm$ 0.1  & 0.7 $\pm$ 0.5 $\pm$ 0.1  & 1.4 $\pm$ 0.7 $\pm$ 0.3  & 0.6 $\pm$ 0.4 $\pm$ 0.2  \\
Total prediction & 14.3 $\pm$ 2.5 $\pm$ 4.7  & 12.8 $\pm$ 2.8 $\pm$ 1.2 & 5.1 $\pm$ 1.5 $\pm$ 0.3  & 6.3 $\pm$ 1.6 $\pm$ 1.8  & 2.0 $\pm$ 1.0 $\pm$ 0.6  & 3.2 $\pm$ 1.1 $\pm$ 0.4  \\
Observed & 11  & 24 & 7 & 9  & 3  & 3   \\
$m(\staul)=100\GeV$ & 0.9  $\pm$ 0.2  $\pm$ 0.1  & 0.2  $\pm$ 0.1  $\pm$ 0.1 & 1.0  $\pm$ 0.2  $\pm$ 0.2  & 0.3  $\pm$ 0.1  $\pm$ 0.1 & 1.0  $\pm$ 0.2  $\pm$ 0.2  & 0.4  $\pm$ 0.1  $\pm$ 0.1 \\
\hline
\end{tabular}
}
\label{tab:results2017}
\end{table*}

\begin{table*}[tbh]
\centering
\topcaption{Predicted background yields and observed event counts in the most sensitive last bins of the BDT distributions in the \etau and \mutau final states, in data collected in 2016. The numbers in parentheses in the first row are the \PSGt and \PSGczDo masses corresponding to the signal model for left-handed \PSGt pair production that is used to train the BDT. In the bottom row, we list the corresponding predicted signal yields in the last bin of the BDT distribution. The first and second uncertainties given are statistical and systematic, respectively.}
\cmsTable{
\begin{tabular}{l  c  c  c  c  c  c }
\hline
BDT training & BDT(\mutau,100,1) & BDT(\mutau,150,1) & BDT(\mutau,200,1) & BDT(\etau,100,1) & BDT(\etau,150,1) & BDT(\etau,200,1) \\
\hline
Misidentified \tauh & 1.6 $\pm$ 0.8 $\pm$ 0.3  & 2.3 $\pm$ 1.0 $\pm$ 0.4  & 1.5 $\pm$ 0.8 $\pm$ 0.3  & 3.3 $\pm$ 1.1 $\pm$ 0.5  & 0.2 $\pm$ 0.4 $\pm$ 0.1  & 0.5 $\pm$ 0.7 $\pm$ 0.3 \\
DY+jets & $<$0.1  & 0.8 $\pm$ 0.8 $\pm$ 0.1  & $<$0.1    & $<$0.1   & $<$0.1   & 0.1 $\pm$ 0.1 $\pm$ 0.1  \\
Top quark &  0.3 $\pm$ 0.3 $\pm$ 0.1  & 1.8 $\pm$ 1.2 $\pm$ 0.2  & 1.7 $\pm$ 1.2 $\pm$ 0.6  & 0.2 $\pm$ 0.2 $\pm$ 0.1  & 0.2 $\pm$ 0.2 $\pm$ 0.1  & 1.4 $\pm$ 0.8 $\pm$ 2.0 \\
Other SM & 0.3 $\pm$ 0.3 $\pm$ 0.1  & 1.4 $\pm$ 0.6 $\pm$ 0.5  & 1.5 $\pm$ 0.6 $\pm$ 0.4  & 0.9 $\pm$ 0.5 $\pm$ 0.4  & 0.6 $\pm$ 0.4 $\pm$ 0.5  & 2.0 $\pm$ 0.7 $\pm$ 1.0 \\
Total prediction & 2.1 $\pm$ 0.9 $\pm$ 0.4  & 6.4 $\pm$ 1.8 $\pm$ 1.0  & 4.6 $\pm$ 1.6 $\pm$ 0.9  & 4.5 $\pm$ 1.3 $\pm$ 0.8  & 1.0 $\pm$ 0.6 $\pm$ 0.5  & 4.2 $\pm$ 1.3 $\pm$ 1.8 \\
Observed & 1  & 6  & 7  & 5  & 2  & 7 \\
Signal & 1.3 $\pm$ 0.4 $\pm$ 0.2  & 0.9 $\pm$ 0.2 $\pm$ 0.1  & 0.7 $\pm$ 0.1 $\pm$ 0.5  & 1.5 $\pm$ 0.4 $\pm$ 0.2  & 0.4 $\pm$ 0.1 $\pm$ 0.1  & 1.0 $\pm$ 0.1 $\pm$ 0.2 \\
\hline
\end{tabular}
\label{tab:resultsSemilep2016}
}
\end{table*}

\begin{table*}[tbh]
\centering
\topcaption{Predicted background yields and observed event counts in the most sensitive last bins of the BDT distributions in the \etau and \mutau final states, in data collected in 2017. The numbers in parentheses in the first row are the \PSGt and \PSGczDo masses corresponding to the signal model for left-handed \PSGt pair production that is used to train the BDT. In the bottom row, we list the corresponding predicted signal yields in the last bin of the BDT distribution. The first and second uncertainties given are statistical and systematic, respectively.}
\cmsTable{
\begin{tabular}{l  c  c  c  c  c  c}
\hline
BDT training & BDT(\mutau,100,1) & BDT(\mutau,150,1) & BDT(\mutau,200,1) & BDT(\etau,100,1) & BDT(\etau,150,1) & BDT(\etau,200,1) \\
\hline
Misidentified \tauh & 0.9 $\pm$ 0.5 $\pm$ 0.4  & $<$0.1  & $<$0.1  & 2.5 $\pm$ 0.9 $\pm$ 1.3  & 0.3 $\pm$ 0.3 $\pm$ 0.1  & $<$0.1  \\
DY+jets & 2.1 $\pm$ 2.1 $\pm$ 3.3  & $<$0.1   & $<$0.1  & $<$0.1  & $<$0.1   & $<$0.1  \\
Top quark & $<$0.1  & 0.9 $\pm$ 0.4 $\pm$ 0.8  & 0.6 $\pm$ 0.5 $\pm$ 0.5  & 0.3 $\pm$ 0.3 $\pm$ 0.1  & $<$0.1   & 0.2 $\pm$ 0.2 $\pm$ 0.2 \\
Other SM & $<$0.1  & 1.0 $\pm$ 0.7 $\pm$ 1.6  & 0.6 $\pm$ 0.6 $\pm$ 1.1  & 1.0 $\pm$ 0.7 $\pm$ 1.5  & 0.2 $\pm$ 0.2 $\pm$ 0.5  & 1.0 $\pm$ 0.6 $\pm$ 1.6 \\
Total prediction & 3.0 $\pm$ 2.2 $\pm$ 3.1  & 2.0 $\pm$ 1.0 $\pm$ 2.0  & 1.2 $\pm$ 0.7 $\pm$ 1.3  & 3.7 $\pm$ 1.1 $\pm$ 2.3  & 0.4 $\pm$ 0.4 $\pm$ 0.5  & 1.2 $\pm$ 0.7 $\pm$ 1.6 \\
Observed & 2  & 6  & 2  & 2  & 1  & 1 \\
Signal & 0.6 $\pm$ 0.3 $\pm$ 0.1  & 0.4 $\pm$ 0.1 $\pm$ 0.8  & 0.6 $\pm$ 0.1 $\pm$ 0.3  & 1.0 $\pm$ 0.4 $\pm$ 0.1  & 0.2 $\pm$ 0.1 $\pm$ 0.1  & 0.2 $\pm$ 0.1 $\pm$ 0.1  \\
\hline
\end{tabular}
}
\label{tab:resultsSemilep2017}
\end{table*}

\section {Summary}
\label{sec:summary}
A search for direct \PGt slepton (\PSGt) pair production has been performed in proton-proton collisions at a center-of-mass energy of 13\TeV in events with a \PGt lepton pair and significant missing transverse momentum. Search regions are defined using kinematic observables that exploit expected differences in discriminants between signal and background. The data used for this search correspond to an integrated luminosity of 77.2\fbinv collected in 2016 and 2017 with the CMS detector. No excess above the expected standard model background has been observed. Upper limits have been set on the cross section for direct \PSGt pair production for simplified models in which each \PSGt decays to a \PGt lepton and the lightest neutralino, with the latter being assumed to be the lightest supersymmetric particle. For purely left-handed \PSGt pair production, the analysis is most sensitive to a \PSGt mass of 125\GeV when the neutralino is nearly massless. The observed limit is a factor of 1.14 larger than the expected production cross section in this model. The limits observed for left-handed \PSGt pair production are the strongest obtained thus far for low values of the \PSGt mass. In a more optimistic, degenerate production model, in which both left- and right-handed \PSGt pairs are produced, we exclude \PSGt masses up to 150\GeV, again under the assumption of a nearly massless neutralino. These results represent the first exclusion reported for this model for low values of the \PSGt mass between 90 and 120\GeV.

\ifthenelse{\boolean{cms@external}}{\clearpage}{}

\begin{acknowledgments}
We congratulate our colleagues in the CERN accelerator departments for the excellent performance of the LHC and thank the technical and administrative staffs at CERN and at other CMS institutes for their contributions to the success of the CMS effort. In addition, we gratefully acknowledge the computing centers and personnel of the Worldwide LHC Computing Grid for delivering so effectively the computing infrastructure essential to our analyses. Finally, we acknowledge the enduring support for the construction and operation of the LHC and the CMS detector provided by the following funding agencies: BMBWF and FWF (Austria); FNRS and FWO (Belgium); CNPq, CAPES, FAPERJ, FAPERGS, and FAPESP (Brazil); MES (Bulgaria); CERN; CAS, MoST, and NSFC (China); COLCIENCIAS (Colombia); MSES and CSF (Croatia); RPF (Cyprus); SENESCYT (Ecuador); MoER, ERC IUT, PUT and ERDF (Estonia); Academy of Finland, MEC, and HIP (Finland); CEA and CNRS/IN2P3 (France); BMBF, DFG, and HGF (Germany); GSRT (Greece); NKFIA (Hungary); DAE and DST (India); IPM (Iran); SFI (Ireland); INFN (Italy); MSIP and NRF (Republic of Korea); MES (Latvia); LAS (Lithuania); MOE and UM (Malaysia); BUAP, CINVESTAV, CONACYT, LNS, SEP, and UASLP-FAI (Mexico); MOS (Montenegro); MBIE (New Zealand); PAEC (Pakistan); MSHE and NSC (Poland); FCT (Portugal); JINR (Dubna); MON, RosAtom, RAS, RFBR, and NRC KI (Russia); MESTD (Serbia); SEIDI, CPAN, PCTI, and FEDER (Spain); MOSTR (Sri Lanka); Swiss Funding Agencies (Switzerland); MST (Taipei); ThEPCenter, IPST, STAR, and NSTDA (Thailand); TUBITAK and TAEK (Turkey); NASU and SFFR (Ukraine); STFC (United Kingdom); DOE and NSF (USA).

\hyphenation{Rachada-pisek} Individuals have received support from the Marie-Curie program and the European Research Council and Horizon 2020 Grant, contract Nos.\ 675440, 752730, and 765710 (European Union); the Leventis Foundation; the A.P.\ Sloan Foundation; the Alexander von Humboldt Foundation; the Belgian Federal Science Policy Office; the Fonds pour la Formation \`a la Recherche dans l'Industrie et dans l'Agriculture (FRIA-Belgium); the Agentschap voor Innovatie door Wetenschap en Technologie (IWT-Belgium); the F.R.S.-FNRS and FWO (Belgium) under the ``Excellence of Science -- EOS" -- be.h project n.\ 30820817; the Beijing Municipal Science \& Technology Commission, No. Z181100004218003; the Ministry of Education, Youth and Sports (MEYS) of the Czech Republic; the Lend\"ulet (``Momentum") Program and the J\'anos Bolyai Research Scholarship of the Hungarian Academy of Sciences, the New National Excellence Program \'UNKP, the NKFIA research grants 123842, 123959, 124845, 124850, 125105, 128713, 128786, and 129058 (Hungary); the Council of Science and Industrial Research, India; the HOMING PLUS program of the Foundation for Polish Science, cofinanced from European Union, Regional Development Fund, the Mobility Plus program of the Ministry of Science and Higher Education, the National Science Center (Poland), contracts Harmonia 2014/14/M/ST2/00428, Opus 2014/13/B/ST2/02543, 2014/15/B/ST2/03998, and 2015/19/B/ST2/02861, Sonata-bis 2012/07/E/ST2/01406; the National Priorities Research Program by Qatar National Research Fund; the Ministry of Science and Education, grant no. 3.2989.2017 (Russia); the Programa Estatal de Fomento de la Investigaci{\'o}n Cient{\'i}fica y T{\'e}cnica de Excelencia Mar\'{\i}a de Maeztu, grant MDM-2015-0509 and the Programa Severo Ochoa del Principado de Asturias; the Thalis and Aristeia programs cofinanced by EU-ESF and the Greek NSRF; the Rachadapisek Sompot Fund for Postdoctoral Fellowship, Chulalongkorn University and the Chulalongkorn Academic into Its 2nd Century Project Advancement Project (Thailand); the Welch Foundation, contract C-1845; and the Weston Havens Foundation (USA).
\end{acknowledgments}

\bibliography{auto_generated}

\providecommand{\href}[2]{#2}\begingroup\raggedright\begin{thebibliography}{10}%
\makeatletter
\providecommand{\hrefCMSnoop }[0]{\@secondoftwo}%
\makeatother
\providecommand{\doi}{\texttt{doi:}\begingroup \urlstyle{tt}\Url}

\bibitem{Ramond:1971gb}
\hrefCMSnoop {}{P.~Ramond, ``Dual theory for free fermions'',} \textit{ Phys.
  Rev. D} \textbf{ 3} (1971) 2415,
\href{http://dx.doi.org/10.1103/PhysRevD.3.2415}{\doi{10.1103/PhysRevD.3.2415}}.

\bibitem{Golfand:1971iw}
\href {http://www.jetpletters.ac.ru/ps/1584/article_24309.pdf}{Y.~A. Gol'fand
  and E.~P. Likhtman, ``Extension of the algebra of {P}oincar\'{e} group
  generators and violation of {P} invariance'',} \textit{ JETP Lett.} \textbf{
  13} (1971)
323.

\bibitem{Neveu:1971rx}
\hrefCMSnoop {}{A.~Neveu and J.~H. Schwarz, ``Factorizable dual model of
  pions'',} \textit{ Nucl. Phys. B} \textbf{ 31} (1971) 86,
\href{http://dx.doi.org/10.1016/0550-3213(71)90448-2}{\doi{10.1016/0550-3213(71)90448-2}}.

\bibitem{Volkov:1972jx}
\hrefCMSnoop {}{D.~V. Volkov and V.~P. Akulov, ``Possible universal neutrino
  interaction'',} \textit{ JETP Lett.} \textbf{ 16} (1972)
438.

\bibitem{Wess:1973kz}
\hrefCMSnoop {}{J.~Wess and B.~Zumino, ``A {Lagrangian} model invariant under
  supergauge transformations'',} \textit{ Phys. Lett. B} \textbf{ 49} (1974)
  52,
\href{http://dx.doi.org/10.1016/0370-2693(74)90578-4}{\doi{10.1016/0370-2693(74)90578-4}}.

\bibitem{Wess:1974tw}
\hrefCMSnoop {}{J.~Wess and B.~Zumino, ``{Supergauge transformations in four
  dimensions}'',} \textit{ Nucl. Phys. B} \textbf{ 70} (1974) 39,
\href{http://dx.doi.org/10.1016/0550-3213(74)90355-1}{\doi{10.1016/0550-3213(74)90355-1}}.

\bibitem{Fayet:1974pd}
\hrefCMSnoop {}{P.~Fayet, ``{Supergauge invariant extension of the {H}iggs
  mechanism and a model for the electron and its neutrino}'',} \textit{ Nucl.
  Phys. B} \textbf{ 90} (1975) 104,
\href{http://dx.doi.org/10.1016/0550-3213(75)90636-7}{\doi{10.1016/0550-3213(75)90636-7}}.

\bibitem{Nilles:1983ge}
\hrefCMSnoop {}{H.~P. Nilles, ``{Supersymmetry, supergravity and particle
  physics}'',} \textit{ Phys. Rep.} \textbf{ 110} (1984) 1,
\href{http://dx.doi.org/10.1016/0370-1573(84)90008-5}{\doi{10.1016/0370-1573(84)90008-5}}.

\bibitem{Gildener:1976ai}
\hrefCMSnoop {}{E.~Gildener, ``Gauge symmetry hierarchies'',} \textit{ Phys.
  Rev. D} \textbf{ 14} (1976) 1667,
\href{http://dx.doi.org/10.1103/PhysRevD.14.1667}{\doi{10.1103/PhysRevD.14.1667}}.

\bibitem{Veltman:1976rt}
\hrefCMSnoop {}{M.~J.~G. Veltman, ``Second threshold in weak interactions'',}
  \textit{ Acta Phys. Polon. B} \textbf{ 8} (1977)
475.

\bibitem{'tHooft:1979bh}
\hrefCMSnoop {}{G.~'t~Hooft, ``Naturalness, chiral symmetry, and spontaneous
  chiral symmetry breaking'',} \textit{ NATO Sci. Ser. B} \textbf{ 59} (1980)
135.

\bibitem{Witten:1981nf}
\hrefCMSnoop {}{E.~Witten, ``Dynamical breaking of supersymmetry'',} \textit{
  Nucl. Phys. B} \textbf{ 188} (1981) 513,
\href{http://dx.doi.org/10.1016/0550-3213(81)90006-7}{\doi{10.1016/0550-3213(81)90006-7}}.

\bibitem{Farrar:1978xj}
\hrefCMSnoop {}{G.~R. Farrar and P.~Fayet, ``Phenomenology of the production,
  decay, and detection of new hadronic states associated with supersymmetry'',}
  \textit{ Phys. Lett. B} \textbf{ 76} (1978) 575,
\href{http://dx.doi.org/10.1016/0370-2693(78)90858-4}{\doi{10.1016/0370-2693(78)90858-4}}.

\bibitem{Goldberg:1983nd}
\hrefCMSnoop {}{H.~Goldberg, ``{Constraint on the Photino Mass from
  Cosmology}'',} \textit{ Phys. Rev. Lett.} \textbf{ 50} (1983) 1419,
  \href{http://dx.doi.org/10.1103/PhysRevLett.50.1419}{\doi{10.1103/PhysRevLett.50.1419}}.
[Erratum: Phys. Rev. Lett.103,099905(2009)].

\bibitem{Ellis:1983ew}
J.~R. Ellis\hrefCMSnoop {}{ {et~al.}, ``{Supersymmetric Relics from the Big
  Bang}'',} \textit{ Nucl. Phys.} \textbf{ B238} (1984) 453--476,
  \href{http://dx.doi.org/10.1016/0550-3213(84)90461-9}{\doi{10.1016/0550-3213(84)90461-9}}.
[,223(1983)].

\bibitem{darkmatter}
\hrefCMSnoop {}{G.~Jungman, M.~Kamionkowski, and K.~Griest, ``{Supersymmetric
  dark matter}'',} \textit{ Phys. Rept.} \textbf{ 267} (1996) 195,
  \href{http://dx.doi.org/10.1016/0370-1573(95)00058-5}{\doi{10.1016/0370-1573(95)00058-5}},
\href{http://www.arXiv.org/abs/hep-ph/9506380}{\texttt{arXiv:hep-ph/9506380}}.

\bibitem{WMAP}
\hrefCMSnoop {}{G.~Hinshaw {et~al.}, ``Nine-year {Wilkinson Microwave
  Anisotropy Probe (WMAP)} observations: cosmological parameter results'',}
  \textit{ Astrophys. J. Suppl.} \textbf{ 208} (2013) 19,
  \href{http://dx.doi.org/10.1088/0067-0049/208/2/19}{\doi{10.1088/0067-0049/208/2/19}},
\href{http://www.arXiv.org/abs/1212.5226}{\texttt{arXiv:1212.5226}}.

\bibitem{Griest:1990kh}
\hrefCMSnoop {}{K.~Griest and D.~Seckel, ``{Three exceptions in the calculation
  of relic abundances}'',} \textit{ Phys. Rev. D} \textbf{ 43} (1991) 3191,
\href{http://dx.doi.org/10.1103/PhysRevD.43.3191}{\doi{10.1103/PhysRevD.43.3191}}.

\bibitem{Vasquez}
\hrefCMSnoop {}{D.~A. Vasquez, G.~B\'elanger, and C.~Boehm, ``{Revisiting light
  neutralino scenarios in the MSSM}'',} \textit{ Phys. Rev. D} \textbf{ 84}
  (2011) 095015,
  \href{http://dx.doi.org/10.1103/PhysRevD.84.095015}{\doi{10.1103/PhysRevD.84.095015}},
\href{http://www.arXiv.org/abs/1108.1338}{\texttt{arXiv:1108.1338}}.

\bibitem{King}
\hrefCMSnoop {}{S.~F. King, J.~P. Roberts, and D.~P. Roy, ``Natural dark matter
  in {SUSY GUTs} with non-universal gaugino masses'',} \textit{ JHEP} \textbf{
  10} (2007) 106,
  \href{http://dx.doi.org/10.1088/1126-6708/2007/10/106}{\doi{10.1088/1126-6708/2007/10/106}},
\href{http://www.arXiv.org/abs/0705.4219}{\texttt{arXiv:0705.4219}}.

\bibitem{Battaglia:2001zp}
M.~Battaglia\hrefCMSnoop {}{ {et~al.}, ``{Proposed post-LEP benchmarks for
  supersymmetry}'',} \textit{ Eur. Phys. J. C} \textbf{ 22} (2001) 535,
  \href{http://dx.doi.org/10.1007/s100520100792}{\doi{10.1007/s100520100792}},
\href{http://www.arXiv.org/abs/hep-ph/0106204}{\texttt{arXiv:hep-ph/0106204}}.

\bibitem{Arnowitt:2008bz}
R.~L. Arnowitt\hrefCMSnoop {}{ {et~al.}, ``Determining the dark matter relic
  density in the minimal supergravity stau-neutralino coannihilation region at
  the {Large Hadron Collider}'',} \textit{ Phys. Rev. Lett.} \textbf{ 100}
  (2008) 231802,
  \href{http://dx.doi.org/10.1103/PhysRevLett.100.231802}{\doi{10.1103/PhysRevLett.100.231802}},
\href{http://www.arXiv.org/abs/0802.2968}{\texttt{arXiv:0802.2968}}.

\bibitem{Belanger:2012jn}
\hrefCMSnoop {}{G.~B\'elanger, S.~Biswas, C.~Boehm, and B.~Mukhopadhyaya,
  ``Light neutralino dark matter in the {MSSM} and its implication for {LHC}
  searches for staus'',} \textit{ JHEP} \textbf{ 12} (2012) 076,
  \href{http://dx.doi.org/10.1007/JHEP12(2012)076}{\doi{10.1007/JHEP12(2012)076}},
\href{http://www.arXiv.org/abs/1206.5404}{\texttt{arXiv:1206.5404}}.

\bibitem{Arganda:2018hdn}
\hrefCMSnoop {}{E.~Arganda, V.~Martin-Lozano, A.~D. Medina, and N.~Mileo,
  ``{Potential discovery of staus through heavy Higgs boson decays at the
  LHC}'',} \textit{ JHEP} \textbf{ 09} (2018) 056,
  \href{http://dx.doi.org/10.1007/JHEP09(2018)056}{\doi{10.1007/JHEP09(2018)056}},
\href{http://www.arXiv.org/abs/1804.10698}{\texttt{arXiv:1804.10698}}.

\bibitem{Alwall:2008ag}
\hrefCMSnoop {}{J.~Alwall, P.~Schuster, and N.~Toro, ``Simplified models for a
  first characterization of new physics at the {LHC}'',} \textit{ Phys. Rev. D}
  \textbf{ 79} (2009) 075020,
  \href{http://dx.doi.org/10.1103/PhysRevD.79.075020}{\doi{10.1103/PhysRevD.79.075020}},
\href{http://www.arXiv.org/abs/0810.3921}{\texttt{arXiv:0810.3921}}.

\bibitem{Alwall:2008va}
\hrefCMSnoop {}{J.~Alwall, M.-P. Le, M.~Lisanti, and J.~Wacker,
  ``Model-independent jets plus missing energy searches'',} \textit{ Phys. Rev.
  D} \textbf{ 79} (2009) 015005,
  \href{http://dx.doi.org/10.1103/PhysRevD.79.015005}{\doi{10.1103/PhysRevD.79.015005}},
\href{http://www.arXiv.org/abs/0809.3264}{\texttt{arXiv:0809.3264}}.

\bibitem{Alves:2011wf}
\hrefCMSnoop {}{{LHC New Physics Working Group}, ``Simplified models for {LHC}
  new physics searches'',} \textit{ J. Phys. G} \textbf{ 39} (2012) 105005,
  \href{http://dx.doi.org/10.1088/0954-3899/39/10/105005}{\doi{10.1088/0954-3899/39/10/105005}},
\href{http://www.arXiv.org/abs/1105.2838}{\texttt{arXiv:1105.2838}}.

\bibitem{Heister:2001nk}
\hrefCMSnoop {}{{ALEPH} Collaboration, ``{Search for scalar leptons in
  e$^{+}$e$^{-}$ collisions at center-of-mass energies up to 209 GeV}'',}
  \textit{ Phys. Lett. B} \textbf{ 526} (2002) 206,
  \href{http://dx.doi.org/10.1016/S0370-2693(01)01494-0}{\doi{10.1016/S0370-2693(01)01494-0}},
\href{http://www.arXiv.org/abs/hep-ex/0112011}{\texttt{arXiv:hep-ex/0112011}}.

\bibitem{Abdallah:2003xe}
\hrefCMSnoop {}{{DELPHI} Collaboration, ``{Searches for supersymmetric
  particles in e$^{+}$e$^{-}$ collisions up to 208 GeV and interpretation of
  the results within the MSSM}'',} \textit{ Eur. Phys. J. C} \textbf{ 31}
  (2003) 421,
  \href{http://dx.doi.org/10.1140/epjc/s2003-01355-5}{\doi{10.1140/epjc/s2003-01355-5}},
\href{http://www.arXiv.org/abs/hep-ex/0311019}{\texttt{arXiv:hep-ex/0311019}}.

\bibitem{Achard:2003ge}
\hrefCMSnoop {}{{L3} Collaboration, ``{Search for scalar leptons and scalar
  quarks at LEP}'',} \textit{ Phys. Lett. B} \textbf{ 580} (2004) 37,
  \href{http://dx.doi.org/10.1016/j.physletb.2003.10.010}{\doi{10.1016/j.physletb.2003.10.010}},
\href{http://www.arXiv.org/abs/hep-ex/0310007}{\texttt{arXiv:hep-ex/0310007}}.

\bibitem{Abbiendi:2003ji}
\hrefCMSnoop {}{{OPAL} Collaboration, ``{Search for anomalous production of
  dilepton events with missing transverse momentum in e$^{+}$e$^{-}$ collisions
  at $\sqrt{s}$ = 183 GeV to 209 GeV}'',} \textit{ Eur. Phys. J. C} \textbf{
  32} (2004) 453,
  \href{http://dx.doi.org/10.1140/epjc/s2003-01466-y}{\doi{10.1140/epjc/s2003-01466-y}},
\href{http://www.arXiv.org/abs/hep-ex/0309014}{\texttt{arXiv:hep-ex/0309014}}.

\bibitem{Aad:2014yka}
\hrefCMSnoop {}{{ATLAS Collaboration}, ``{Search for the direct production of
  charginos, neutralinos and staus in final states with at least two
  hadronically decaying taus and missing transverse momentum in pp collisions
  at $\sqrt{s}$ = 8\TeV with the ATLAS detector}'',} \textit{ JHEP} \textbf{
  10} (2014) 96,
  \href{http://dx.doi.org/10.1007/JHEP10(2014)096}{\doi{10.1007/JHEP10(2014)096}},
\href{http://www.arXiv.org/abs/1407.0350}{\texttt{arXiv:1407.0350}}.

\bibitem{Aad:2015eda}
\hrefCMSnoop {}{{ATLAS Collaboration}, ``{Search for the electroweak production
  of supersymmetric particles in $\sqrt{s} = 8$ TeV pp collisions with the
  ATLAS detector}'',} \textit{ Phys. Rev. D} \textbf{ 93} (2016) 052002,
  \href{http://dx.doi.org/10.1103/PhysRevD.93.052002}{\doi{10.1103/PhysRevD.93.052002}},
\href{http://www.arXiv.org/abs/1509.07152}{\texttt{arXiv:1509.07152}}.

\bibitem{SUS14022}
\hrefCMSnoop {}{{CMS Collaboration}, ``{Search for electroweak production of
  charginos in final states with two tau leptons in pp collisions at $
  \sqrt{s}=8 $ TeV}'',} \textit{ JHEP} \textbf{ 04} (2017) 018,
  \href{http://dx.doi.org/10.1007/JHEP04(2017)018}{\doi{10.1007/JHEP04(2017)018}},
\href{http://www.arXiv.org/abs/1610.04870}{\texttt{arXiv:1610.04870}}.

\bibitem{Sirunyan:2018vig}
\hrefCMSnoop {}{{CMS Collaboration}, ``{Search for supersymmetry in events with
  a $\tau$ lepton pair and missing transverse momentum in proton-proton
  collisions at $\sqrt{s} =$ 13 TeV}'',} \textit{ JHEP} \textbf{ 11} (2018)
  151,
  \href{http://dx.doi.org/10.1007/JHEP11(2018)151}{\doi{10.1007/JHEP11(2018)151}},
\href{http://www.arXiv.org/abs/1807.02048}{\texttt{arXiv:1807.02048}}.

\bibitem{geant4}
\hrefCMSnoop {}{{{GEANT4}} Collaboration, ``{\GEANTfour} --- a simulation
  toolkit'',} \textit{ Nucl. Instrum. Meth. A} \textbf{ 506} (2003) 250,
\href{http://dx.doi.org/10.1016/S0168-9002(03)01368-8}{\doi{10.1016/S0168-9002(03)01368-8}}.

\bibitem{Fuks:2013lya}
\hrefCMSnoop {}{B.~Fuks, M.~Klasen, D.~R. Lamprea, and M.~Rothering,
  ``{Revisiting slepton pair production at the Large Hadron Collider}'',}
  \textit{ JHEP} \textbf{ 01} (2014) 168,
  \href{http://dx.doi.org/10.1007/JHEP01(2014)168}{\doi{10.1007/JHEP01(2014)168}},
\href{http://www.arXiv.org/abs/1310.2621}{\texttt{arXiv:1310.2621}}.

\bibitem{Khachatryan:2016bia}
\hrefCMSnoop {}{{CMS Collaboration}, ``{The CMS trigger system}'',} \textit{
  JINST} \textbf{ 12} (2017) P01020,
  \href{http://dx.doi.org/10.1088/1748-0221/12/01/P01020}{\doi{10.1088/1748-0221/12/01/P01020}},
\href{http://www.arXiv.org/abs/1609.02366}{\texttt{arXiv:1609.02366}}.

\bibitem{Chatrchyan:2008zzk}
\hrefCMSnoop {}{{CMS Collaboration}, ``The {CMS} experiment at the {CERN}
  {LHC}'',} \textit{ JINST} \textbf{ 3} (2008) S08004,
\href{http://dx.doi.org/10.1088/1748-0221/3/08/S08004}{\doi{10.1088/1748-0221/3/08/S08004}}.

\bibitem{Sirunyan:2017ulk}
\hrefCMSnoop {}{{CMS Collaboration}, ``{Particle-flow reconstruction and global
  event description with the CMS detector}'',} \textit{ JINST} \textbf{ 12}
  (2017) P10003,
  \href{http://dx.doi.org/10.1088/1748-0221/12/10/P10003}{\doi{10.1088/1748-0221/12/10/P10003}},
\href{http://www.arXiv.org/abs/1706.04965}{\texttt{arXiv:1706.04965}}.

\bibitem{Sirunyan:2019kia}
\hrefCMSnoop {}{{CMS Collaboration}, ``{Performance of missing transverse
  momentum reconstruction in proton-proton collisions at $\sqrt{s} =$ 13 TeV
  using the CMS detector}'',} \textit{ JINST} \textbf{ 14} (2019) P07004,
  \href{http://dx.doi.org/10.1088/1748-0221/14/07/P07004}{\doi{10.1088/1748-0221/14/07/P07004}},
\href{http://www.arXiv.org/abs/1903.06078}{\texttt{arXiv:1903.06078}}.

\bibitem{Cacciari:2008gp}
\hrefCMSnoop {}{M.~Cacciari, G.~P. Salam, and G.~Soyez, ``The anti-\kt jet
  clustering algorithm'',} \textit{ JHEP} \textbf{ 04} (2008) 063,
  \href{http://dx.doi.org/10.1088/1126-6708/2008/04/063}{\doi{10.1088/1126-6708/2008/04/063}},
  \href{http://www.arXiv.org/abs/0802.1189}{\texttt{arXiv:0802.1189}}.

\bibitem{Cacciari:2011ma}
\hrefCMSnoop {}{M.~Cacciari, G.~P. Salam, and G.~Soyez, ``{FastJet user
  manual}'',} \textit{ Eur. Phys. J. C} \textbf{ 72} (2012) 1896,
  \href{http://dx.doi.org/10.1140/epjc/s10052-012-1896-2}{\doi{10.1140/epjc/s10052-012-1896-2}},
\href{http://www.arXiv.org/abs/1111.6097}{\texttt{arXiv:1111.6097}}.

\bibitem{pileup}
\href {https://cds.cern.ch/record/1751454}{{CMS Collaboration}, ``{Study of
  pileup removal algorithms for jets}'',} CMS Physics Analysis Summary
  CMS-PAS-JME-14-001, 2014.

\bibitem{Sirunyan:2017ezt}
\hrefCMSnoop {}{{CMS Collaboration}, ``{Identification of heavy-flavour jets
  with the CMS detector in pp collisions at 13 TeV}'',} \textit{ JINST}
  \textbf{ 13} (2018) P05011,
  \href{http://dx.doi.org/10.1088/1748-0221/13/05/P05011}{\doi{10.1088/1748-0221/13/05/P05011}},
\href{http://www.arXiv.org/abs/1712.07158}{\texttt{arXiv:1712.07158}}.

\bibitem{Khachatryan:2015hwa}
\hrefCMSnoop {}{{CMS Collaboration}, ``{Performance of electron reconstruction
  and selection with the CMS detector in proton-proton collisions at $\sqrt{s}
  = 8$\TeV}'',} \textit{ JINST} \textbf{ 10} (2015) P06005,
  \href{http://dx.doi.org/10.1088/1748-0221/10/06/P06005}{\doi{10.1088/1748-0221/10/06/P06005}},
\href{http://www.arXiv.org/abs/1502.02701}{\texttt{arXiv:1502.02701}}.

\bibitem{Sirunyan:2018}
\hrefCMSnoop {}{{CMS Collaboration}, ``{Performance of the CMS muon detector
  and muon reconstruction with proton-proton collisions at $\sqrt{s} =
  13$\TeV}'',} \textit{ JINST} \textbf{ 13} (2018) P06015,
  \href{http://dx.doi.org/10.1088/1748-0221/13/06/P06015}{\doi{10.1088/1748-0221/13/06/P06015}},
\href{http://www.arXiv.org/abs/1804.04528}{\texttt{arXiv:1804.04528}}.

\bibitem{Sirunyan:2018pgf}
\hrefCMSnoop {}{{CMS Collaboration}, ``{Performance of reconstruction and
  identification of $\tau$ leptons decaying to hadrons and $\nu_\tau$ in pp
  collisions at $\sqrt{s}=$ 13 TeV}'',} \textit{ JINST} \textbf{ 13} (2018)
  P10005,
  \href{http://dx.doi.org/10.1088/1748-0221/13/10/P10005}{\doi{10.1088/1748-0221/13/10/P10005}},
\href{http://www.arXiv.org/abs/1809.02816}{\texttt{arXiv:1809.02816}}.

\bibitem{Lecun98gradient-basedlearning}
\hrefCMSnoop {}{Y.~Lecun, L.~Bottou, Y.~Bengio, and P.~Haffner,
  ``Gradient-based learning applied to document recognition'',} in \textit{
  Proceedings of the IEEE}, p.~2278.
\newblock 1998.
\newblock \href{http://dx.doi.org/10.1109/5.726791}{\doi{10.1109/5.726791}}.

\bibitem{Alwall:2014hca}
J.~Alwall\hrefCMSnoop {}{ {et~al.}, ``{The automated computation of tree-level
  and next-to-leading order differential cross sections, and their matching to
  parton shower simulations}'',} \textit{ JHEP} \textbf{ 07} (2014) 079,
  \href{http://dx.doi.org/10.1007/JHEP07(2014)079}{\doi{10.1007/JHEP07(2014)079}},
\href{http://www.arXiv.org/abs/1405.0301}{\texttt{arXiv:1405.0301}}.

\bibitem{Nason:2004rx}
\hrefCMSnoop {}{P.~Nason, ``{A new method for combining NLO QCD with shower
  Monte Carlo algorithms}'',} \textit{ JHEP} \textbf{ 11} (2004) 040,
  \href{http://dx.doi.org/10.1088/1126-6708/2004/11/040}{\doi{10.1088/1126-6708/2004/11/040}},
\href{http://www.arXiv.org/abs/hep-ph/0409146}{\texttt{arXiv:hep-ph/0409146}}.

\bibitem{Frixione:2007vw}
\hrefCMSnoop {}{S.~Frixione, P.~Nason, and C.~Oleari, ``{Matching NLO QCD
  computations with parton shower simulations: the POWHEG method}'',} \textit{
  JHEP} \textbf{ 11} (2007) 070,
  \href{http://dx.doi.org/10.1088/1126-6708/2007/11/070}{\doi{10.1088/1126-6708/2007/11/070}},
\href{http://www.arXiv.org/abs/0709.2092}{\texttt{arXiv:0709.2092}}.

\bibitem{Alioli:2010xd}
\hrefCMSnoop {}{S.~Alioli, P.~Nason, C.~Oleari, and E.~Re, ``{A general
  framework for implementing NLO calculations in shower Monte Carlo programs:
  the POWHEG BOX}'',} \textit{ JHEP} \textbf{ 06} (2010) 043,
  \href{http://dx.doi.org/10.1007/JHEP06(2010)043}{\doi{10.1007/JHEP06(2010)043}},
\href{http://www.arXiv.org/abs/1002.2581}{\texttt{arXiv:1002.2581}}.

\bibitem{Re:2010bp}
\hrefCMSnoop {}{E.~Re, ``{Single-top $Wt$-channel production matched with
  parton showers using the POWHEG method}'',} \textit{ Eur. Phys. J. C}
  \textbf{ 71} (2011) 1547,
  \href{http://dx.doi.org/10.1140/epjc/s10052-011-1547-z}{\doi{10.1140/epjc/s10052-011-1547-z}},
\href{http://www.arXiv.org/abs/1009.2450}{\texttt{arXiv:1009.2450}}.

\bibitem{Sjostrand:2014zea}
T.~Sj{\"o}strand\hrefCMSnoop {}{ {et~al.}, ``An introduction to {PYTHIA}
  8.2'',} \textit{ Comput. Phys. Commun.} \textbf{ 191} (2015) 159,
  \href{http://dx.doi.org/10.1016/j.cpc.2015.01.024}{\doi{10.1016/j.cpc.2015.01.024}},
\href{http://www.arXiv.org/abs/1410.3012}{\texttt{arXiv:1410.3012}}.

\bibitem{Kalogeropoulos:2018cke}
\hrefCMSnoop {}{A.~Kalogeropoulos and J.~Alwall, ``The {SysCalc} code: A tool
  to derive theoretical systematic uncertainties'',} (2018).
\href{http://www.arXiv.org/abs/1801.08401}{\texttt{arXiv:1801.08401}}.

\bibitem{Khachatryan:2015pea}
\hrefCMSnoop {}{{CMS Collaboration}, ``{Event generator tunes obtained from
  underlying event and multiparton scattering measurements}'',} \textit{ Eur.
  Phys. J. C} \textbf{ 76} (2016) 155,
  \href{http://dx.doi.org/10.1140/epjc/s10052-016-3988-x}{\doi{10.1140/epjc/s10052-016-3988-x}},
\href{http://www.arXiv.org/abs/1512.00815}{\texttt{arXiv:1512.00815}}.

\bibitem{CMS-PAS-TOP-16-021}
\href {https://cds.cern.ch/record/2235192}{{CMS Collaboration},
  ``{Investigations of the impact of the parton shower tuning in Pythia 8 in
  the modelling of $\mathrm{t\overline{t}}$ at $\sqrt{s}=8$ and 13 TeV}'',} CMS
  Physics Analysis Summary CMS-PAS-TOP-16-021, 2016.

\bibitem{Sirunyan:2019dfx}
\hrefCMSnoop {}{{CMS Collaboration}, ``{Extraction and validation of a new set
  of CMS PYTHIA8 tunes from underlying-event measurements}'',} \textit{ Eur.
  Phys. J. C} \textbf{ 80} (2020) 4,
  \href{http://dx.doi.org/10.1140/epjc/s10052-019-7499-4}{\doi{10.1140/epjc/s10052-019-7499-4}},
\href{http://www.arXiv.org/abs/1903.12179}{\texttt{arXiv:1903.12179}}.

\bibitem{Ball:2014uwa}
\hrefCMSnoop {}{{NNPDF} Collaboration, ``{Parton distributions for the LHC Run
  II}'',} \textit{ JHEP} \textbf{ 04} (2015) 040,
  \href{http://dx.doi.org/10.1007/JHEP04(2015)040}{\doi{10.1007/JHEP04(2015)040}},
\href{http://www.arXiv.org/abs/1410.8849}{\texttt{arXiv:1410.8849}}.

\bibitem{Chatrchyan:2013xna}
\hrefCMSnoop {}{{CMS Collaboration}, ``{Search for top-squark pair production
  in the single-lepton final state in pp collisions at $\sqrt{s}$ = 8 TeV}'',}
  \textit{ Eur. Phys. J. C} \textbf{ 73} (2013) 2677,
  \href{http://dx.doi.org/10.1140/epjc/s10052-013-2677-2}{\doi{10.1140/epjc/s10052-013-2677-2}},
\href{http://www.arXiv.org/abs/1308.1586}{\texttt{arXiv:1308.1586}}.

\bibitem{MT2variable}
\hrefCMSnoop {}{C.~G. Lester and D.~J. Summers, ``Measuring masses of
  semi-invisibly decaying particle pairs produced at hadron colliders'',}
  \textit{ Phys. Lett. B} \textbf{ 463} (1999) 99,
  \href{http://dx.doi.org/10.1016/S0370-2693(99)00945-4}{\doi{10.1016/S0370-2693(99)00945-4}},
\href{http://www.arXiv.org/abs/hep-ph/9906349}{\texttt{arXiv:hep-ph/9906349}}.

\bibitem{MT2variable2}
\hrefCMSnoop {}{A.~Barr, C.~Lester, and P.~Stephens, ``{$m_{\mathrm{T2}}$: the
  truth behind the glamour}'',} \textit{ J. Phys. G} \textbf{ 29} (2003) 2343,
  \href{http://dx.doi.org/10.1088/0954-3899/29/10/304}{\doi{10.1088/0954-3899/29/10/304}},
\href{http://www.arXiv.org/abs/hep-ph/0304226}{\texttt{arXiv:hep-ph/0304226}}.

\bibitem{MT2variable3}
\hrefCMSnoop {}{C.~G. Lester and B.~Nachman, ``Bisection-based asymmetric
  {$m_{\mathrm{T2}}$} computation: a higher precision calculator than existing
  symmetric methods'',} \textit{ JHEP} \textbf{ 03} (2015) 100,
  \href{http://dx.doi.org/10.1007/JHEP03(2015)100}{\doi{10.1007/JHEP03(2015)100}},
\href{http://www.arXiv.org/abs/1411.4312}{\texttt{arXiv:1411.4312}}.

\bibitem{Tovey:2008ui}
\hrefCMSnoop {}{D.~Tovey, ``On measuring the masses of pair-produced
  semi-invisibly decaying particles at hadron colliders'',} \textit{ JHEP}
  \textbf{ 04} (2008) 034,
  \href{http://dx.doi.org/10.1088/1126-6708/2008/04/034}{\doi{10.1088/1126-6708/2008/04/034}},
\href{http://www.arXiv.org/abs/0802.2879}{\texttt{arXiv:0802.2879}}.

\bibitem{Polesello:2009rn}
\hrefCMSnoop {}{G.~Polesello and D.~Tovey, ``Supersymmetric particle mass
  measurement with boost-corrected contransverse mass'',} \textit{ JHEP}
  \textbf{ 03} (2010) 030,
  \href{http://dx.doi.org/10.1007/JHEP03(2010)030}{\doi{10.1007/JHEP03(2010)030}},
\href{http://www.arXiv.org/abs/0910.0174}{\texttt{arXiv:0910.0174}}.

\bibitem{CuencaAlmenar:2008zza}
\hrefCMSnoop {}{C.~Cuenca~Almenar, ``Search for the neutral {MSSM} {Higgs}
  bosons in the ditau decay channels at {CDF Run II}''}.
\newblock PhD thesis, Valencia U., IFIC, 2008.
\newblock
\href{http://dx.doi.org/10.2172/953708}{\doi{10.2172/953708}}.

\bibitem{Khachatryan:2014wca}
\hrefCMSnoop {}{{CMS Collaboration}, ``{Search for neutral MSSM Higgs bosons
  decaying to a pair of tau leptons in pp collisions}'',} \textit{ JHEP}
  \textbf{ 10} (2014) 160,
  \href{http://dx.doi.org/10.1007/JHEP10(2014)160}{\doi{10.1007/JHEP10(2014)160}},
\href{http://www.arXiv.org/abs/1408.3316}{\texttt{arXiv:1408.3316}}.

\bibitem{pvalue}
\hrefCMSnoop {}{L.~Demortier, ``{P} values and nuisance parameters'',} in
  \textit{ Statistical issues for {LHC} physics. {Proceedings, Workshop,
  PHYSTAT-LHC, Geneva, Switzerland, June} 27-29, 2007}, p.~23.
\newblock 2008.
\newblock
\href{http://dx.doi.org/10.5170/CERN-2008-001}{\doi{10.5170/CERN-2008-001}}.

\bibitem{Sirunyan:2018owv}
\hrefCMSnoop {}{{CMS Collaboration}, ``{Measurement of the differential
  Drell-Yan cross section in proton-proton collisions at $ \sqrt{s} $ = 13
  TeV}'',} \textit{ JHEP} \textbf{ 12} (2019) 059,
  \href{http://dx.doi.org/10.1007/JHEP12(2019)059}{\doi{10.1007/JHEP12(2019)059}},
\href{http://www.arXiv.org/abs/1812.10529}{\texttt{arXiv:1812.10529}}.

\bibitem{Sirunyan:2018ucr}
\hrefCMSnoop {}{{CMS Collaboration}, ``{Measurements of
  $\mathrm{t\overline{t}}$ differential cross sections in proton-proton
  collisions at $\sqrt{s}=$ 13 TeV using events containing two leptons}'',}
  \textit{ JHEP} \textbf{ 02} (2019) 149,
  \href{http://dx.doi.org/10.1007/JHEP02(2019)149}{\doi{10.1007/JHEP02(2019)149}},
\href{http://www.arXiv.org/abs/1811.06625}{\texttt{arXiv:1811.06625}}.

\bibitem{CMS-PAS-LUM-17-001}
\href {http://cds.cern.ch/record/2257069}{{CMS Collaboration}, ``{CMS}
  luminosity measurements for the 2016 data taking period'',} CMS Physics
  Analysis Summary CMS-PAS-LUM-17-001, 2017.

\bibitem{CMS-PAS-LUM-17-004}
\href {http://cds.cern.ch/record/2621960}{{CMS Collaboration}, ``{CMS
  luminosity measurement for the 2017 data-taking period at $\sqrt{s} =
  13~\mathrm{TeV}$}'',} CMS Physics Analysis Summary CMS-PAS-LUM-17-004, 2018.

\bibitem{Simp}
\hrefCMSnoop {}{{CMS Collaboration}, ``Interpretation of searches for
  supersymmetry with simplified models'',} \textit{ Phys. Rev. D} \textbf{ 88}
  (2013) 052017,
  \href{http://dx.doi.org/10.1103/PhysRevD.88.052017}{\doi{10.1103/PhysRevD.88.052017}},
\href{http://www.arXiv.org/abs/1301.2175}{\texttt{arXiv:1301.2175}}.

\bibitem{Junk:1999kv}
\hrefCMSnoop {}{T.~Junk, ``{Confidence level computation for combining searches
  with small statistics}'',} \textit{ Nucl. Instrum. Meth. A} \textbf{ 434}
  (1999) 435,
  \href{http://dx.doi.org/10.1016/S0168-9002(99)00498-2}{\doi{10.1016/S0168-9002(99)00498-2}},
\href{http://www.arXiv.org/abs/hep-ex/9902006}{\texttt{arXiv:hep-ex/9902006}}.

\bibitem{Read:2002hq}
\hrefCMSnoop {}{A.~L. Read, ``Presentation of search results: the
  $\text{CL}_\text{s}$ technique'',} \textit{ J. Phys. G} \textbf{ 28} (2002)
  2693,
\href{http://dx.doi.org/10.1088/0954-3899/28/10/313}{\doi{10.1088/0954-3899/28/10/313}}.

\bibitem{CMS-NOTE-2011-005}
\href {https://cds.cern.ch/record/1379837}{{The ATLAS Collaboration, The CMS
  Collaboration, The LHC Higgs Combination Group}, ``Procedure for the {LHC}
  {Higgs} boson search combination in {Summer} 2011'',} Technical Report
  CMS-NOTE-2011-005, ATL-PHYS-PUB-2011-11, 2011.

\bibitem{Cowan:2010js}
\hrefCMSnoop {}{G.~Cowan, K.~Cranmer, E.~Gross, and O.~Vitells, ``Asymptotic
  formulae for likelihood-based tests of new physics'',} \textit{ Eur. Phys. J.
  C} \textbf{ 71} (2011) 1554,
  \href{http://dx.doi.org/10.1140/epjc/s10052-011-1554-0}{\doi{10.1140/epjc/s10052-011-1554-0}},
  \href{http://www.arXiv.org/abs/1007.1727}{\texttt{arXiv:1007.1727}}.
  [Erratum: \DOI{10.1140/epjc/s10052-013-2501-z}].

\end{thebibliography}\endgroup

\cleardoublepage \appendix\section{The CMS Collaboration \label{app:collab}}\begin{sloppypar}\hyphenpenalty=5000\widowpenalty=500\clubpenalty=5000\vskip\cmsinstskip
\textbf{Yerevan Physics Institute, Yerevan, Armenia}\\*[0pt]
A.M.~Sirunyan$^{\textrm{\dag}}$, A.~Tumasyan
\vskip\cmsinstskip
\textbf{Institut f\"{u}r Hochenergiephysik, Wien, Austria}\\*[0pt]
W.~Adam, F.~Ambrogi, T.~Bergauer, J.~Brandstetter, M.~Dragicevic, J.~Er\"{o}, A.~Escalante~Del~Valle, M.~Flechl, R.~Fr\"{u}hwirth\cmsAuthorMark{1}, M.~Jeitler\cmsAuthorMark{1}, N.~Krammer, I.~Kr\"{a}tschmer, D.~Liko, T.~Madlener, I.~Mikulec, N.~Rad, J.~Schieck\cmsAuthorMark{1}, R.~Sch\"{o}fbeck, M.~Spanring, D.~Spitzbart, W.~Waltenberger, C.-E.~Wulz\cmsAuthorMark{1}, M.~Zarucki
% ...
\vskip\cmsinstskip
\dag: Deceased\\
1:  Also at Vienna University of Technology, Vienna, Austria\\
% ...
\end{sloppypar}
\end{document}